\documentclass[a4paper,11pt]{article}
\pdfoutput=1 

\usepackage{jcappub} 

\usepackage{graphicx}
\usepackage{dcolumn}
\usepackage{bm}
\usepackage{amssymb}
\usepackage{amsmath}
\usepackage{epsfig}
\usepackage{hyperref}
\usepackage{hhline}
\usepackage{color}
\usepackage{multirow}
\usepackage[normalem]{ulem}
\usepackage{slashed}
\usepackage[section]{placeins}
\definecolor{Zsug}{RGB}{0, 145, 33} 
\definecolor{Zcor}{RGB}{210, 0, 210}
\definecolor{Zque}{RGB}{0, 180, 190} 
 
\definecolor{jd}{rgb}{0.858, 0.188, 0.478}

\def\lapp{\mathrel{\rlap{\raise.5ex\hbox{$<$}}
                    {\lower.5ex\hbox{$\sim$}}}}
\def\gapp{\mathrel{\rlap{\raise.5ex\hbox{$>$}}
                    {\lower.5ex\hbox{$\sim$}}}}

{
{

\newcommand{\bmt}{\begin{pmatrix}}
\newcommand{\emt}{\end{pmatrix}}
\newcommand{\ba}{\begin{array}{c}}
\newcommand{\ea}{\end{array}}
\newcommand{\be}{\begin{equation}}
\newcommand{\ee}{\end{equation}}
\newcommand{\bea}{\begin{eqnarray}}
\newcommand{\eea}{\end{eqnarray}}

\newcommand{\bi}{\begin{itemize}}
\newcommand{\ei}{\end{itemize}}

\newcommand{\baz}{\begin{array}{cc}}

\newcommand{\mathsym}[1]{{}}

\newcommand{\bt}{\begin{tabular}}
\newcommand{\et}{\end{tabular}}

\newcommand{\benu}{\begin{enumerate}}
\newcommand{\eenu}{\end{enumerate}}

\newcommand{\bav}{\begin{array}{cccc}}

\affiliation[a]{Department of Physics, Indian Institute of Technology Guwahati, North Guwahati, Assam- 781039, India}
\affiliation[b]{State Key Laboratory of Theoretical Physics, Institute of Theoretical Physics, Chinese Academy of Sciences, Beijing 100190, P.R. China}

\author[a]{Basabendu Barman,}
\emailAdd{bb1988@iitg.ernet.in}
\author[a]{Subhaditya Bhattacharya,}
\emailAdd{subhab@iitg.ernet.in}
\author[b]{and Mohammadreza Zakeri}
\emailAdd{mzake001@ucr.edu}

\abstract{
An $SU(2)_N$ extension ($N$ stands for neutral) of the Standard Model (SM) is proposed with an additional $U(1)=S^{'}$ global symmetry, which stabilizes the lightest of the vector boson ($X,\bar{X}$) as dark matter (DM) through unbroken $S=T_{3N}+S^{'}$. The field content of the model is motivated to address neutrino mass generation, a possible unification to $SU(7)$, along with spontaneous symmetry breaking of $SU(2)_N$ resulting in massive gauge bosons. None of the SM particles are charged under $SU(2)_N$ and therefore $X,\bar{X}$ do not have a direct coupling to the visible sector besides a Higgs portal, which is tiny to avoid any conflict with Higgs data. We show that, a large kinematic region of this model allows the neutral component of $SU(2)_N$ scalar triplet and heavy neutrinos introduced here to become additional DM components. In this paper we explore the viability of such multipartite DM parameter space, including non-zero DM-DM interactions, to comply with relic density and direct search constraints. We also demonstrate that the model may yield hadronically quiet single lepton and two lepton signatures with missing energy at the Large Hadron Collider (LHC) that can be accessed with high luminosity.}

\begin{document}

\renewcommand*{\thefootnote}{\fnsymbol{footnote}}
\title{\bf Multipartite Dark Matter in $SU(2)_N$ extension of Standard Model and signatures at the LHC\\} 


\maketitle
\flushbottom


\setcounter{footnote}{0}
\renewcommand*{\thefootnote}{\arabic{footnote}}

\section{Introduction}
\label{sec:intro}

Motivation for a particle dark matter (DM) comes from different astrophysical/cosmological evidences like rotation curves of galaxies~\cite{zwicky,rubin}, anisotropies in CMBR~\cite{cmbr}, observations in Bullet cluster~\cite{bullet} etc., which triggers physics beyond the Standard Model (SM). DM as fundamental particles necessarily lack electromagnetic interactions, but can have different properties depending on the masses (cold, warm or hot) and interaction strength. The major classification goes as $(i)$ weakly interacting massive particle (WIMP)~\cite{Kolb:1990vq,Jungman:1995df}, $(ii)$ feebly interacting massive particles (FIMP)~\cite{Hall:2009bx} and $(iii)$ strongly interacting massive particles (SIMP)~\cite{Hochberg:2014dra}. Stabilization of DM (or the decay life time as large as the age of the universe) is also required to fit the observed DM relic density ($\Omega h^2\sim 0.1$)~\cite{WMAP,Ade:2015xua} and is achieved by an additional unbroken symmetry under which the dark sector particles transform non-trivially while the SM particles do not. DM can also have any intrinsic spin and therefore can be a scalar, fermion or a vector boson. Vector boson dark matter (VBDM) models are not abundant in literature as it is more involved with the necessity of extending SM gauge group: $SU(3)_c\times SU(2)_L \times U(1)_Y$. The paper aims to discuss one such possibility of a non-abelian vector boson as a DM and its consequences in relic density, direct and collider search prospects.

The additional gauge bosons must be electromagnetic charge neutral to be qualified as DM. The simplest possibility is to assume an abelian $U(1)_X$ extension and make sure it remains hypercharge zero~\cite{Farzan:2012hh,Baek:2013nr,Duch:2015jta,Davoudiasl:2013jma}. Simplest non-abelian extension can then be assumed as $SU(2)$ (see for example,~\cite{Hambye:2008bq,Arcadi:2017kky}). How to save the gauge bosons from SM hypercharge is a matter of group theoretic manipulation and is not unique. One way of achieving so, is described in this paper, following the analysis in~\cite{Fraser:2014yga}. However, the requirement of a VBDM, also demands the breaking of the additional gauge group completely through spontaneous symmetry breaking (SSB). Therefore the symmetry required to keep DM stable is often as additional one and we assume it to be an unbroken $U(1)$ in this analysis. The particle content and their transformation properties provide the phenomenology through which DM can interact and therefore freeze out or freeze-in. The guiding principle for choosing the additional fields here is motivated by (i) neutrino mass generation, (ii) successful SSB to generate massive gauge bosons and (iii) a possible high-scale realization of the model in $SU(7)$~\cite{Ma:2013nga}. Together, they point out to a completely different DM phenomenology from the case of VBDM framework addressed in~\cite{Bhattacharya:2011tr,Barman:2017yzr}.

The key feature of this model is to assume that SM particles do not transform under additional $SU(2)_N$ symmetry unlike the case in~\cite{DiazCruz:2010dc}. Therefore, the VBDM lacks a direct search cross-section except for the Higgs portal which is constrained from Higgs data to avoid large mixing. This indeed helps the model to be allowed in a large parameter space from non-observation of DM in direct searches, for example in PANDA data~\cite{Cui:2017nnn}. Another interesting aspect of this analysis is to show the presence of scalar triplet as additional DM component apart from the VBDM as pointed out in~\cite{Fraser:2014yga}. The scalar DM will again have interactions to SM via Higgs portal (not necessarily small) and has direct search prospect. The analysis explores such a two-component DM parameter space of the model poised with non-zero DM-DM interactions. 

The model also assumes the presence of not-so-heavy neutrinos to generate light neutrino masses through {\it inverse seesaw} mechanism. This allows, in one hand, the heavy neutrinos to be stable and contribute as DM, while on the other hand, they can be produced in the Large Hadron Collider (LHC) hadronically quiet single lepton and hadronically quiet opposite sign dilepton (OSD) channel, with missing energy. This serves as one of the important directions of this analysis, which was not addressed in the earlier proposal of the model~\cite{Fraser:2014yga}. The SM background can be tamed down to some extent by large missing energy cut ($\slashed{E_T}$) and $H_T$ cut ($\slashed{H_T}$). The discovery potential thus can be reached with a high luminosity. Generation of light neutrino masses (with not-so-heavy neutrinos ($\sim \mathcal{O}(500)\rm{GeV}$)) also necessitates the VBDM ($X,\bar{X}$) to be degenerate with the third gauge boson component $X_3$. Therefore, co-annihilations play a crucial part on top of annihilation  for VBDM (this was also not taken into account in the earlier analysis~\cite{Fraser:2014yga}) and bridges a connection between the neutrino and the dark sector. For more illuminating discussions on this, see for example~\cite{Boehm:2006mi,Ma:2006km}.        

The paper is organised as follows: we discuss the model in Sec.~\ref{sec:model}, neutrino mass generation mechanism in Sec.~\ref{sec:neutrino mass}, followed by the vector boson DM analysis in Sec.~\ref{sec:DM pheno}. The multipartite DM features are elaborated in subsection.~\ref{sec:degenerate DM} and~\ref{sec:x-delta DM}. Collider signatures are analysed in Sec.~\ref{sec:collider pheno}. Finally we conclude in Sec.~\ref{sec:conclusion}.

\section{The Model}
\label{sec:model}

The model under consideration has an extended gauge group $SU(2)_N$, where $N$ stands for neutral\footnote{The electromagnetic charge neutrality of the vector bosons under this 
gauge group is ensured through spontaneous symmetry breaking, as discussed in~\cite{Fraser:2014yga}.}. The main idea is to have the lightest of the gauge bosons as a DM candidate. 
The particle content is chosen here minimally to have a spontaneous symmetry breaking (SSB) of $SU(2)_N$ to yield massive gauge bosons and also to have a successful neutrino 
mass generation as proposed in~\cite{Fraser:2014yga}. An important difference from the $SU(2)_N$ model proposed in~\cite{DiazCruz:2010dc,Bhattacharya:2011tr}, is that all of the 
SM fermions here are singlet under $SU(2)_N$. The stability of DM is ensured by an added global $U(1)$ symmetry ($S^{'}$), imposed on the new particles (as in~\cite{Fraser:2014yga}), 
so that $S=S^{'}+T_{3N}$ remains unbroken. The stability of DM under an unbroken global continuous symmetry may however be broken by the presence of a possible quantum theory 
of  gravity~\cite{Mambrini:2015sia}, which will therefore have observable effects in gamma ray, X-ray, neutrino and CMB data through DM decay, thus constraining such a case. 
The analysis in ~\cite{Mambrini:2015sia} shows that the limits on DM mass scale can be as stringent as few MeVs, by assuming SM gauge non-invariant dimension five effective 
operators suppressed by Planck scale\footnote {Gauge invariance requires higher dimensional effective operators, where the limit on DM mass becomes much more relaxed.}, which explicitly breaks the DM symmetry. However, due to the lack of our knowledge of a possible quantum theory of gravity, and the fact that $S$ is generated by a combination of global symmetry $S^{'}$ 
together with $T_{3N}$ (isospin of a broken gauge symmetry), we assume $S$ to be unbroken up to Planck scale and avoid such constraints.

The new particles and their charges under $SU(3)_C \,\otimes$ $SU(2)_L\, \otimes$ $U(1)_Y \otimes$ $SU(2)_N \otimes\, S^{'}$ are given as: 

\begin{align*}
\text{Three SU}(2)_N \text{ gauge bosons: }&\qquad\qquad X_{1,2,3}\equiv (1, 1, 0, 3, 0),\\ \\
\text{Three Dirac fermion doublets: }&\qquad\qquad  n=(n_1, n_2)_{L,R}\equiv (1, 1, 0, 2, \frac{1}{2}),\\ \\
\text{One scalar doublet: }&\qquad\qquad  \chi=(\chi_1, \chi_2) \equiv (1, 1, 0, 2, \frac{1}{2}),\\ \\
\text{One scalar bi-doublet: }&\qquad\qquad  \zeta=\begin{pmatrix}
\zeta_1^0 & \zeta_2^0\\
\zeta_1^- & \zeta_2^-
\end{pmatrix} \equiv (1, 2, -\frac{1}{2}, 2, -\frac{1}{2}),
\end{align*}

where $\zeta$ transforms (vertically) under $SU(2)_L$ and (horizontally) under $SU(2)_N$. Furthermore, an $SU(2)_N$  scalar triplet ($\Delta$) is introduced: 

\begin{align*}
\Delta = \begin{pmatrix}
\Delta_2/\sqrt{2} & \Delta_3\\
\Delta_1 & -\Delta_2/\sqrt{2}
\end{pmatrix} \equiv (1, 1, 0, 3, -1),
\end{align*}

for generating neutrino masses, which will be discussed in the next section. The crucial construct of the model lies in the choice of $S^{'}$  charges, which will be clear in a moment. Note that the only additional fermions introduced here are three families of a vector like $SU(2)_N$ doublet $n$. This mediates the interactions of the dark sector (non-zero $S$ charged particles as noted below) with the SM sector. This was the reason that the authors in~\cite{Fraser:2014yga} proposed the model as vector boson dark matter with {\it leptonic} connection. The field content of this model is essentially motivated by a unified $SU(7)$ prescription to generate neutrino mass and to have a stable DM as described in~\cite{Ma:2013nga}. Also, note that, the presence of left chiral heavy neutrinos $(n_1,n_2)_L$ plays an important role in achieving light neutrino masses through {\it inverse seesaw mechanism}, resulting in $m_n \sim \mathcal{O}$(TeV) and therefore allowing to explore them at the colliders.

Spontaneous symmetry breaking of $SU(2)_N\otimes S^{'}$ to $S=S'+T_{3N}$ happens via the non-zero vacuum expectation value (VEV) of $SU(2)_N$ scalar doublet: $\langle\chi_2\rangle = u_2$. $S$ charge assignment for the new particles is given as:

\begin{align*}
n_1, \chi_1 \sim +1, \quad &\quad n_2, \chi_2, \zeta_2, \Delta_3 \sim 0,\quad\quad
\zeta_1, \Delta_2 \sim -1, \quad\quad \Delta_1 \sim -2,\\
&X(\overline{X}) = \frac{X_1 \mp iX_2}{
\sqrt{2}} \sim \pm 1, \quad\quad Z' = X_3 \sim 0.
\end{align*}

All the SM particles have zero $S$ charge. Therefore, particles with non-zero $S$ charge will be protected from decaying into the SM. We can assume $X$ to be the lightest of the particles with non-zero $S$ charge, and therefore a possible DM candidate. Furthermore, $\Delta_{1,2,3}$ scalars can become kinematically stable in certain regions of parameter space~\cite{Fraser:2014yga}, and be part of a multi-component DM framework. We will investigate this possibility in details.

The three other scalars which acquire VEV are: $\langle \zeta_2^0\rangle =v_2$, $\langle\Delta_3\rangle =u_3$, and $\langle \phi^0\rangle=v_1$. Note that this assignment is different from that in~\cite{Bhattacharya:2011tr} where $\langle \Delta_1^0 \rangle$ is also non-zero. Therefore, the $X_{1,2}$ bosons will have equal masses in this model, and more importantly $S=S'+T_{3N}$ global symmetry remains unbroken unlike in~\cite{Bhattacharya:2011tr}. The masses of the gauge bosons are given by:

\begin{equation}
\begin{split}
m_W^2 = \frac{1}{2} g_2^2\left(v_1^2+v_2^2\right), \quad\quad
m_{X}^2 = \frac{1}{2}g_N^2\left(u_2^2+v_2^2+2u_3^2\right),  \quad\quad
m_{Z'}^2 \simeq  \frac{1}{2}g_N^2\left(u_2^2+v_2^2+4u_3^2\right),
\end{split}
\end{equation}

where $Z-Z'$ mixing matrix is given by:

\begin{equation}
m_{Z,Z'}^2 = \frac{1}{2}\begin{pmatrix}
\left(g_1^2+g_2^2\right)\left(v_1^2+v_2^2\right) & -g_N\sqrt{g_1^2+g_2^2}\, v_2^2\\
-g_N\sqrt{g_1^2+g_2^2}\, v_2^2 & g_N^2\left(u_2^2 + v_2^2 + 4u_3^2\right)
\end{pmatrix}.
\end{equation}

To ensure small $Z$-$Z^{'}$ mixing~\cite{Andreev:2014fwa}, we assume $v_2\ll u_2$. Furthermore, $u_3$ is assumed to be small which breaks the lepton number global symmetry ($L$) to lepton parity ($(-1)^L$) as explained in Sec.~\ref{sec:neutrino mass}. Therefore, the $X$ boson masses are nearly degenerate, i.e. $m_{Z'}(m_{X_3})\simeq m_X$. This still makes the model phenomenologically viable in a large parameter space as $Z'$ doesn't have a tree level coupling to SM. This hides $Z'$ of this model from being observed at the LHC, and adds to the the freedom of choosing $m_{Z'}$ as a free parameter. This should again be contrasted to the case in~\cite{Barman:2017yzr}, where there is a minimum limit on $M_{X_{1,2,3}} \geqslant 1$ TeV, for the degenerate vector boson DM case to respect the bound from $Z^{'}$ search data.

The scalar potential of this model remains the same as in the original proposal~\cite{Fraser:2014yga} and noted in Appendix-A of the paper. We also do not address the details of SSB and the physical scalars appearing in this framework. We would however, provide the approximate SM-like Higgs eigenstate:

\begin{equation}
h = -\phi_{2R}^0 + \left(\frac{f_5\, v_1}{\lambda_4\, u_2}\right)\,\chi_{2R} - \left( \frac{2 f_5^2\, v_1}{f_4\lambda_4\, v_2}\right)\, \zeta_{2R}^0,
\label{eq:higgsmass}
\end{equation} 

with

\begin{equation}
m_h^2 \simeq \frac{2v_1^2 \left(\lambda_2\lambda_4 - f_5^2\right)}{\lambda_4}\label{eq:M-Hig}.
\end{equation}

\begin{figure}[htb!]
\centering
\includegraphics[scale=0.4]{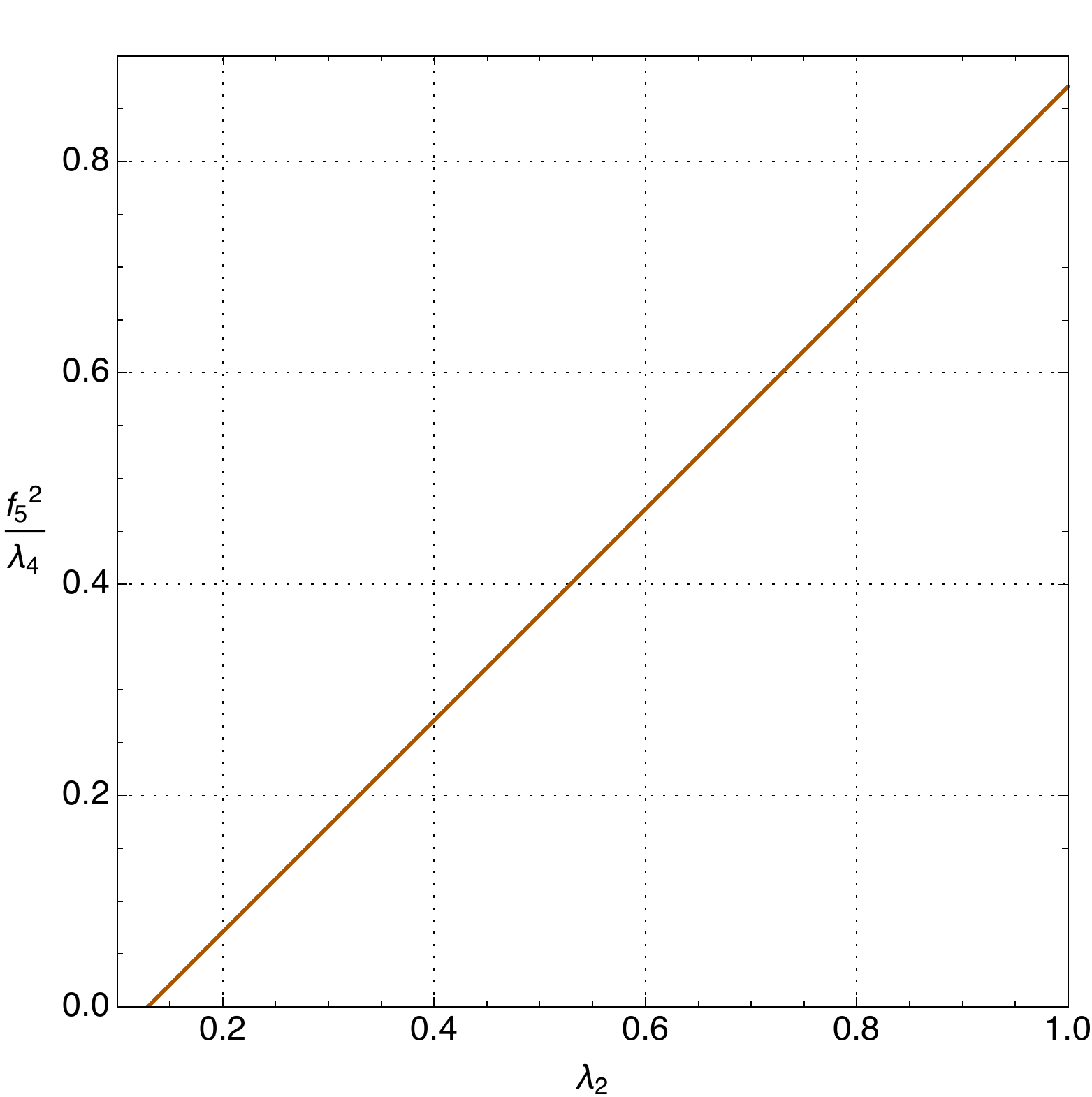}
\caption{$\frac{f_5^2}{\lambda_4}$ plotted against $\lambda_2$ using the Higgs mass constraint in Eq.~\ref{eq:M-Hig}. Note that $\lambda_{2,4} > 0$ in order to ensure the stability of the scalar potential. 
\label{fig:f5}}
\end{figure}

All the dimensionless couplings are borrowed from the scalar potential. One important point is, to note that, from the current knowledge of the Higgs mass (125 GeV) we will get a relation of $\frac{f_5^2}{\lambda_4}$ with $\lambda_2$ (using Eq.~\ref{eq:higgsmass}) as shown in Fig.~\ref{fig:f5}.  Note that Fig.~\ref{fig:f5} does not strictly constrain $f_5^2/\lambda_4$ (can be large with larger $\lambda_2$). $f_5^2/\lambda_4$ essentially determines $SU(2)_N$ Higgs components $\left(\chi_{2R},\zeta_{2R}\right)$ to be present in SM-like Higgs and this will be limited from the production and decay of Higgs observed at the LHC. In the limit of heavier $SU(2)_N$ fields, we choose a moderate limit on $f_5/\lambda_4$: \{0.1-0.6\} for further analysis.

\section{Neutrino Mass}
\label{sec:neutrino mass}

One of the important features of the model is to generate neutrino mass successfully and thus addressing dark matter and neutrinos under one umbrella. The scalar bi-doublet ($\zeta$), which acts as a mediator between the dark and visible sectors, also generates masses for neutrinos. The Yukawa terms responsible for neutrino mass generation are given by:

\bea
f_{\zeta} &\left[ \left(\overline{\nu}_L \zeta_1^0 + \overline{e}_L \zeta_1^- \right) n_{1R} + \left( \overline{\nu}_L \zeta_2^0 + \overline{e}_L \zeta_2^- \right) n_{2R} \right] \label{f-zeta}\\
f_{\Delta} &\left[ n_1 n_1 \Delta_1 + \left(n_1 n_2 + n_2 n_1\right) \Delta_2/\sqrt{2} - n_2 n_2 \Delta_3  \right],\label{f-delta}
\label{eq:yukawaint}
\eea

where in the second line $nn$ includes both of $n_Ln_L$ and $n_Rn_R$. The lepton number is conserved in~\eqref{f-zeta} with $n$ carrying $L=1$, and is broken to lepton parity, i.e. $(-1)^L$ by the $nn$ terms in~\eqref{f-delta}. After SSB, we have the following mass terms for the neutrinos:

\begin{align}
f_{\zeta}\, v_2\, \overline{\nu}_L n_{2R} - f_{\Delta}^L\, u_3\, n_{2L} n_{2L} - f_{\Delta}^R\, u_3\, n_{2R} n_{2R}+ \text{h.c.}
\end{align} 
where $f_{\zeta}$ and $f_{\Delta}$ are $3\times 3$ matrices, and the neutrino mass matrix in the $\left(\overline{\nu}_L, n_{2R}, \overline{n}_{2L}\right)$ basis is given by:

\begin{align}
M_{\nu} = \begin{pmatrix}
0 & m_D & 0\\
m_D & m_2' & M \\
0 & M & m_2
\end{pmatrix},
\end{align}

where each entry is a $3\times 3$ matrix with $m_D = f_{\zeta}\, v_2$, $m_2' = f_{\Delta}^R\, u_3$, $m_2 = {f_{\Delta}^{L}}^*\, u_3$, and $M$ is a free Dirac mass term in $M \left(\overline{n}_{2L} n_{2R} + \overline{n}_{2R} n_{2L}\right)$.  The inverse seesaw neutrino mass is thus generated and given by:

\bea
m_{\nu} \simeq \frac{m_D^2\, m_2}{M^2} = f_{\zeta}^2 f_{\Delta}\, \left(\frac{v_2}{M}\right)^2 u_3.
\label{eq:correctnumass}
\eea

\begin{figure}[htb!]
$$
\includegraphics[height=6.cm]{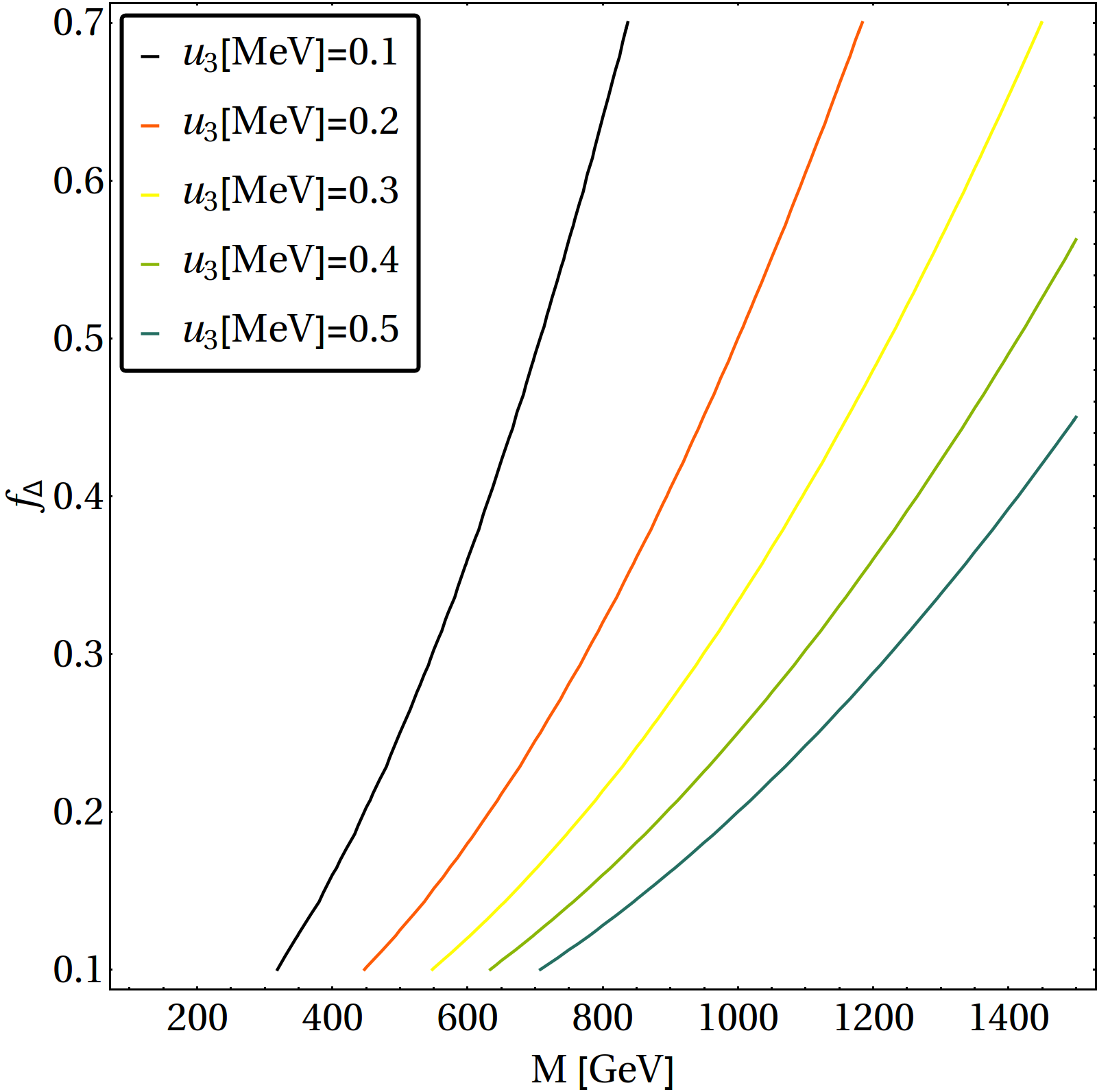}
\includegraphics[height=6.cm]{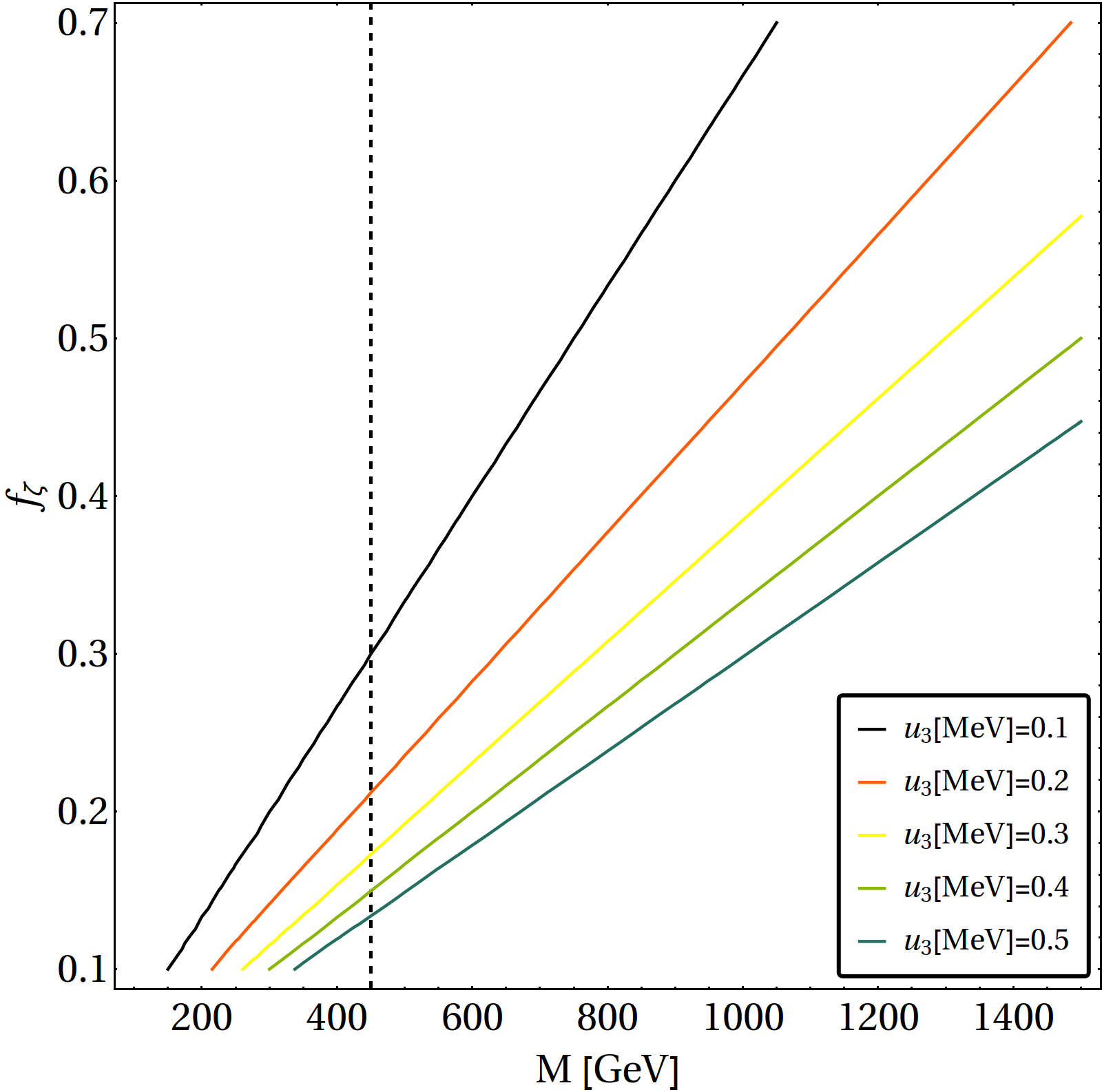}
$$
\centering
$$
\includegraphics[height=6.cm]{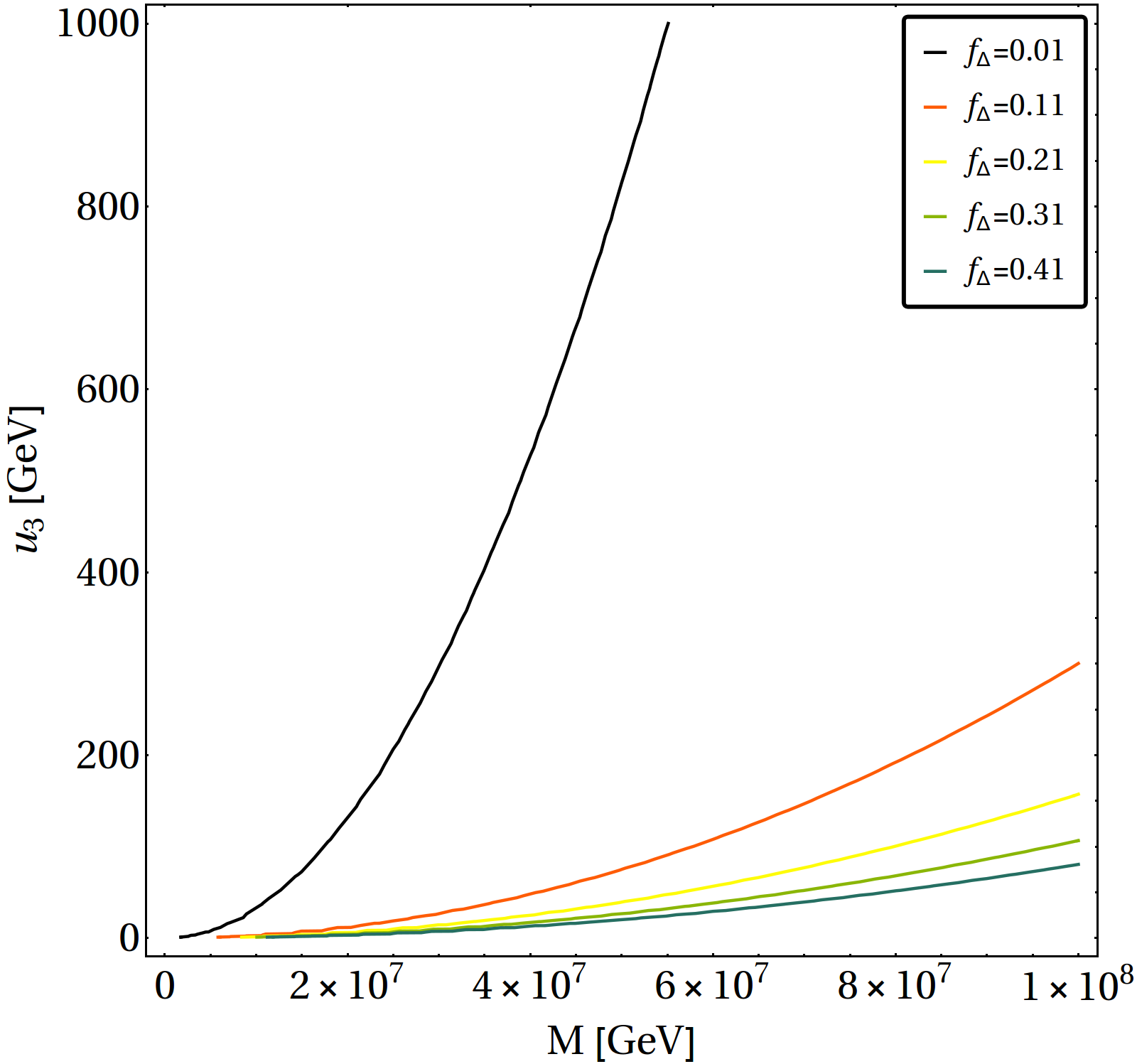}
$$
\caption{Top Left: $f_\Delta$ versus heavy neutrino mass $M$ ($\sim$ $\mathcal{O}$ (hundreds of GeVs)) for different choices of $u_3$ ($\sim$ MeV) to keep $m_{\nu}\sim~0.1~eV$ with $f_\zeta\sim\mathcal{O}(1)$; Top Right: $f_{\zeta}$ versus heavy neutrino mass $M$($\sim\mathcal{O}$ (hundreds of GeVs)) corresponding to different values of the VEV $u_3$ to obtain right neutrino mass for $f_{\Delta}\sim\mathcal{O}(1)$. The black dashed line shows the mass of the heavy neutral chosen for selecting the BPs (Table.~\ref{tab:bp}). Bottom: $u_3$ (in GeV) versus $M$  ($\sim\mathcal{O}(10^7$) GeV) for different values of coupling $f_{\Delta}$, where each contour satisfies $m_{\nu}\sim0.1~eV$.}
\label{fig:rhnmass}
\end{figure}

Assuming $m_2, m_2', m_D \ll M$, $n$ remains pseudo-dirac with $m_n \simeq M$. Since, $\zeta$ is the portal between the SM and the hidden sector, the collider signatures of this model involve processes with $n$ in the final states. Therefore, a phenomenologically interesting choice of parameters would be $M\sim \mathcal{O}(\text{TeV})$, with $f_{\zeta}\sim 1$. Furthermore, we assume $v_2 \simeq 1$ GeV in order to have a small $Z-Z'$ mixing. Using $\sum m_{\nu} < 0.17$ eV~\cite{Couchot:2017pvz}, we take $m_{\nu}\simeq \mathcal{O}(0.1 \text{ eV})$ such that:

\bea
u_3 \sim \frac{0.1}{f_{\Delta}} \text{ MeV}.
\eea

A contour plot for correct neutrino mass $m_{\nu}\simeq\mathcal{O}(0.1 \text{ eV})$, following Eq.~\ref{eq:correctnumass}, is depicted in Fig.~\ref{fig:rhnmass}. The contours in $M-f_\Delta$ plane has been shown for $f_\zeta\sim\mathcal{O}(1)$ for different choices of $u_3$ in the LHS of top panel in Fig.~\ref{fig:rhnmass}. The same exercise is done in $M-f_\zeta$ plane for $f_\Delta\sim\mathcal{O}(1)$ {\footnote{While a large Yukawa may cause trouble to vacuum stability, the extended scalar sector is expected to save it.}} in top RHS graph for different $u_3$. We choose a few benchmark points at the scale of heavy neutrino mass $450$ GeV, shown by the vertical dashed line in this plot. In both of these cases, $X$ is nearly degenerate with $X_3$ due to very small values of $u_3$ ($\sim$ MeV). Therefore, co-annihilations play an important role in determining the relic abundance of the $X$ DM. We will explore this in details in the DM section.

The other possible regime is to assume $M\sim\mathcal{O}(10^7)$ GeV, which allows larger $u_3$ ($\sim$ hundreds of GeVs). This is shown on the bottom panel of Fig.~\ref{fig:rhnmass} for $f_\zeta\sim\mathcal{O}(1)$ and $f_\Delta:\{0.01, 0.9\}$. The mass degeneracy between $X, X_3$ is broken in such a scenario and thus co-annihilations become subdominant to the annihilation processes for $X$ DM. We will also show that when $M\sim 500$ GeV, the heavy neutrinos are stable and can be DM candidate, while heavy neutrinos with $M\sim 10^7$ GeV will decay and will not contribute as DM. Therefore such heavy neutrinos are also viable from neutrino mass and DM constraints, but will complicate the model in collider detection. We will therefore choose lighter $n_{1,2}$ scenario (as in the top panel of Fig.~\ref{fig:rhnmass}) and show that it plays a crucial role in yielding possible leptonic signature at the LHC. 

\section{Dark Matter Phenomenology}
\label{sec:DM pheno}

In this analysis, we highlight a couple of interesting features regarding DM phenomenology of the model : (i) The alteration to the single component vector boson DM freeze-out and its relic density due to co-annihilation contribution, which was not taken into account in the earlier analysis~\cite{Fraser:2014yga} and (ii) the presence of a second DM candidate ($\Delta$) in a large region of parameter space of the model, which is significantly influenced by DM-DM interactions. The heavy neutrinos $(n_1,n_2)$, assumed in this framework, can also be kinematically stable and serve as DM. However, for generating correct neutrino masses, the relic density of these particles will be very small. We will discuss this separately in subsection.~\ref{sec:rhn}.

\subsection{Possible DM candidates of the model}
\label{sec:dmcandidate}

At the very outset, we will sketch the parameter space of the model, where we can have different DM components coexisting together. 

\begin{figure}[htb]
$$
\includegraphics[scale=0.72]{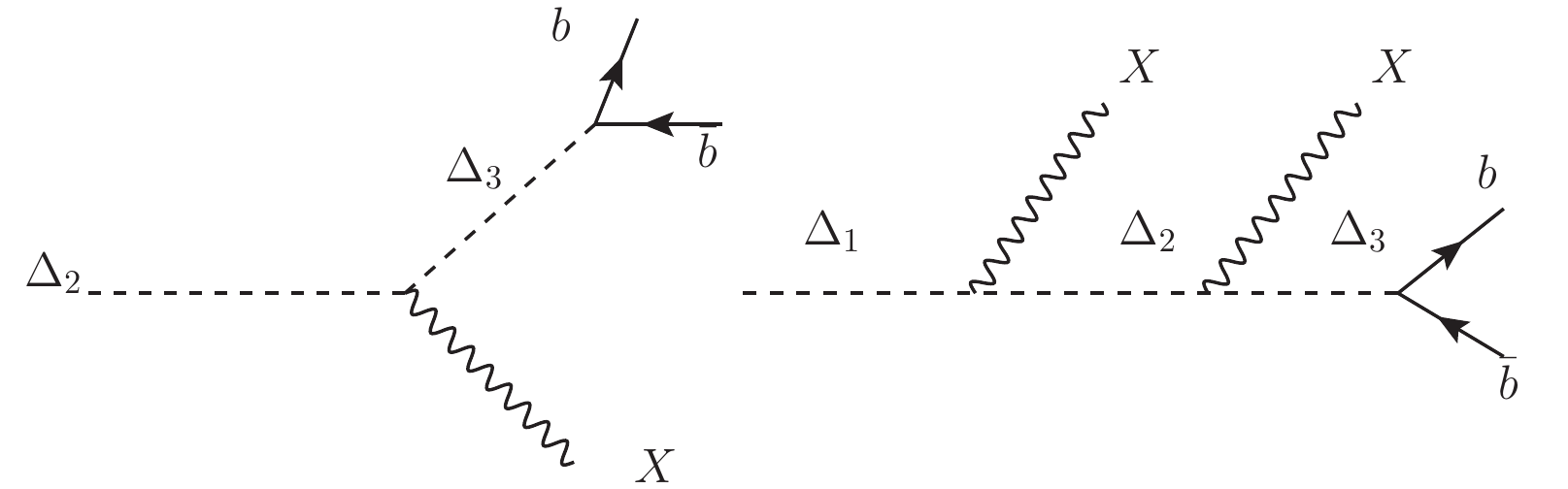}
 $$
\caption{Decay of the triplet scalars to vector boson $X$ for $m_{\Delta_{1,2,3}} > m_X$. Left: Decay of $\Delta_2$ to SM via $\Delta_3$; Right: Decay of $\Delta_1$ to SM and $X$ via off-shell $\Delta_3$ and $\Delta_2$.}
\label{fig:del2decay}
 \end{figure}

$\Delta_1$ and $\Delta_2$ component of the $SU(2)_N$ scalar triplet have non-zero $S$ charges (as mentioned in Sec.~\ref{sec:model}). As they are charge neutral, they can qualify as DM if their stability is ensured. $\Delta_3$ having zero $S$ charge, mixes with the SM Higgs due to non-zero VEV (instigated by $f_8\Phi^{\dagger}\Phi Tr(\Delta^{\dagger}\Delta)$ term in the scalar potential) and decaying to SM. Therefore, $\Delta_3$ does not qualify as DM. On the other hand, $\Delta_1$ and $\Delta_2$ have the following interaction vertices with the vector boson $X$:$\Delta_1\Delta_2^{*}X$,~$\Delta_2\Delta_3^{*}X$,~$\Delta_1XX$,~$\Delta_2 X X_3$. As a result, possible decay of $\Delta_2$ to SM can occur via off-shell $\Delta_3$ as shown in left hand side (LHS) of Fig.~\ref{fig:del2decay}. Similarly, $\Delta_1$ can also decay to SM via off-shell $\Delta_2$ and $\Delta_3$, shown in right hand side (RHS) of Fig.~\ref{fig:del2decay}. So, $\Delta_1$ and/or $\Delta_2$ can be potential DM candidates if we can stop the decays shown in Fig.~\ref{fig:del2decay}. The viability of $\Delta_1$ and $\Delta_2$ as DM are discussed in two possible scenarios: (i) Degenerate triplet scalar ($m_{\Delta_1}=m_{\Delta_2}=m_{\Delta_3}=m_{\Delta}$), (ii) Non-degenerate triplet scalar ($m_{\Delta_1} \ne m_{\Delta_2} \ne m_{\Delta_3}$). \\

\begin{figure}
\centering
\includegraphics[scale=0.45]{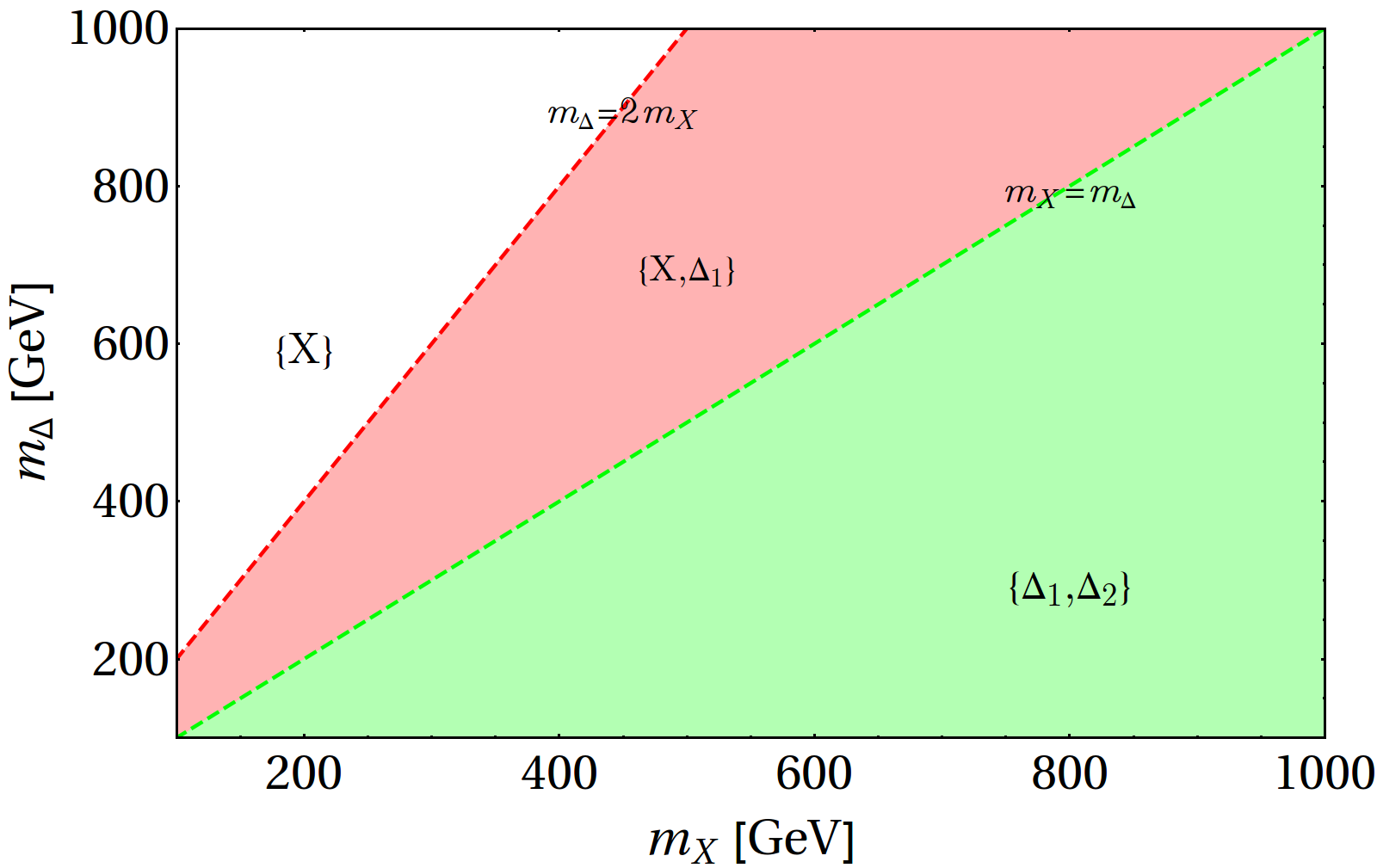}
\caption{Regions of $m_X-m_{\Delta}$ (in GeV) parameter space, where single component and multi-component DM frameworks can be realised for degenerate scalar triplet masses $m_{\Delta_1}=m_{\Delta_2}=m_{\Delta_3}=m_{\Delta}$. In the white region ($2m_X<m_\Delta$), only $X$ can be a single component DM. In the pink region ($m_\Delta/2 <m_X<m_\Delta$), two component DM with $\{X,\Delta_1\}$ is operative. In the green region ($m_X>m_\Delta$), $\{\Delta_1,\Delta_2\}$ forms degenerate two-component DM.}
\label{fig:regions}
\end{figure}

(a) {\bf Degenerate triplet scalar}: The triplet scalar components can be degenerate in the limit of $f_7=0$~\cite{Fraser:2014yga}. In this limit, 

 \begin{itemize}
 \item when $m_{\Delta}>m_X$: 
 \begin{itemize}
  \item[i)] $X$ is stable and a DM.
  \item[ii)] $\Delta_2 \rightarrow X b\bar{b}$ is always possible with $m_X<m_{\Delta}$, hence $\Delta_2$ can never be a DM.
  \item [iii)] If $m_{\Delta}<2 m_X$ then $\Delta_1$ is stable and becomes second DM component.
  \item[iv)] If $m_{\Delta}>2 m_X$, then $\Delta_1$ decays and is not a DM candidate.  
  \end{itemize}
  
 \item when $m_{\Delta}<m_X$:
 \begin{itemize}
  \item[i)] By default this implies $m_{\Delta}<2 m_X$ and hence $\Delta_1$ is stable and a DM. 
  \item[ii)] $\Delta_2$ is also stable and acts as second degenerate DM component with $\Delta_1$.
  \item[iii)] $X$ can decay into $\Delta_2$ (and subsequently to $\bar{b}b$) so it can not be a DM candidate.
 \end{itemize}
\end{itemize}

Therefore, when $m_{\Delta}>m_X$, we can have both two-component (for $m_{\Delta}<2 m_X: \{X, \Delta_1\}$) and one-component DM scenario (for $m_{\Delta}>2 m_X: \{X\}$). On the other hand, when $m_{\Delta}<m_X$, we will have a degenerate 2-component DM scenario comprising of $\Delta_1$ and $\Delta_2$. The above situation for degenerate scalar triplet case is summarised in Fig.~\ref{fig:regions}.

\begin{figure}
\centering
\includegraphics[scale=0.3]{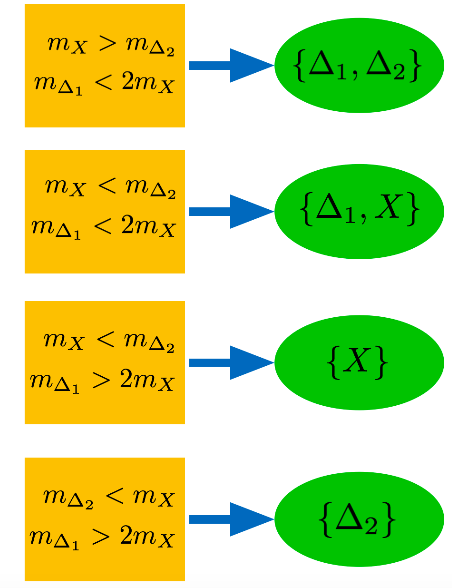}
\caption{Main kinematic regions for single and two component DMs for non-degenerate scalar triplet scenario. They are: $\{\Delta_1,\Delta_2\},\{\Delta_1,X\},\{X\},\{\Delta_2\}$.}
\label{fig:regions2}
\end{figure}

(b) {\bf Non-Degenerate triplet scalar}: Non-degenerate scalar triplet scenario ($f_7 \ne 0$) can have four possible DM framework depending on the hierarchy of $m_{\Delta_1},~m_{\Delta_2}, ~m_X$. It is quite understood from Figure \ref{fig:del2decay}, between $\Delta_2~\rm{and}~X$, we can have one of them as a DM, while the possibility of $\Delta_1$ as DM will be guided by the hierarchy between $m_{\Delta_1}~vs~2 m_X$. Therefore, the situations of interest are:

\begin{enumerate}
\item $\Delta_1, \Delta_2$ forming non-degenerate DM  components : when $m_X>m_{\Delta_2}$ and $m_{\Delta_1}<2 m_X$,
\item  $\Delta_1, X$ forming non-degenerate DM components : when $m_X<m_{\Delta_2}$ and $m_{\Delta_1}<2 m_X$
\item $X$ as single component DM: when $m_{\Delta_2}>m_X,~ m_{\Delta_1}>2m_X$
\item $\Delta_2$ as single component DM: when $m_{\Delta_2}<m_X,~ m_{\Delta_1}>2m_X$
\end{enumerate}

This is also summarised in Fig.~\ref{fig:regions2}. Here, we would like to mention that, the decay lifetime of $\Delta_3$ to SM $b\bar{b}$ can be comparable to the age of the Universe (or in other words, $\Delta_3$ can be made stable compared to the Universe's life time), if we consider the coupling of $\Delta_3$ to SM (which is given by $f_8$) to be vanishingly small $\sim\mathcal{O}(10^{-22})$ as estimated in Appendix-C. In that case $\Delta_3$ can also be a DM along with $\Delta_{1,2} ~\rm{and/or}~ X$. However given the fact that annihilation of the scalars to SM is controlled by $f_8$, such a tiny value of the coupling will yield overabundance of scalar DM through freeze-out mechanism. Therefore, we restrict ourselves in elaborating such prospects. 

 We will analyze the degenerate scalar triplet model here for simplicity and economy of parameters. This itself offers a variety of single component ($X$) or a multi-component interacting DM set-up (in the form of \{$\Delta_1,X$\} or \{$\Delta_1,\Delta_2$\}).


\subsection{$X$ as single component vector boson DM}
\label{sec:X DM}

\begin{figure}[htb!]
$$
\includegraphics[height=4cm]{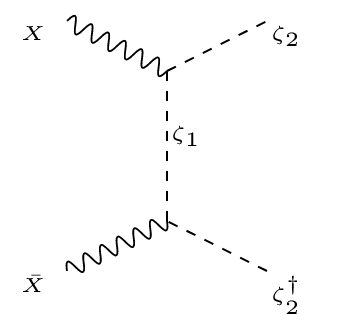}
\includegraphics[height=3.8cm]{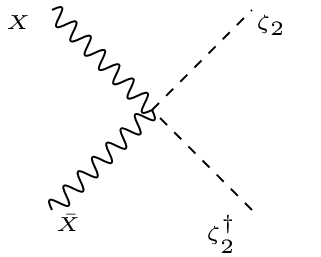}
$$
$$
\includegraphics[height=3.0cm]{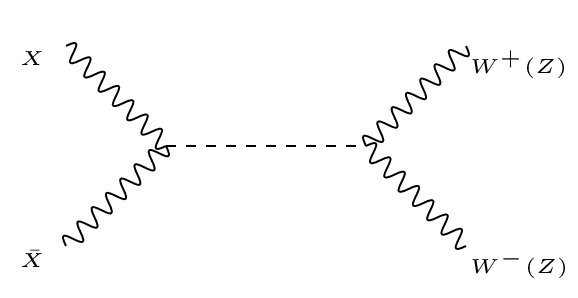}
\includegraphics[height=3.0cm]{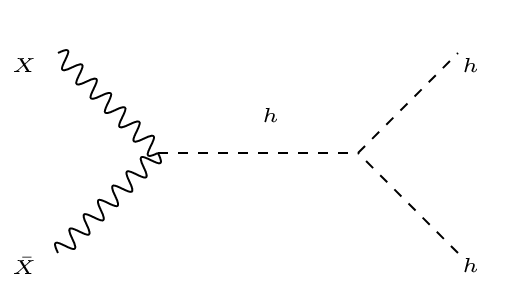}
\includegraphics[height=3.0cm]{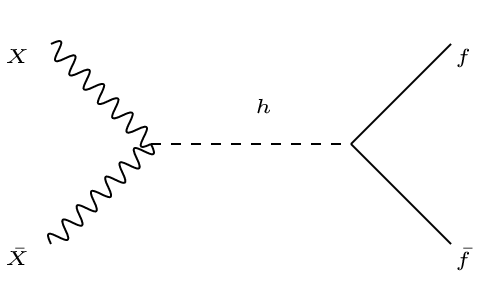}
$$
\caption{Top: Annihilation of $X$ DM into heavy scalar $\zeta_2, {\zeta_2}^{\dagger}$ via $t$-channel mediation of $\zeta_1$ and four pint interaction assuming $m_{\zeta_2} < m_X < m_{\zeta_1}$. Bottom: Annihilation of $X$ into SM via Higgs mediation in $s$-channel.}
\label{fig:annihilx1}
\end{figure}

$X$ can appear as a single component DM in degenerate scalar triplet case when $m_\Delta>m_X$ and $m_\Delta>2m_X$. It can also be a single component DM for non-degenerate scalar triplet case when $m_{\Delta_2}>m_X$ and $m_{\Delta_1}>2m_X$. The dominant annihilation channels for $X$, shown in Fig.~\ref{fig:annihilx1}, can be classified into two categories: (i) annihilation to heavy scalars ($\zeta$), (shown in the upper panel) and (ii) annihilation to the SM through Higgs portal (shown in the lower panel). The latter was not considered in the previous work~\cite{Fraser:2014yga}. Throughout the analysis, all the annihilation cross sections are calculated on threshold: $s_0=4 m_X^2$, assuming only dominant $s$-wave contribution. The total annihilation cross section of $X$ is then given by:

\begin{align}
\langle\sigma v_{rel}\rangle &= \frac{g_N^4}{576\pi m_X^2}\sqrt{1-\frac{m_{\zeta_2}^2}{m_X^2}}\left(2+\left[1+\frac{4\left(m_X^2-m_{\zeta_2}^2\right)}{m_{\zeta_1}^2+m_X^2-m_{\zeta_2}^2}\right]^2\right)\nonumber\\
&+ \frac{m_{W,Z}^4}{48\pi m_X^2}\sqrt{1-\frac{m_{W,Z}^2}{m_X^2}}\left[\frac{g_N^4  \left(f_5/\lambda_4\right)^2}{\left(4 m_X^2-m_h^2\right)^2+\Gamma_h^2 m_h^2}\right]\left[3+4\Big\{\left(\frac{m_X}{m_{W,Z}}\right)^4-\left(\frac{m_X}{m_{W,Z}}\right)^2\Big\}\right]\nonumber\\
 &+ \frac{m_f^2}{24\pi}\left(1-\frac{m_f^2}{m_X^2}\right)^{3/2}\left(\frac{g_N^4 \left(f_5/\lambda_4\right)^2}{\left(4 m_X^2-m_h^2\right)^2+\Gamma_h^2 m_h^2}\right)+\frac{3 m_h^4}{128\pi m_X^2}\sqrt{1-\frac{m_h^2}{m_X^2}}\nonumber\\
  & \left[\frac{g_N^4  \left(f_5/\lambda_4\right)^2}{\left(4 m_X^2-m_h^2\right)^2+\Gamma_h^2 m_h^2}\right].
\label{eq:sigx}
\end{align}

The first term corresponds to the annihilation of $X$ to lighter exotic scalar $\zeta_2$ via $t$-channel mediation of heavier companion $\zeta_1$, and a four point interaction as shown in the upper panel of Fig.~\ref{fig:annihilx1}. These interaction vertices are solely dependent on the $SU(2)_N$ gauge coupling $g_N$. The next three terms are annihilation to the SM Higgs, SM gauge bosons ($W^{\pm},Z$) and SM fermions respectively, through Higgs portal. These cross sections additionally depend on $f_5/\lambda_4$. The cross sections are obtained for $m_{\zeta_1}>m_X > m_{\zeta_2}$, ensuring the stability of $X$. Otherwise $(m_{\zeta_{1,2}}>m_X)$ the annihilation will occur to SM final states only. 

\begin{figure}[htb!]
$$
\includegraphics[height=3.5cm]{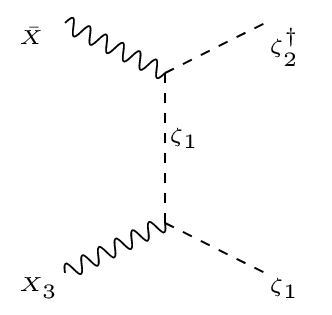}
$$
\caption{Co-annihilation of $X$ with $X_3$ to $\zeta_1 \zeta_2^\dagger$ for $\frac{1}{2}\left(m_{\zeta_1}+m_{\zeta_2}\right) < m_X < m_{\zeta_1}+m_{\zeta_2} $ (see text for details).}
\label{fig:coannihilx}
\end{figure}

Importantly, $X$ can undergo co-annihilation with $X_3$ via the diagram shown in Fig.~\ref{fig:coannihilx}. The effective cross section in this case can be written as:

\be
\langle\sigma \; {\rm v}\rangle_{\text{eff}} = (\sigma\; {\rm v} )_{X\bar{X}\rightarrow SM,~\zeta_2\zeta_2^\dagger}~+~(\sigma\; {\rm v} )_{\bar{X}X_3\rightarrow \zeta_1 \zeta_2^\dagger +hc}\left(1+\frac{\Delta m}{m_X}\right)^{\frac{3}{2}}exp({-\Delta m}~x/{m_X}),
\label{eq:coann}
\ee

where $\Delta m=m_{X_3}-m_{X}$ and $x=\frac{m_X}{T}$. The contribution from co-annihilation has not been considered in the earlier analysis of this model. 

For co-annihilation to occur: 

\begin{equation*}
 m_{\zeta_1}+m_{\zeta_2} < m_X + m_{X_3}\implies m_X > \frac{1}{2}\left(m_{\zeta_1}+m_{\zeta_2}\right),
\end{equation*}

in the limit $m_X\sim m_{X_3}$. Again, for stability of $X$:

\begin{equation*}
m_X < m_{\zeta_1}+m_{\zeta_2}. 
\end{equation*}

Together, we have the following condition for co-annihilation:

\bea
\frac{1}{2}\left(m_{\zeta_1}+m_{\zeta_2}\right) < m_X < m_{\zeta_1}+m_{\zeta_2}.
\eea

We would once again remind that co-annihilation contributions become very important in this model as $\Delta m=m_{X_3}-m_{X} \to 0$. This happens for small $u_3$ ($\sim$ MeV), which is required for neutrino mass generation with heavy neutrinos of the order of hundreds of GeVs, as discussed earlier.

\begin{figure}[htb!]
$$
\includegraphics[height=7cm]{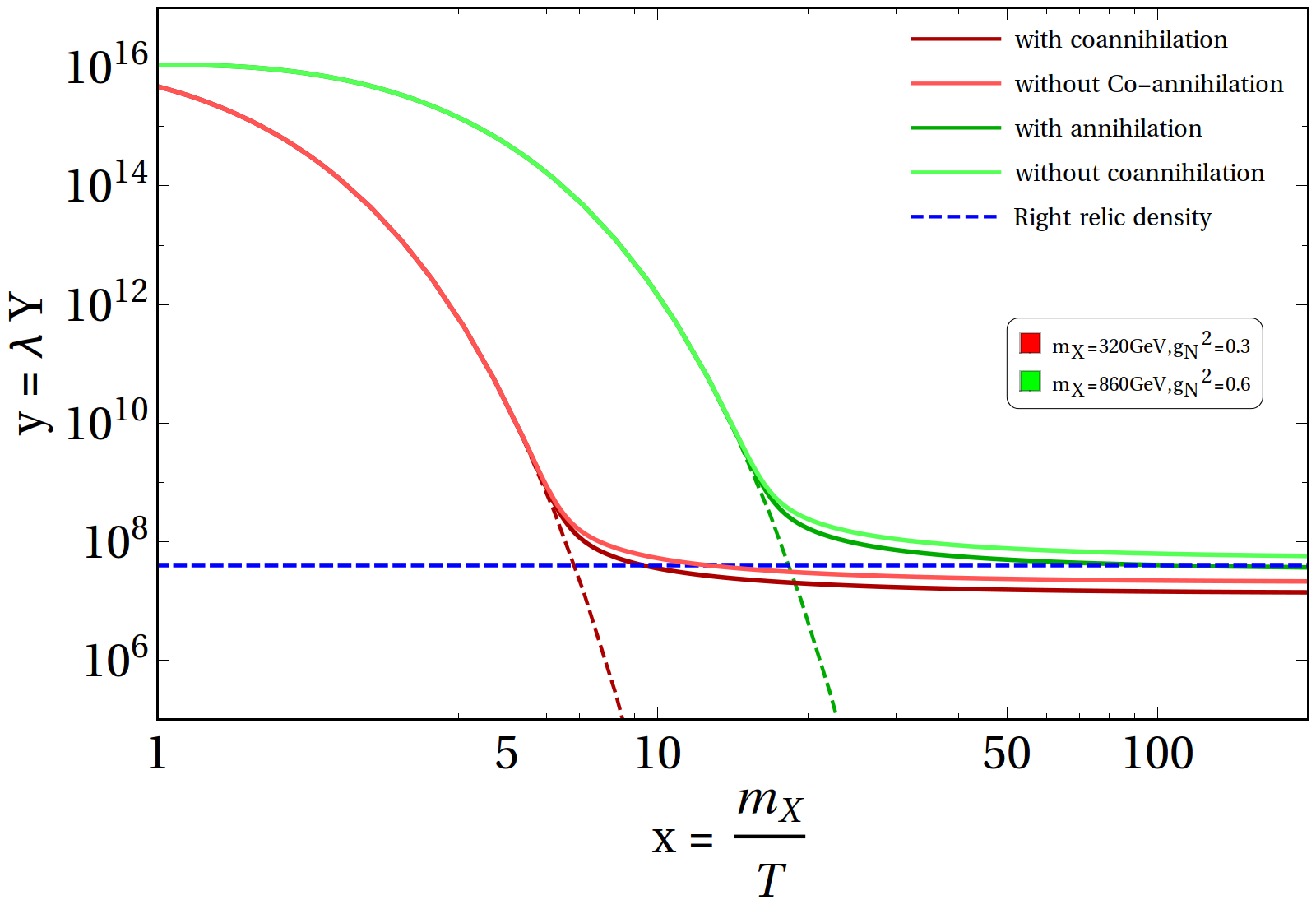}
$$
\caption{Freeze out of vector boson DM ($X$) is shown in $y=\lambda Y$ versus $x=\frac{m}{T}$ plane for two different combinations of DM mass and $SU(2)_N$ couplings: $\{m_X,g_N^2\}=\{320 ~\rm{GeV},0.3\} (\rm{in ~Red}); \{860 ~\rm{GeV},0.6\} (\rm{in~ Green})$. In each case, the equilibrium distributions are shown in dotted lines. The cases with inclusion of co-annihilation contributions are shown through darker thick lines. Right relic density in shown through blue dotted line. }
\label{fig:beq1}
\end{figure}

Boltzmann equation (BEQ) for the single component $X$ DM can be written as :
\be
\frac{dy}{dx}= -\frac{m_{X}}{x^2} \left[\sigma _0(y^2-{y^{EQ}}^2)\right],  
\label{eq:beq4}
\ee

where $\sigma_0=(\sigma\rm{v})_{\rm{eff}}$ given in Eq.~\ref{eq:coann}. The equilibrium co-moving number density is $Y^{EQ}=0.145 \frac{g}{g_{*s}} (\frac{m_{X}}{T})^{\frac{3}{2}}e^{-\frac{m_{X}}{T}}$, where $g=3$ is the degrees of freedom (DoF) associated with the vector boson DM $X$ and $g_{*s}=106.7$ is the total DoF. The solution for BEQ is easier in terms of modified yield $y=\lambda Y$, where $\lambda={(0.264 ~ m_{Pl} \frac{g_{*s}}{\sqrt{g_*}})}$. A typical freeze-out of $X$ is shown in Fig.~\ref{fig:beq1}. 
For brevity, we choose two different combinations of DM mass and $SU(2)_N$ couplings $\{m_X,g_N^2\}=\{320 ~\rm{GeV},0.3\}, \{860 ~\rm{GeV},0.6\}$ as shown in red and green thick lines respectively. Corresponding equilibrium distributions are shown by dashed lines and therefore, the freeze out can easily be identified by the departure of the thick lines from the dashed ones. Inclusion of maximal co-annihilation with $\Delta m \to 0$ as in Eq.~\ref{eq:coann}, is shown by the darker thick lines for both the chosen points and indicate the non-negligible effect. Here, we have kept other masses as:~\{$m_{\zeta_2}$,$m_{\zeta_1}$\}=~\{210 GeV, 360 GeV\};~\{250 GeV, 880 GeV\} corresponding to $\{m_X,g_N^2\}=\{320 ~\rm{GeV},0.3\}, \{860 ~\rm{GeV},0.6\}$ respectively. We also point out the observed relic density using a blue dashed line, which shows that the combination of $\{m_X,g_N^2\}=\{860 ~\rm{GeV},0.6\}$ yields the correct relic density with co-annihilation contribution included. 

\begin{figure}[htb!]
$$
\includegraphics[height=5cm]{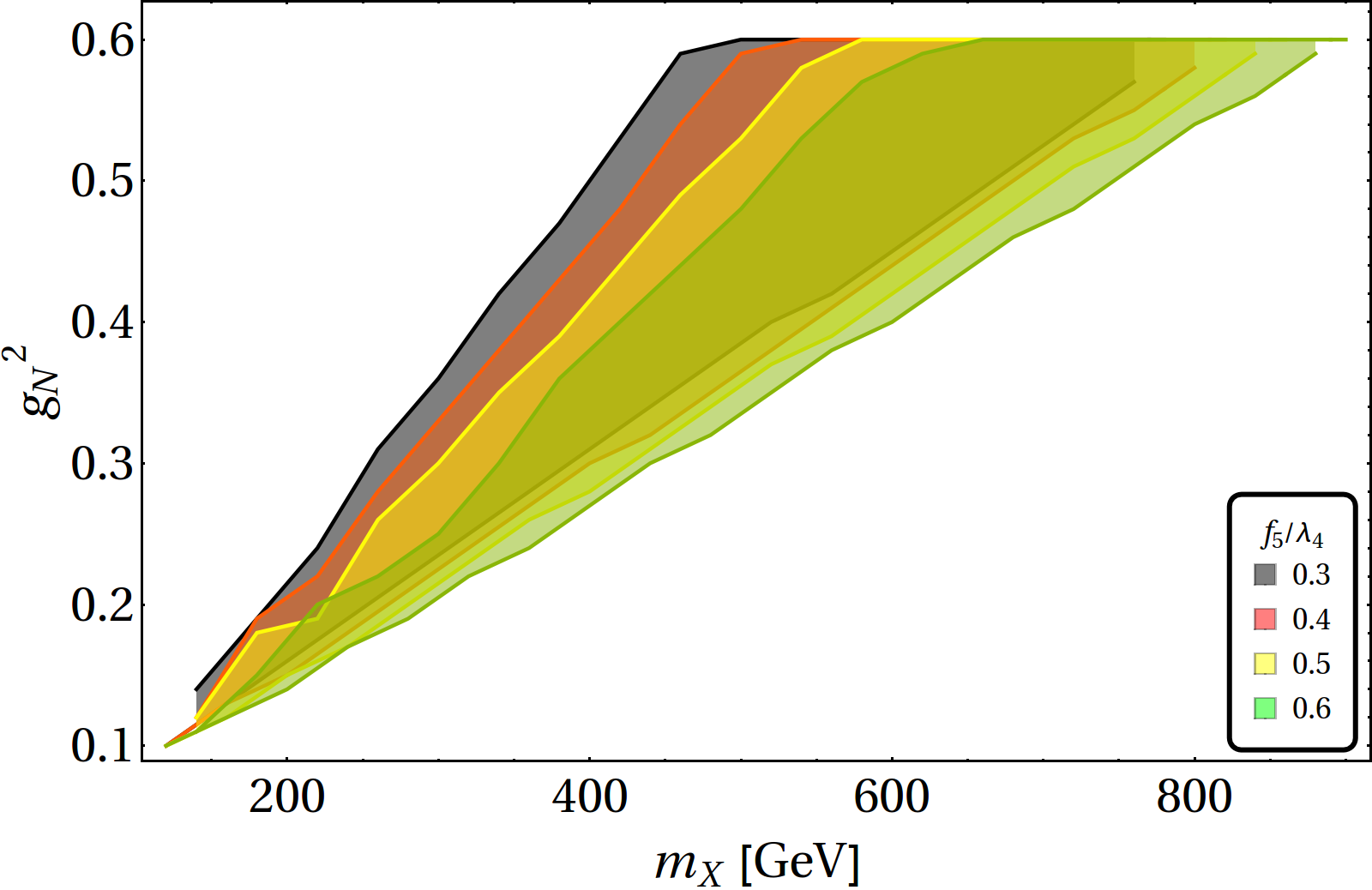}\hspace{0.5cm}
\includegraphics[height=5cm]{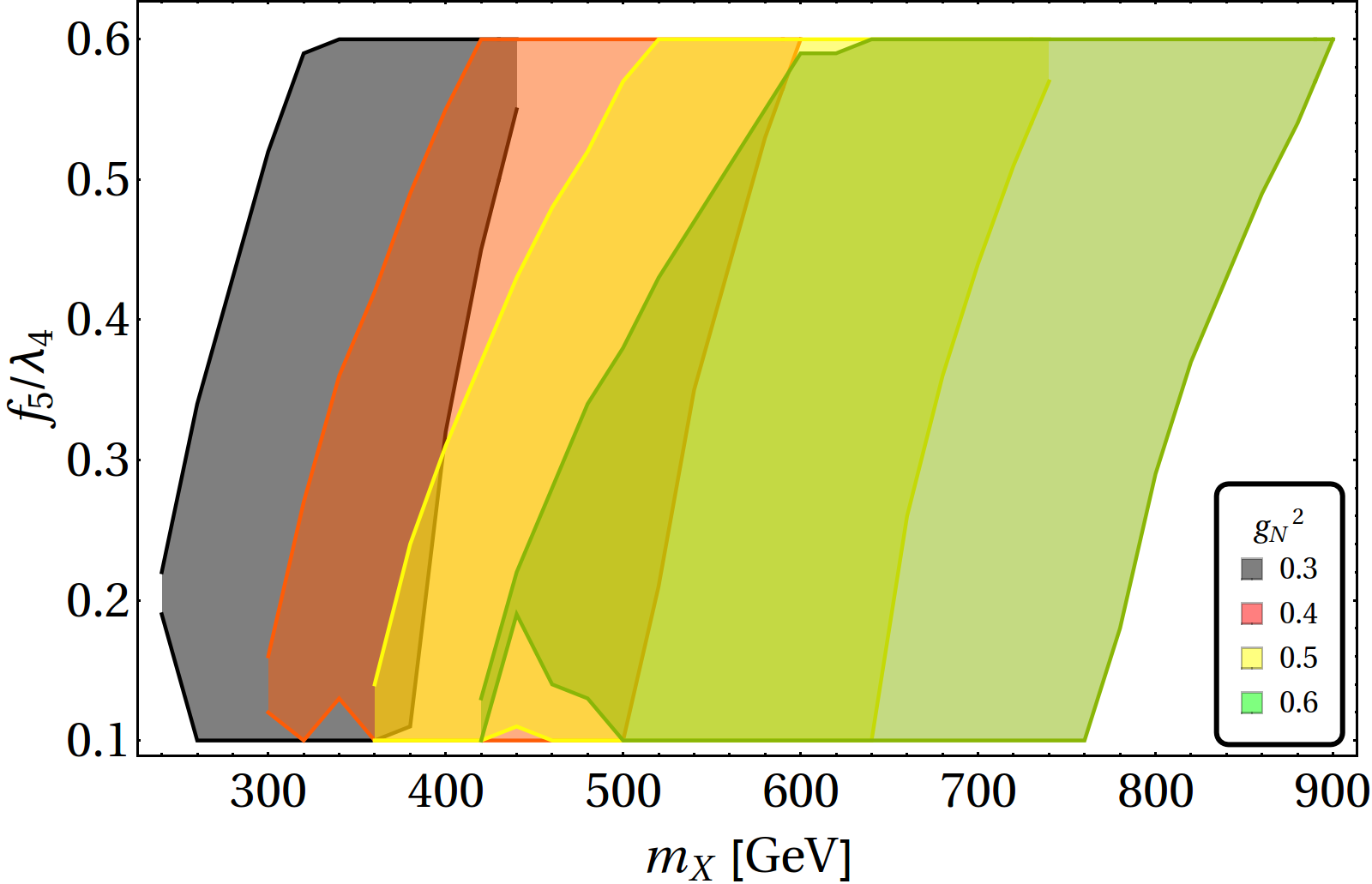}
$$
\caption{Relic density allowed parameter space for $X$ as a single component DM. On the left panel, allowed $m_X-g_N^2$ parameter space is shown for  different $f_5/\lambda_4$; and on the right panel, $m_X-f_5/\lambda_4$ allowed parameter space is shown for different choices of $g_N^2$.}
\label{fig:mxg2relic}
\end{figure}

We will now scan the relic density allowed parameter space of single component X in two different regions: (i) $m_{\zeta_1}>m_X>m_{\zeta_2}$ and (ii) $m_X<m_{\zeta_{2,1}}$. In the first case, annihilation occurs to the heavy scalar $\zeta_2$ and to SM, while in the second the annihilation occurs only to SM. The free parameters for the DM analysis can be chosen as:
\bea
\big\{g_N^2,\frac{f_5}{\lambda_4},m_X,m_{\zeta_1},m_{\zeta_2}\big\}.
\eea

Both the couplings are varied in the range \{$g_N^2$:~0.01-0.6\} and \{$\frac{f_5}{\lambda_4}$:~0.01-0.6\} for scanning the parameter space. The relic density (PLANCK data: $0.1165\le \Omega h^2 \le 0.1227$) allowed parameter space for X is shown in Fig.~\ref{fig:mxg2relic}. On the left panel, we have shown the allowed parameter space in terms of $m_X$ (in GeV) versus $g_N^2$ for different choices of $\frac{f_5}{\lambda_4}$. On the right panel, we show it in terms of $m_X$ (in GeV) versus $\frac{f_5}{\lambda_4}$ for different choices of $g_N^2$. First of all, we see that for larger $g_N^2$, we obtain a larger range of DM mass, that can satisfy relic density constraints. On the other hand, we see that the effect of $\frac{f_5}{\lambda_4}$ is milder than $g_N^2$. Essentially this is due to the fact that the t-channel annihilation of the DM to the heavy scalars ($\zeta_2$) is larger than the s-channel annihilation to SM particles through Higgs mediation. Low DM mass is favored by smaller \{$g_N^2$, $\frac{f_5}{\lambda_4}$\}. For the same reason, $g_N^2$  or $f_5/\lambda_4$ need to be as large as $\sim 0.6$ for DM masses in the range $\sim$ 1 TeV. 

\begin{figure}[htb!]
$$
\includegraphics[scale=0.34]{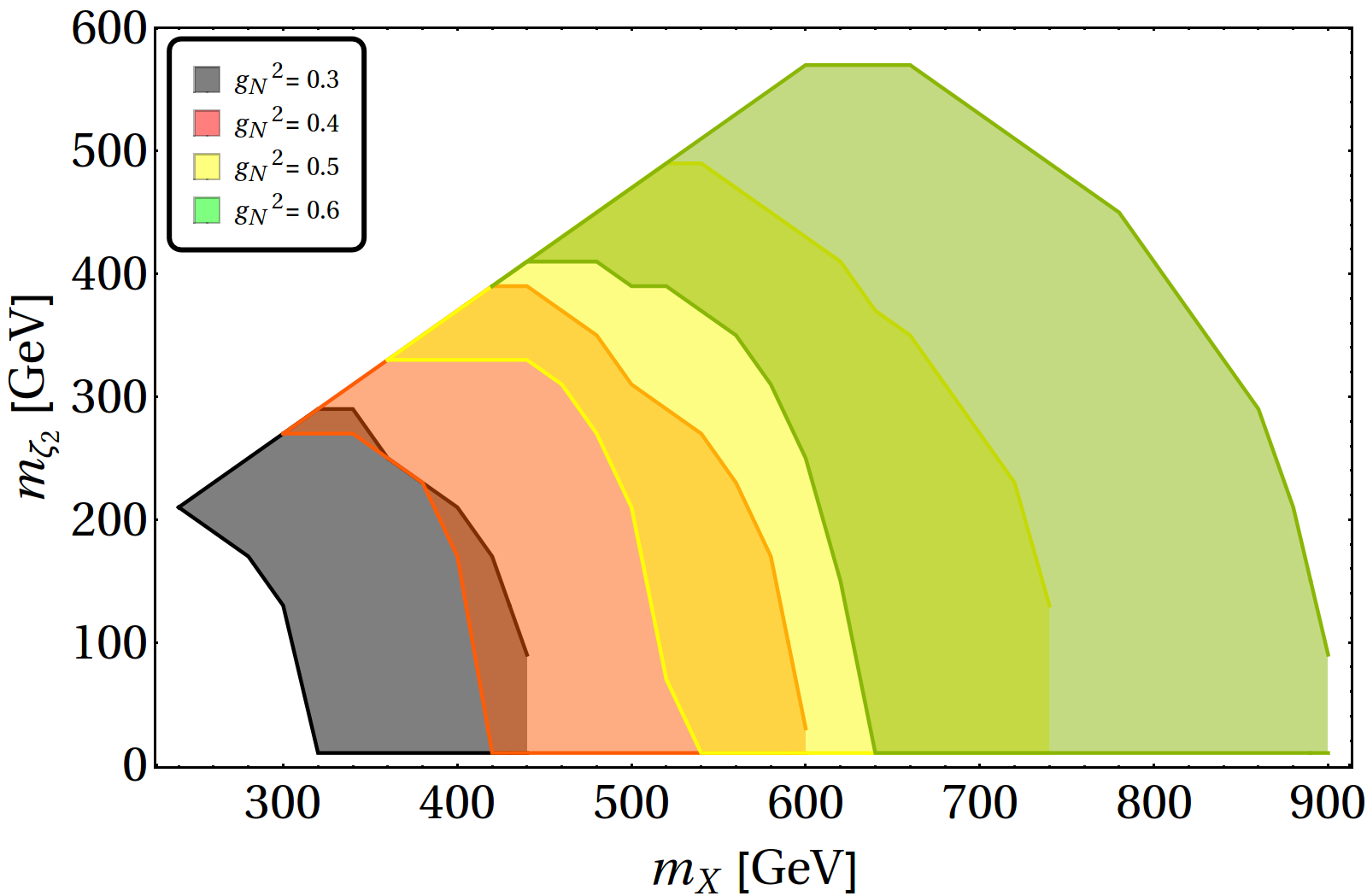}
\includegraphics[scale=0.3]{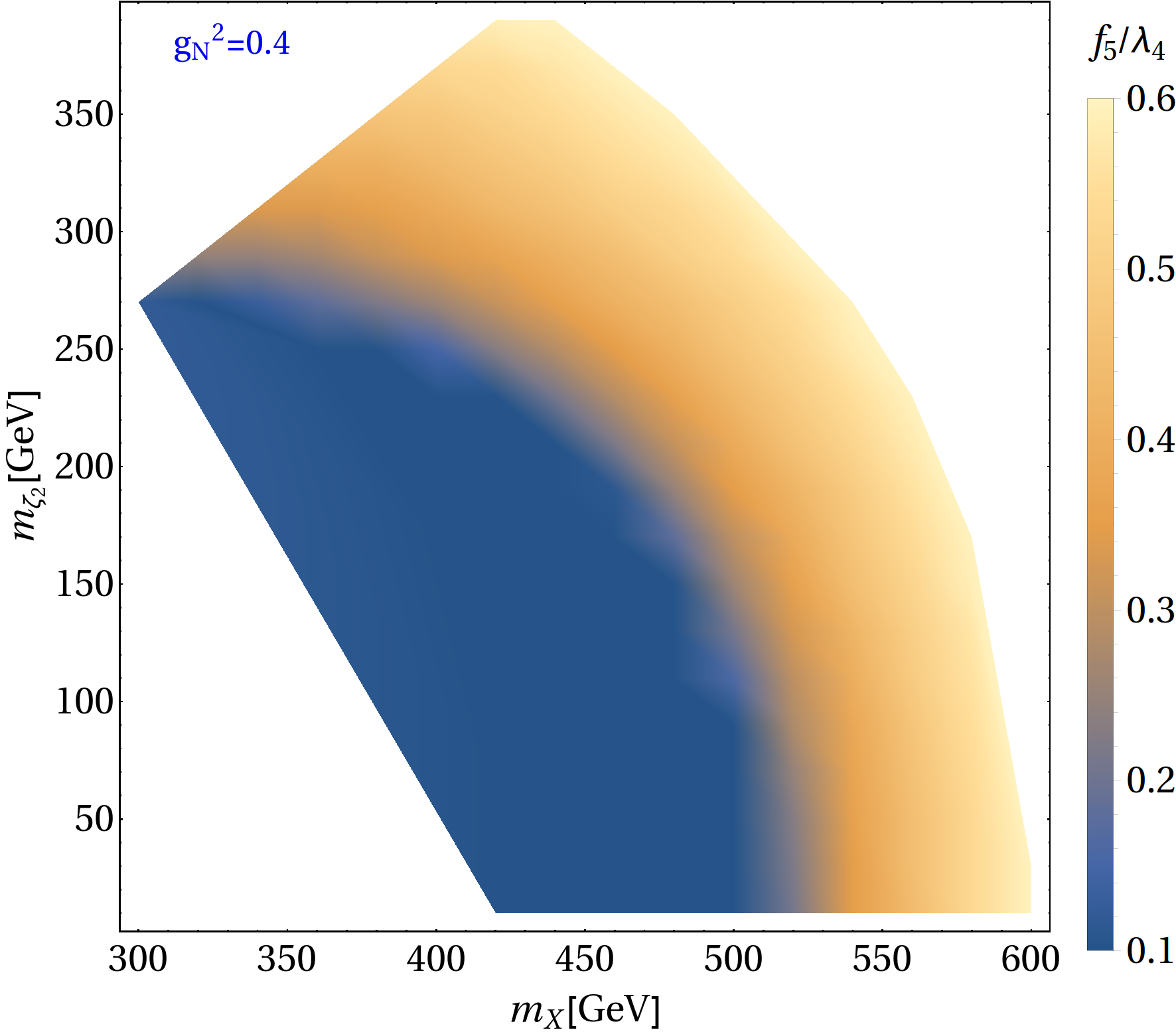}
$$
\caption{LHS: Relic density allowed parameter space in $m_X$-$m_{\zeta_2}$ plane for all possible allowed values of $m_{\zeta_1}$ with different values of $g_N^2$ showed in different colours. RHS: Relic density allowed parameter space for $g_N^2=0.4$ in $m_X$-$m_{\zeta_2}$ plane, where different allowed values of $f_5/\lambda_4$ are shown.}
\label{fig:mxmz2pln}
\end{figure}

Dependence on $m_{\zeta_2}$ for correct relic density is shown in $m_X$-$m_{\zeta_2}$ plane in Fig.~\ref{fig:mxmz2pln} assuming $m_X>m_{\zeta_2}$. Different colour shades indicate different choices of $g_N^2$ in the left plot. Constant $g_N^2$ regions exploit the freedom on $f_5/\lambda_4$ as shown in the right hand side of Fig.~\ref{fig:mxmz2pln}. $f_5/\lambda_4$ is insensitive to the choice of $m_{\zeta_2}$. Larger DM masses have to be adjusted with larger $f_5/\lambda_4$ to keep the total relic density intact. 

%

\begin{figure}[htb!]
$$
\includegraphics[scale=0.4]{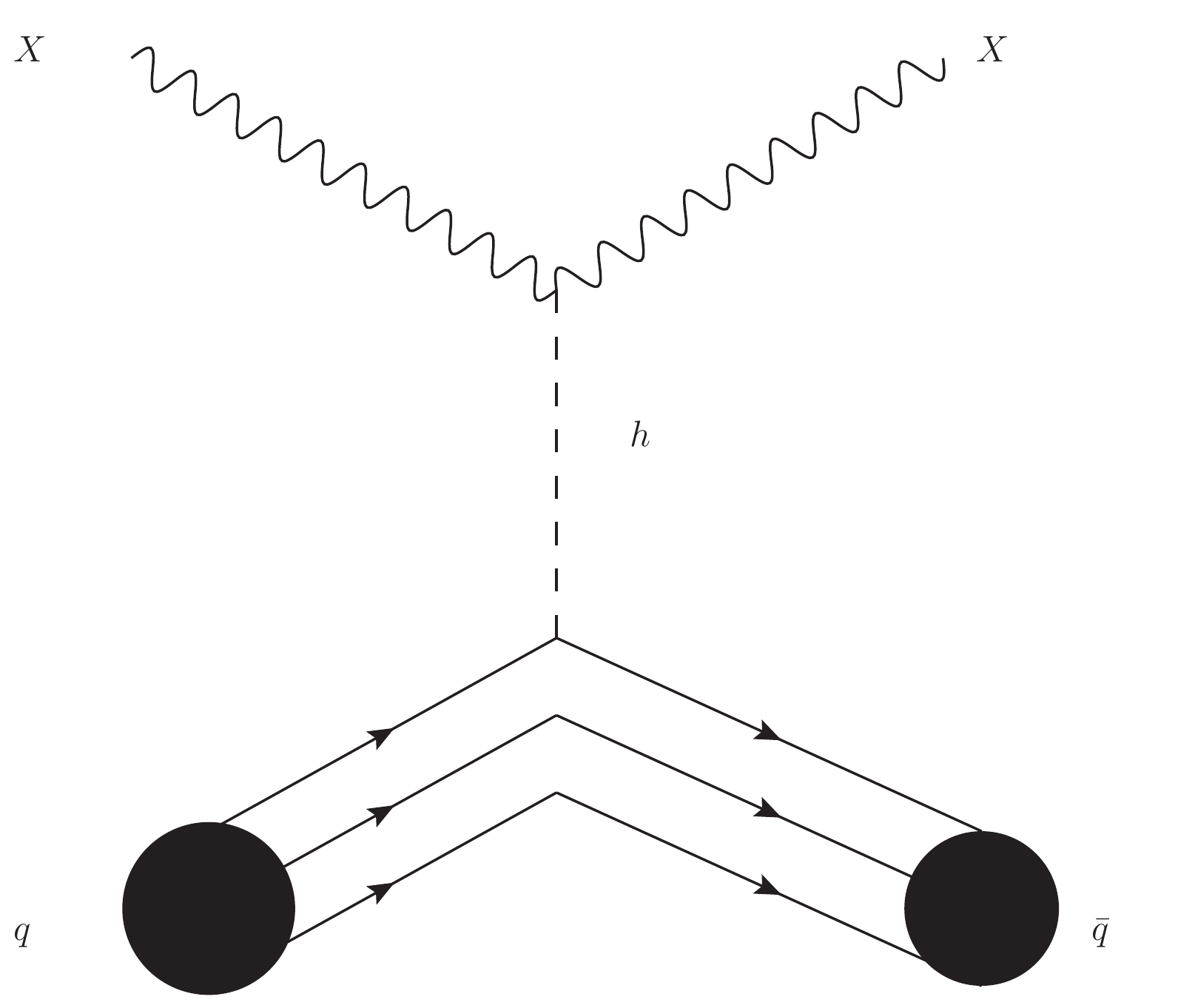} 
$$
\caption{Direct search diagram for vector boson DM X.}
\label{fig:ddX}
\end{figure}

The direct detection interaction for $X$ occurs via $t$-channel Higgs mediation as shown in Fig.~\ref{fig:ddX}. The spin-independent direct detection cross section scattering off a nucleus with $Z$ protons and $A-Z$ neutrons normalized to one nucleon is given by:
\bea
\sigma^{\text{SI}} = \frac{1}{\pi}\left(\frac{m_N}{m_X+A m_N}\right)^2\left(\frac{Z f_p+\left(A-Z\right)f_n}{A}\right)^2,
\eea
where $f_p$ and $f_n$ are the form factors given by~\cite{Durr:2015dna}
\begin{align}
\frac{f_p}{m_p}=& -0.152\left[\frac{g_N^2\left(f_5/\lambda_4\right)}{4 m_h^2}\right]-0.848\left[\frac{g_N^2\left(f_5/\lambda_4\right)}{54 m_h^2}\right]\\
\frac{f_n}{m_n}=& -0.155\left[\frac{g_N^2\left(f_5/\lambda_4\right)}{4 m_h^2}\right]-0.845\left[\frac{g_N^2\left(f_5/\lambda_4\right)}{54 m_h^2}\right]
\end{align}
where we used:
\begin{equation}
\frac{f_N}{m_N} = \left[ \sum_{u,d,s} f_q^N + \frac{2}{27} \left(1-  \sum_{u,d,s} f_q^N \right)\right] \left[\frac{g_N^2\left(f_5/\lambda_4\right)}{4 m_h^2}\right]
\end{equation}

\begin{figure}[htb!]
$$
 \includegraphics[scale=0.33]{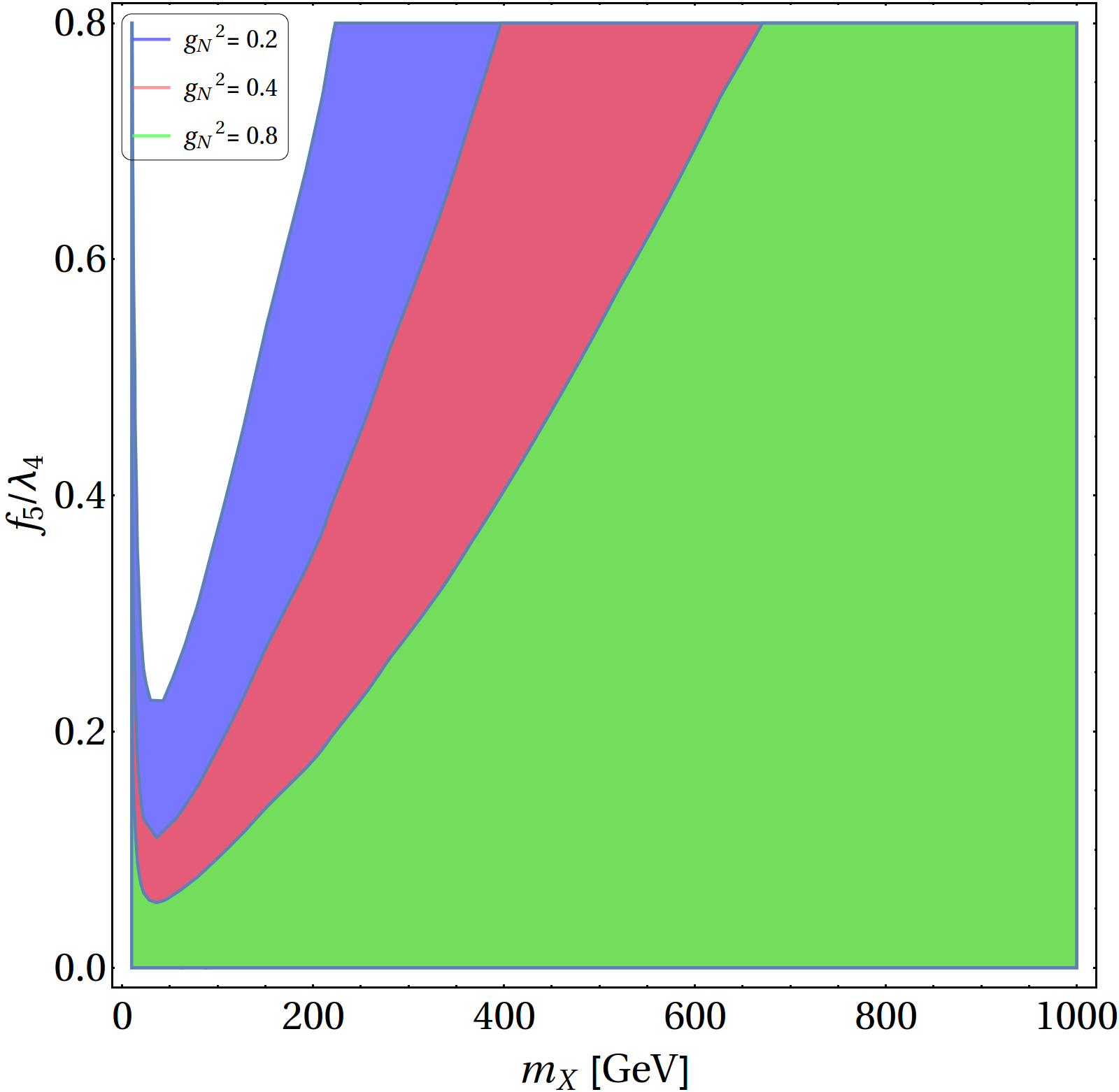}
 $$
 \caption{Contours in $\frac{f_5}{\lambda_4}-m_X$ plane, satisfying direct search bound from PandaX for different values of the gauge coupling $g_N^2$, shown in blue ($g_N^2=0.2$),orange ($g_N^2=0.4$) and green ($g_N^2=0.8$).}
 \label{fig:f5l4bound}
\end{figure}

The above equations yield a bound on $f_5/\lambda_4$ from non-observation of $X$ in direct search experiment for a given $g_N^2$. This can be seen from Fig.~\ref{fig:f5l4bound} where we show different contours for $f_5/\lambda_4$ as function of DM mass, satisfying direct search constraints from PandaX~\cite{Cui:2017nnn} for different choices of $g_N^2$. Any point in the shaded region below the curve is available by direct search data. Here we can see, larger the $g_N^2$, tighter is the limit on $\frac{f_5}{\lambda_4}$. Again, for larger DM mass, $\frac{f_5}{\lambda_4}$ is also large as the direct search cross-section is proportional to the coupling and inversely proportional to the DM mass.

\begin{figure}[htb!]
$$
\includegraphics[scale=0.58]{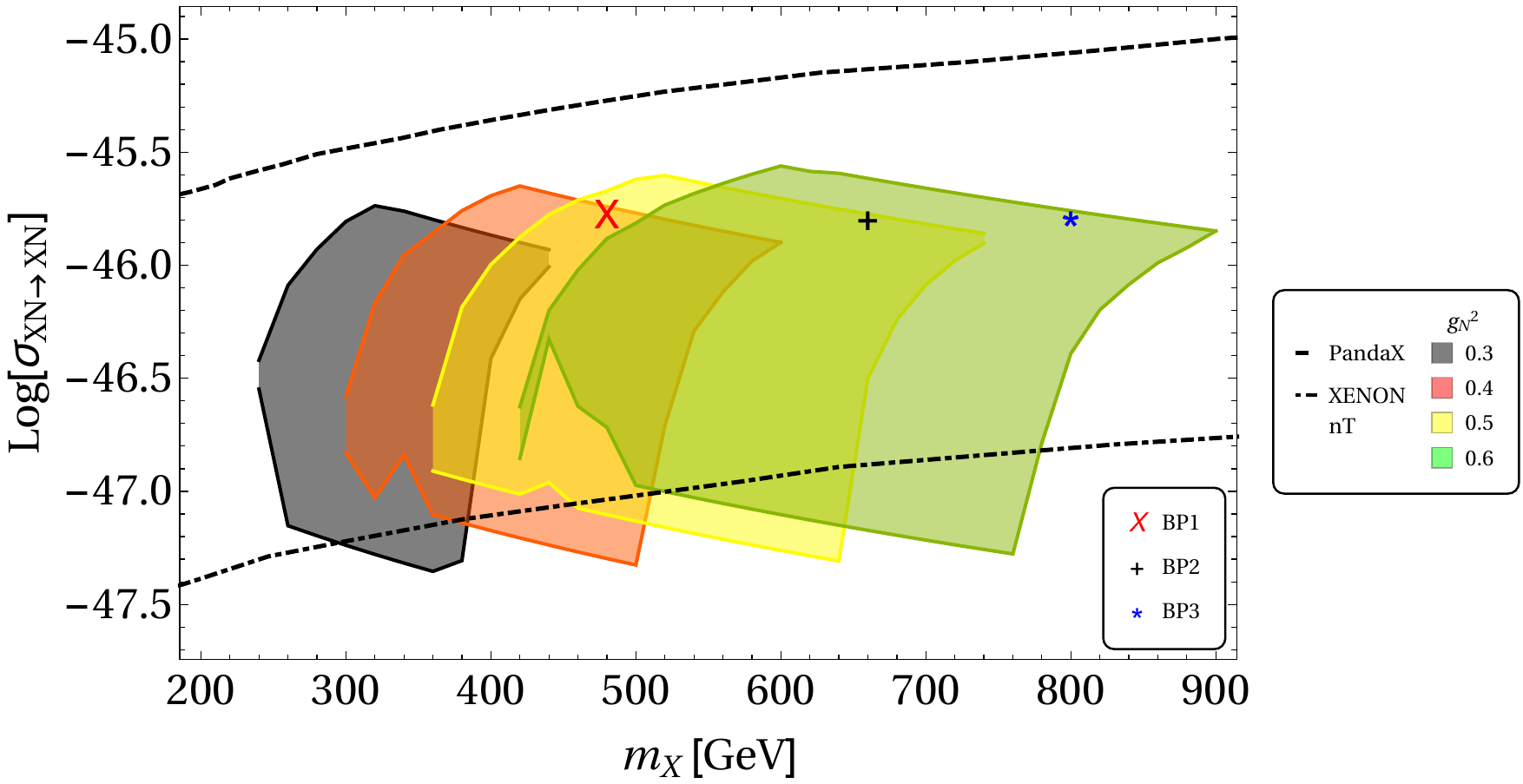}
$$
$$
\includegraphics[scale=0.58]{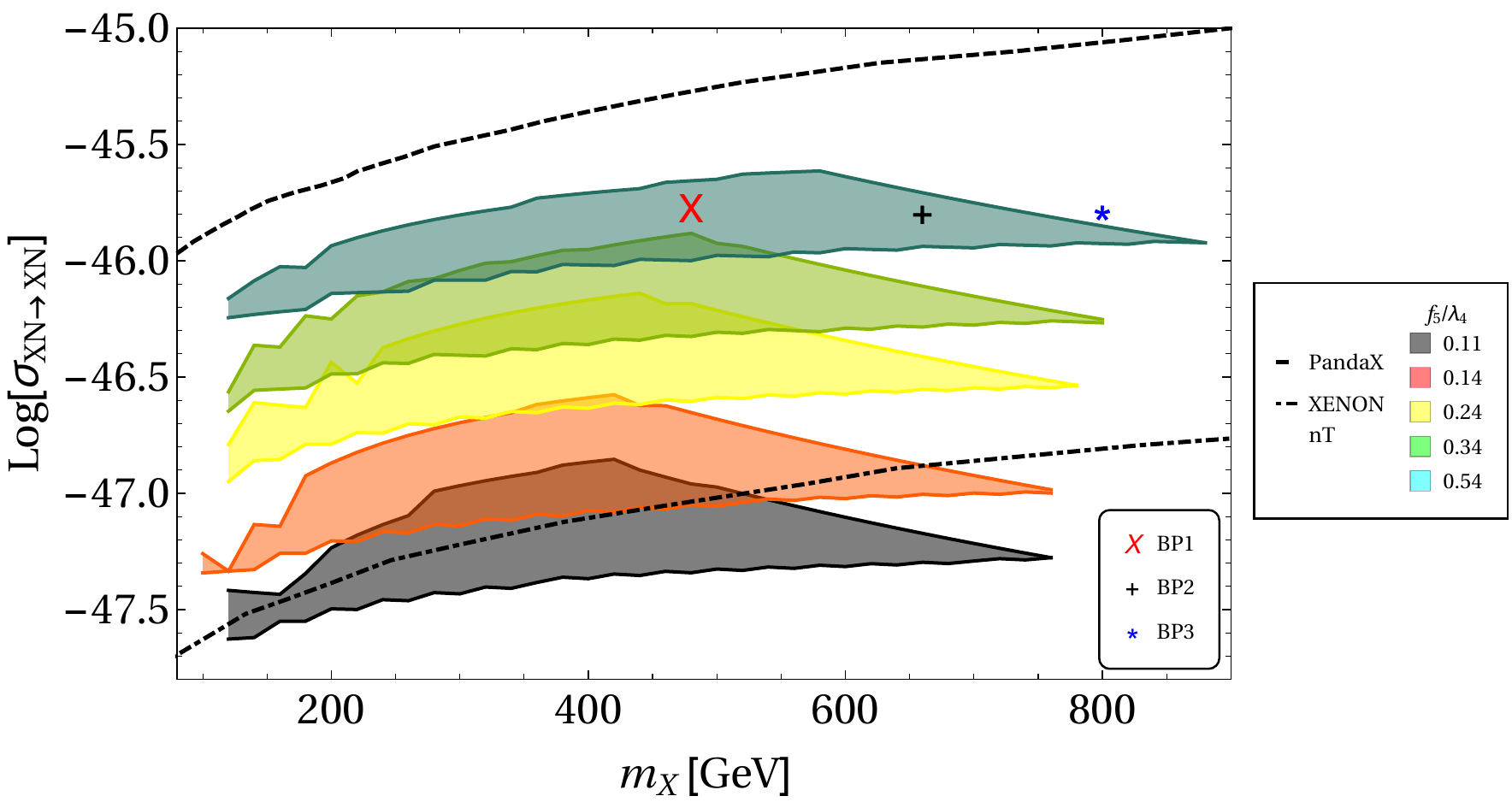}
$$
\caption{Spin independent direct search cross-section for relic density allowed parameter space for a single component $X$. Top: different $g_N^2$ regions are shown with different colours. Bottom: different $f_5/\lambda_4$ regions are shown. The exclusion limit from PandaX and future limit from XENONnT are shown through black dashed  and black dot-dashed lines respectively. }
\label{fig:ddx}
\end{figure}

Spin independent direct search cross-section for relic density allowed parameter space for the single component $X$ DM is shown in Fig.~\ref{fig:ddx}, as a function of DM mass, where $m_{\zeta_1}>m_X>m_{\zeta_2}$. On the upper panel, we show $g_N^2$ dependence through different colour shades, while on the lower panel, we show the dependence on $\frac{f_5}{\lambda_4}$ through different colour shades. The exclusion limit from PandaX is shown by the black dashed line and future limit from XENONnT~\cite{Aprile:2017iyp} is also shown by the black dot-dashed line. We see that single component $X$ DM fits nicely between these two curves, giving this model a chance to be discovered by future direct search experiments. Note that $g_N^2$ has low sensitivity to the direct search cross-sections as the constant $g_N^2$ planes are placed horizontally along larger DM mass. On the other hand, constant $\frac{f_5}{\lambda_4}$ planes are stacked vertically, yielding larger direct search cross-sections for larger $\frac{f_5}{\lambda_4}$.

Finally, we tabulate some benchmark points (BP) in Table.~\ref{tab:bp}, which satisfy both relic density and direct search constraints. Here we note that as we have chosen these points between $2m_\Delta>m_X>m_\Delta$, $X$ is the only DM component between $X$ and $\Delta$ in the degenerate scalar triplet scenario. However, as we are also choosing the heavy neutrino in the ballpark of few hundreds of GeVs and $m_N<m_{\zeta_1}$, heavy neutrinos are stable and contribute to DM relic (although the contribution is negligible). The fate of heavy neutrinos as DM is elaborated in subsection.~\ref{sec:rhn}. These BPs will be used further for the collider analysis in section \ref{sec:collider pheno}.

\begin{table}[htb!]
\small
\setlength\tabcolsep{1.5pt}
\setlength\thickmuskip{1.5mu}
\setlength\medmuskip{1.5mu}
\small
\begin{center}
\begin{tabular}{|c|c|c|c|c|c|c|c|c|c|}
\hline
Benchmark & $g_N$ & $\frac{f_5}{\lambda_4}$ & $m_X$  & $m_{\zeta_2}$ & $m_{\zeta_1}$ & $m_{n_{1R}}$ & $\Omega_X h^2$& $\sigma^{X}_{DD}$ & $\Omega_{n_{1,2}}h^2$ \\ [0.5ex] 
Point &  &  & (GeV) &  (GeV) & (GeV) & (GeV) & & $(cm^2)$  & \\ [0.5ex] 
\hline\hline

BP1 & 0.64 & 0.56 & 480 & 330 & 620 & 450 & 0.12 & $10^{-45.7802}$ & 0.004 \\
\hline
BP2 & 0.70 & 0.60 & 660 & 350 & 700 & 450 & 0.12 & $10^{-45.7918}$ & 0.004 \\
\hline
BP3 & 0.77 & 0.59 & 800 & 410 & 820 & 450 & 0.12 & $10^{-45.7839}$ & 0.004  \\
\hline
\end{tabular}
\end{center}
\caption {Choices of the benchmark points used for collider analysis. Masses, couplings, relic density and direct search cross-sections for the DM candidates are tabulated where $2 m_{\Delta}>m_X>m_{\Delta}$. $X$ has dominant contribution to relic density, while a subdominant contribution comes from $n_{1,2}$.} 
\label{tab:bp}
\end{table}

\begin{figure}[htb!]
$$
\includegraphics[scale=0.35]{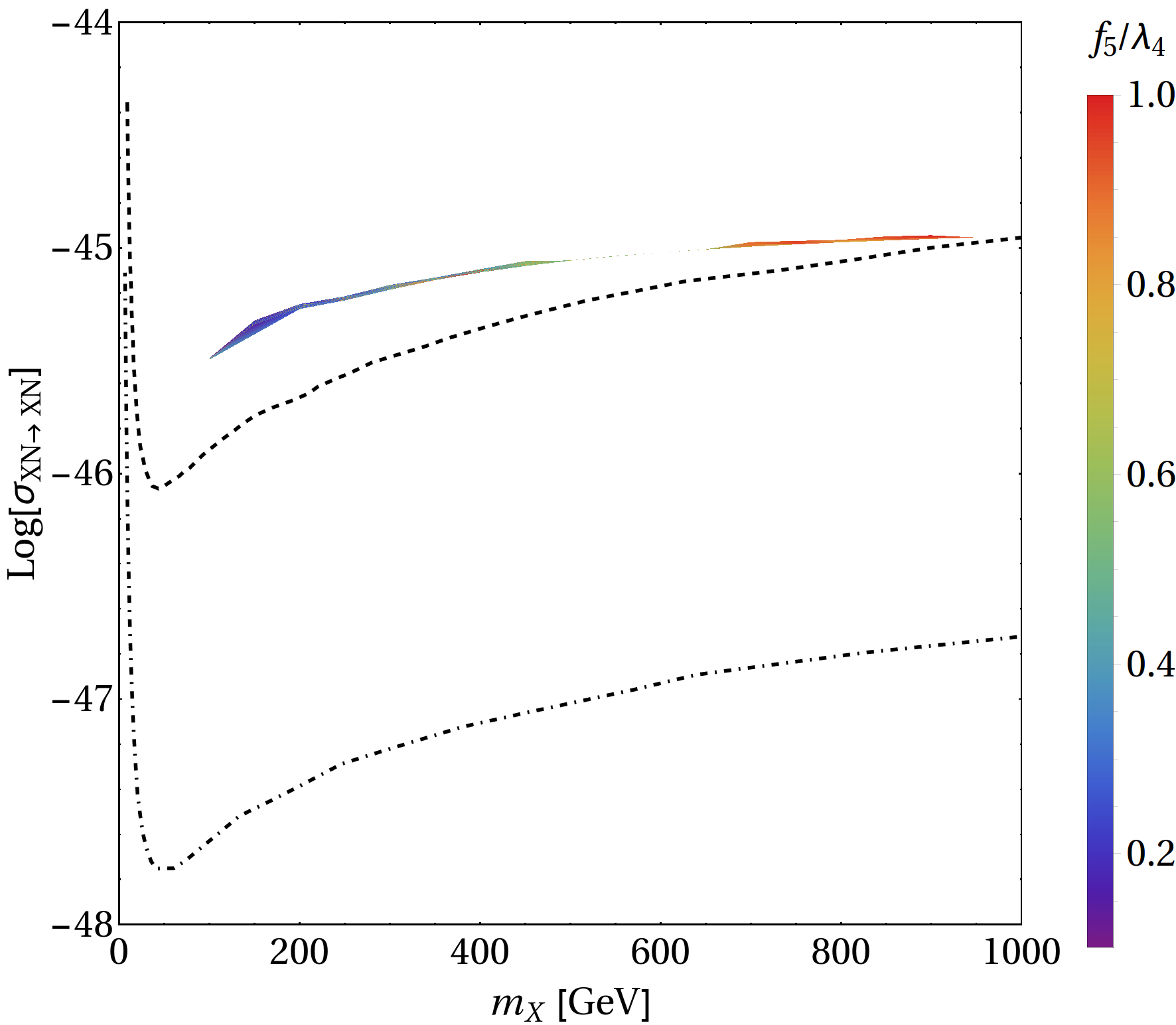}
$$
\caption{Direct search parameter space for single component VBDM $X$ when $m_X<m_{\zeta_2}$. The region allowed by relic density is extremely slim as the only annihilation channel available for $X$ is to SM. The colourbar shows different values of $f_5/\lambda_4$. The exclusion limit from PandaX and future limit from XENONnT are shown through black dashed  and black dot-dashed lines respectively.}
\label{fig:mxltmzeta2}
\end{figure}

Finally in Fig.~\ref{fig:mxltmzeta2} we show the parameter space for case (ii), where $m_X<m_{\zeta_2}$. Under this condition $X$ can only annihilate into SM via Higgs portal. This reduces the annihilation cross-section, thus increasing the relic abundance, which in turn increases the $f_5/\lambda_4$ required to obtain correct relic density. As the same coupling now controls both relic density and direct search, hence all of the relic density allowed region is ruled out by direct detection constraint as can be seen in Fig.~\ref{fig:mxltmzeta2}. 

\subsection{$\Delta_1$ and $\Delta_2$ as degenerate two component scalar DM}
\label{sec:degenerate DM}

\begin{figure}[htb!]
$$
\includegraphics[scale=0.45]{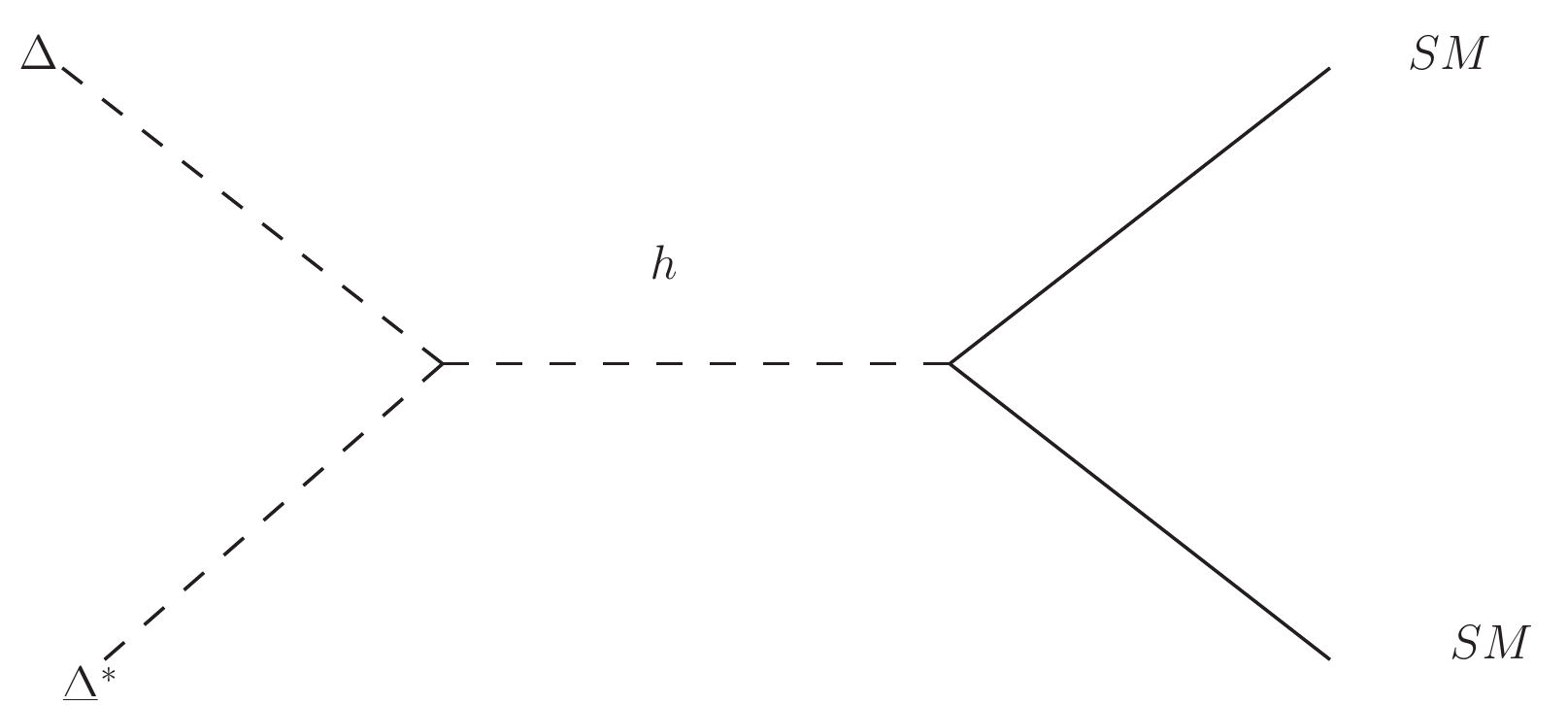}
$$
\caption{Annihilation of $\Delta$ to SM when the components of the scalar triplet are degenerate with $m_{\Delta}< m_X$.}
\label{fig:delannihil1}
\end{figure}

$\Delta_1$ can not be a single component DM in any region of the parameter space as has already been described. When $m_{\Delta}< m_X$, $\Delta_1$ and $\Delta_2$ in the degenerate triplet scenario can yield a two component DM (see Fig.~\ref{fig:regions}). In this case, each of $\Delta_1$ and $\Delta_2$ can annihilate to the SM via Higgs portal as shown in Fig.~\ref{fig:delannihil1} (where `SM' indicates all the SM gauge bosons, scalar and fermions).  The annihilation cross section at threshold is given by:

\bea
\begin{split}
\langle\sigma v_{rel}\rangle_{m_{\Delta}<m_X} &=\frac{f_8^2}{32\pi m_{\Delta}^2} \sqrt{1-\frac{m_h^2}{m_{\Delta}^2}}\left(\frac{\left(4 m_h^2-m_{\Delta}^2\right)^2}{\left(4 m_{\Delta}^2-m_h^2\right)^2+\Gamma^2 m_h^2}\right)+\\& \frac{3 f_8^2}{8\pi}\sqrt{1-\frac{m_f^2}{m_{\Delta}^2}}\frac{m_f^2}{\left(4 m_{\Delta}^2-m_h^2\right)^2+\Gamma^2 m_h^2}\\ & +\frac{f_8^2}{8\pi m_{\Delta}^2}\sqrt{1-\frac{m_W^2}{m_{\Delta}^2}}\frac{m_W^4}{\left(4m_{\Delta}^2-m_h^2\right)^2}\left(2+\frac{\left(2 m_{\Delta}^2-m_W^2\right)^2}{m_W^4}\right)+\\ & \frac{f_8^2}{8\pi m_{\Delta}^2}\sqrt{1-\frac{m_Z^2}{m_{\Delta}^2}}\frac{m_Z^4}{\left(4m_{\Delta}^2-m_h^2\right)^2}\left(2+\frac{\left(2 m_{\Delta}^2-m_Z^2\right)^2}{m_Z^4}\right).
\end{split}
\eea

The free parameters for DM analysis in this region are:

\bea
\{f_8,m_{\Delta}\}.
\eea

The relic density of such a scenario, will be described by 
 \bea
 \Omega_{\text{total}}=2~\Omega_{\Delta},
 \eea
 where the factor of `2' is because $\Delta_1$ and $\Delta_2$ are degenerate.

\begin{figure}[htb!]
\centering
\includegraphics[scale=0.4]{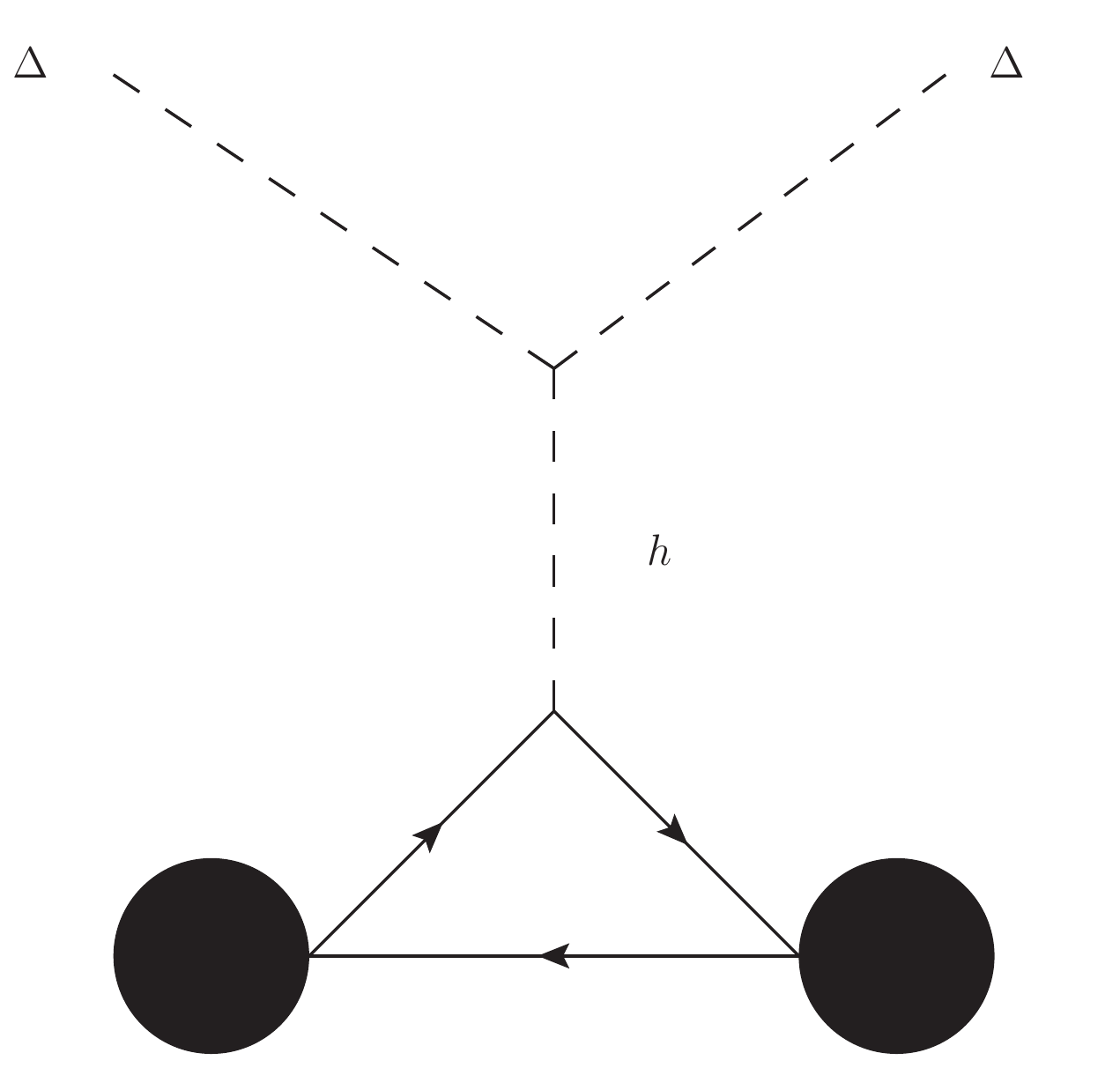} 
\caption{Direct search diagram for scalar DM.}
\label{fig:deldd}
\end{figure}

Direct search for both $\Delta_1$ and $\Delta_2$ again follows through the t-channel Higgs portal graph as shown in Fig.~\ref{fig:deldd}. 
The spin-independent DM-nucleon scattering cross section is given by~\cite{Belanger:2008sj}:
\bea
\sigma_{N_{i}}^{\text{SI}}=\frac{\alpha_N^2 \mu_{N}^2}{4\pi m_{\Delta_{i}}^2},
\label{eq:dd1comp}
\eea
where $\alpha_N$ is the effective DM-nucleon vertex (folded with form factors etc.), which is  given by

\begin{equation}
\alpha_N=\frac{m_N f_8}{m_h^2}\left[f_{T_{u}}^{(N)}+f_{T_{d}}^{(N)}+f_{T_{s}}^{(N)}+\frac{2}{27}\left[1-\left(f_{T_{u}}^{(N)}+f_{T_{d}}^{(N)}+f_{T_{s}}^{(N)}\right)\right]\right],
\end{equation}
$N$ stands for both proton and neutron, and $\mu_N$ is the DM-nucleon reduced mass, and the total cross section per nucleon is given by
 \bea
\sigma^{\text{SI}}_i=\frac{ \mu_{n}^2}{4 \pi \, A^2\,m_{\Delta_{i}}^2}\left[ \alpha_p Z + \alpha_n (A-Z) \right]^2,
\label{eq:dd1comp}
\eea
with $\mu_n$ being the DM-nucleus reduced mass.

\begin{figure}
$$
 \includegraphics[scale=0.3]{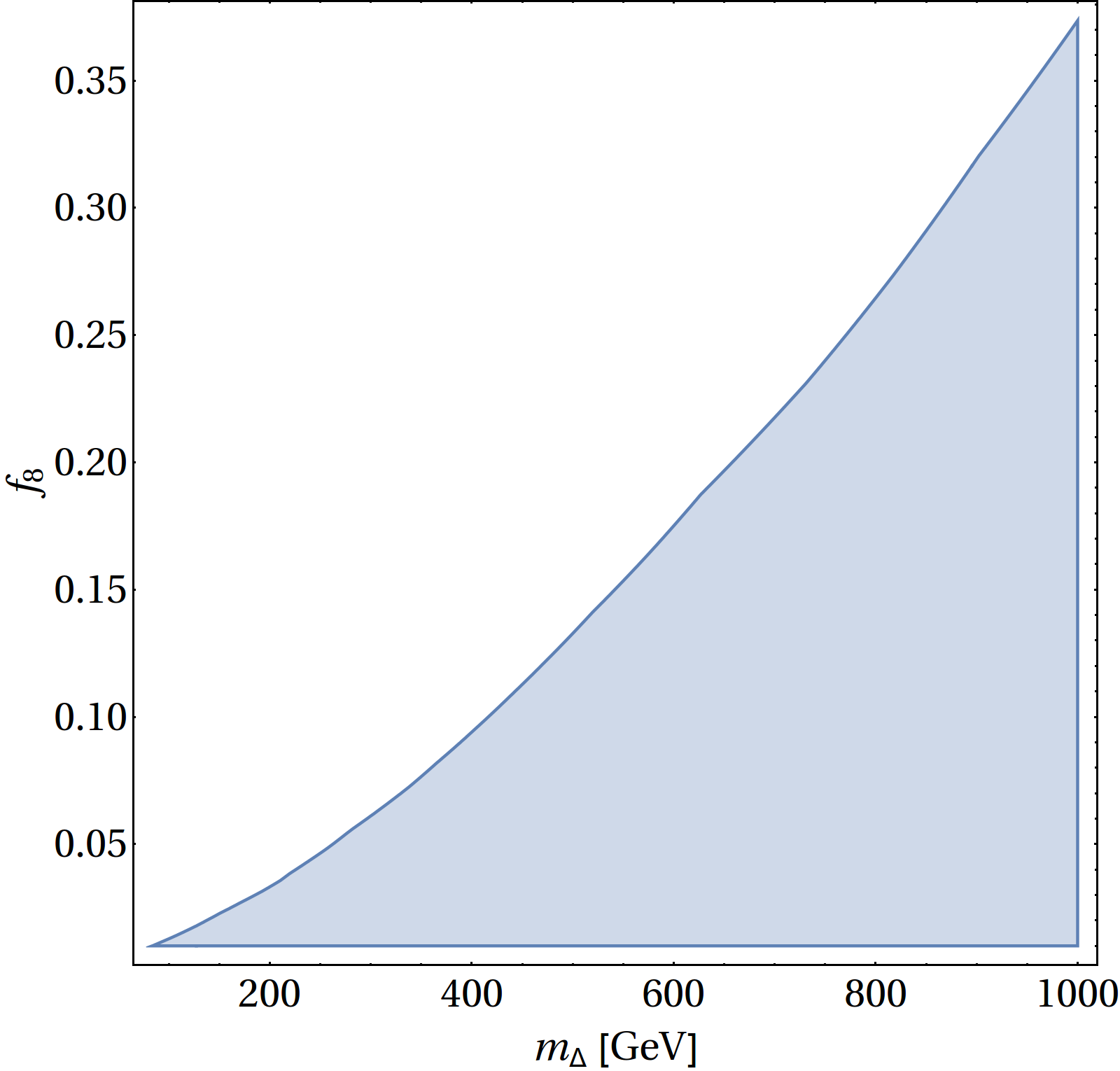}
 \includegraphics[scale=0.3]{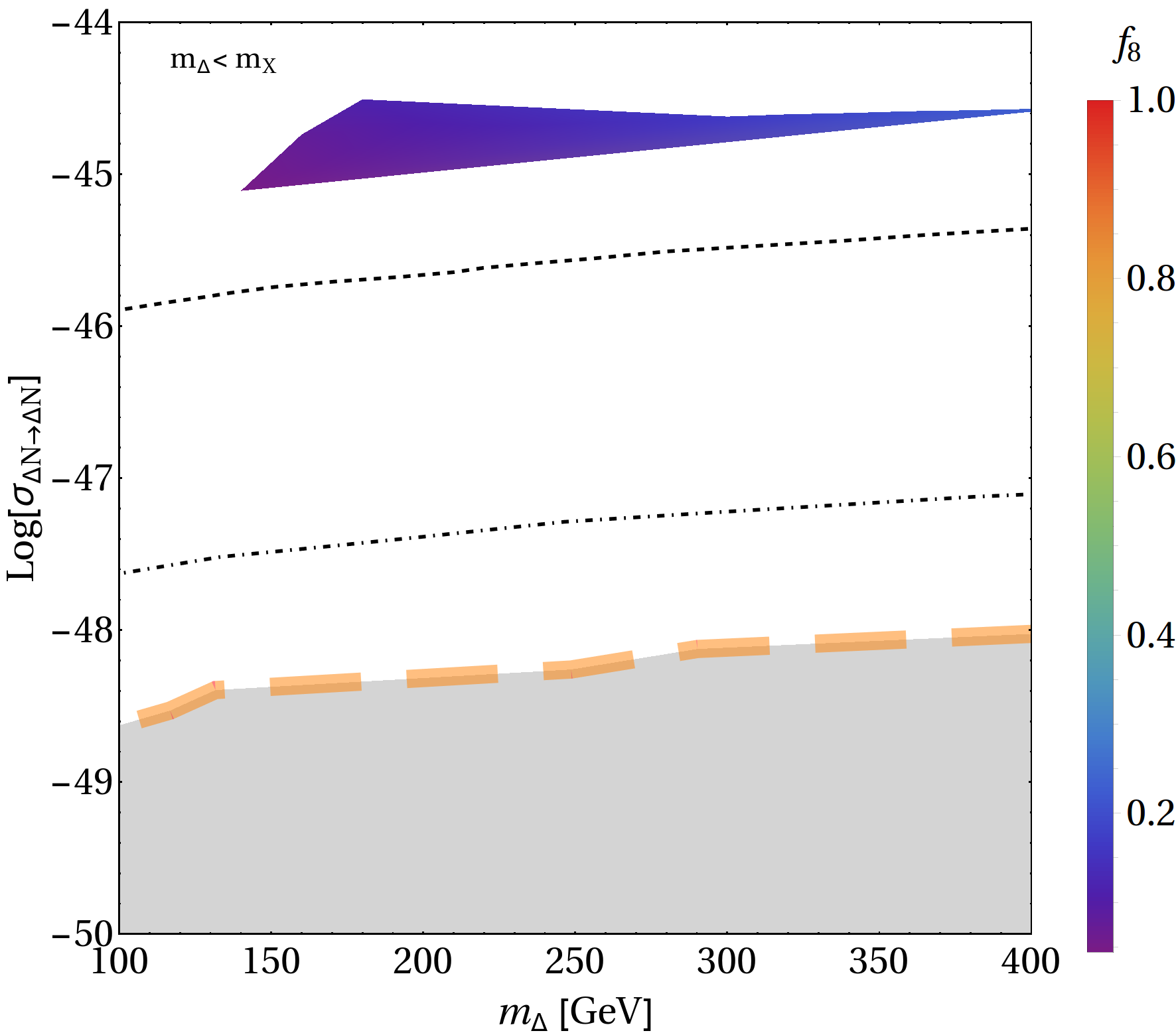}
 $$
 \caption{Left: Allowed values of $f_8$ which satisfy bounds from PandaX for scalar DM scenario. Right: Relic density allowed parameter space for degenerate two component scalar DM. Limits from PandaX, future prediction of XENONnT and neutrino floor are shown in Black dashed, Black dot-dashed and thick orange line respectively.}
 \label{fig:degenare2comp}
\end{figure}

For multi-component DM case, we can express the effective spin independent direct search cross-section of one individual component to be multiplied by roughly the percentage by which it is present~\cite{Cao:2007fy}. For two-component degenerate DM case however, the individual cross-sections can be added together as they are indistinguishable DMs, with same mass and coupling. Therefore, the effective spin-independent direct search cross-section can be expressed as:
\bea
\sigma_{\text{SI}}^{\text{eff}}\left(n_i\right)= 2\times \frac{\Omega_i}{\Omega_T}\sigma_{n_{i}}^{\text{SI}}= \frac{\alpha_n^2 \mu_{n}^2}{4\pi m_{\Delta_{i}}^2}.	
\label{eq:dd2comp}
\eea

We can see the fate of this degenerate two-component DM scenario from relic density and direct search constraints, summarised in Fig.~\ref{fig:degenare2comp}. On the LHS of Fig.~\ref{fig:degenare2comp} we show allowed values of $f_8$ as a function of $m_{\Delta}$ to respect direct search bound from PandaX, while on the RHS we show the direct search parameter space allowed by relic density in the degenerate two-component set-up (\{$\Delta_1$,$\Delta_2$\}). This essentially shows that direct search constraint severely discard this parameter space of the model. This is easy to appreciate as the channel which helps the DM to freeze-out also crucially controls the direct search cross-section of the DM; more importantly the degeneracy of the two components ensure twice as large annihilation which causes the couplings to be increased appropriately to yield an enhancement in direct search cross-sections.

\subsection{$\Delta_1$ and $X$ as two component DM}
\label{sec:x-delta DM}

\begin{figure}[htb!]
$$
\includegraphics[scale=0.6]{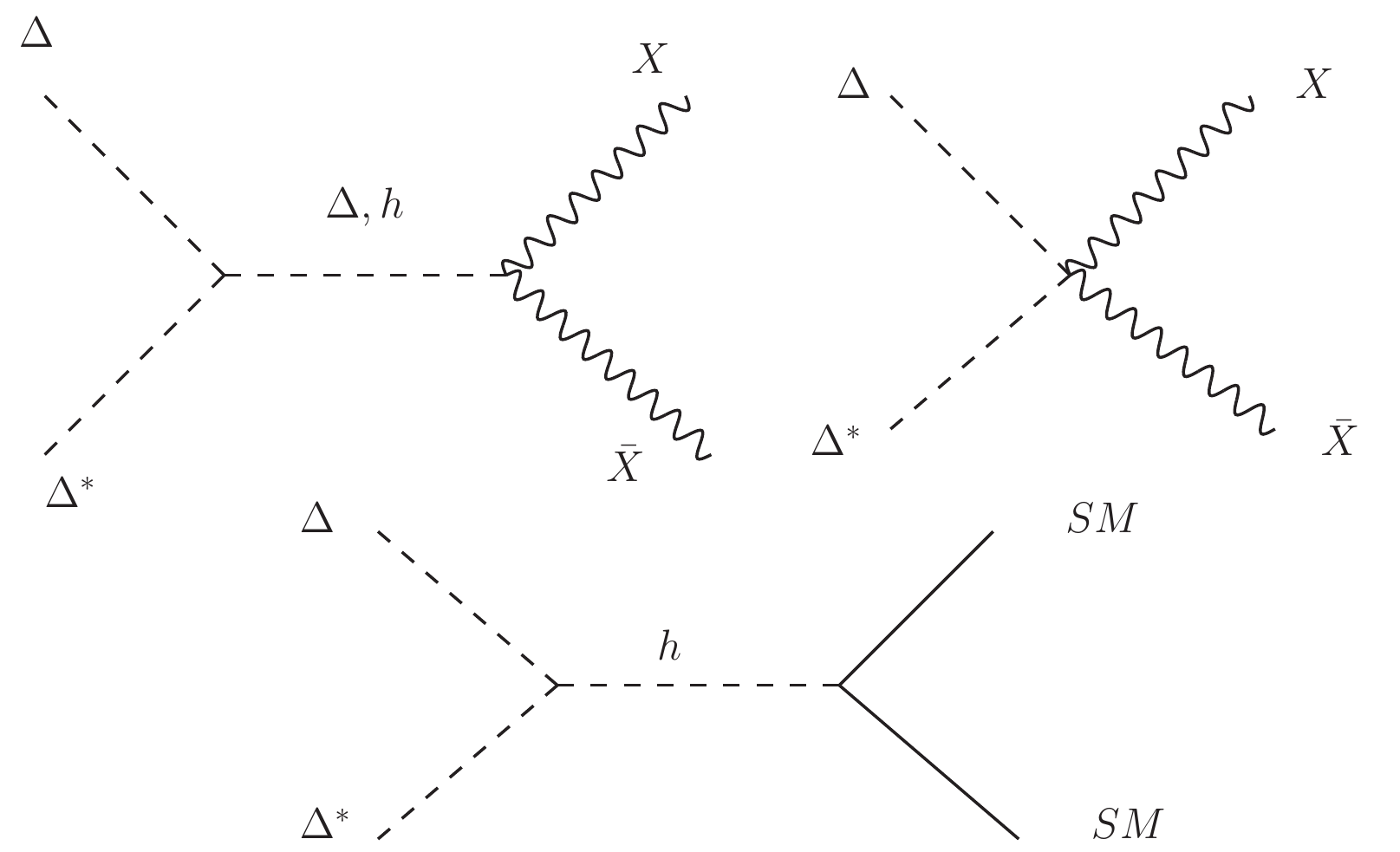}
$$
\caption{Annihilation of $\Delta_1$ to $X$ and SM when $m_X<m_{\Delta}<2 m_X$.}
\label{fig:delannihil}
\end{figure}

 $X$ and $\Delta_1$ can form two component DM when $m_X<m_{\Delta}<2 m_X$ (see Fig.~\ref{fig:regions}) in degenerate triplet scenario. First of all, here $\Delta_1$ can annihilate to $X \bar {X}$ additionally, shown in the upper panel of Fig.~\ref{fig:delannihil}, which was not accessible earlier when $m_{\Delta}<m_X$. This DM-DM conversion will play a crucial role in this region of the parameter space as we will elaborate. 
The annihilation cross-section for $\Delta_1$ will include the contributions from these additional graphs and will read as follows:
\bea
\begin{split}
\langle\sigma v_{rel}\rangle_{m_{\Delta}>m_X} &= \frac{g_N^4}{32\pi m_{\Delta}^2} \sqrt{1-\frac{m_X^2}{m_{\Delta}^2}}\left[2+\left(\frac{2 m_{\Delta}^2}{m_X^2}-1\right)^2\right]\\ & \left[1-\sqrt{2}f_8 \left(\frac{f_5}{\lambda_4}\right) \frac{v^2 \left(4 m_{\Delta}^2-m_h^2\right)}{\left(4 m_{\Delta}^2-m_h^2\right)^2+\Gamma_h^2 m_h^2}+\frac{1}{2} f_8^2 \left(\frac{f_5}{\lambda_4}\right) ^2 \frac{v^4}{\left(4 m_{\Delta}^2-m_h^2\right)^2+\Gamma_h^2 m_h^2}\right]\\ &+ \frac{f_8^2}{32\pi m_{\Delta}^2} \sqrt{1-\frac{m_h^2}{m_{\Delta}^2}}\left(\frac{\left(4 m_h^2-m_{\Delta}^2\right)^2}{\left(4 m_{\Delta}^2-m_h^2\right)^2+\Gamma^2 m_h^2}\right)+\\& \frac{3 f_8^2}{8\pi}\sqrt{1-\frac{m_f^2}{m_{\Delta}^2}}\frac{m_f^2}{\left(4 m_{\Delta}^2-m_h^2\right)^2+\Gamma^2 m_h^2}\\ & +\frac{f_8^2}{8\pi m_{\Delta}^2}\sqrt{1-\frac{m_W^2}{m_{\Delta}^2}}\frac{m_W^4}{\left(4m_{\Delta}^2-m_h^2\right)^2}\left(2+\frac{\left(2 m_{\Delta}^2-m_W^2\right)^2}{m_W^4}\right)+\\ & \frac{f_8^2}{8\pi m_{\Delta}^2}\sqrt{1-\frac{m_Z^2}{m_{\Delta}^2}}\frac{m_Z^4}{\left(4m_{\Delta}^2-m_h^2\right)^2}\left(2+\frac{\left(2 m_{\Delta}^2-m_Z^2\right)^2}{m_Z^4}\right).
\end{split}
\label{eq:deltannihil}
\eea

The parameters for DM analysis in this case are given by:

\bea
\{g_N^2,f_8,m_{\Delta},m_X\}.
\eea

\begin{figure}[htb!]
 $$
 \includegraphics[scale=0.34]{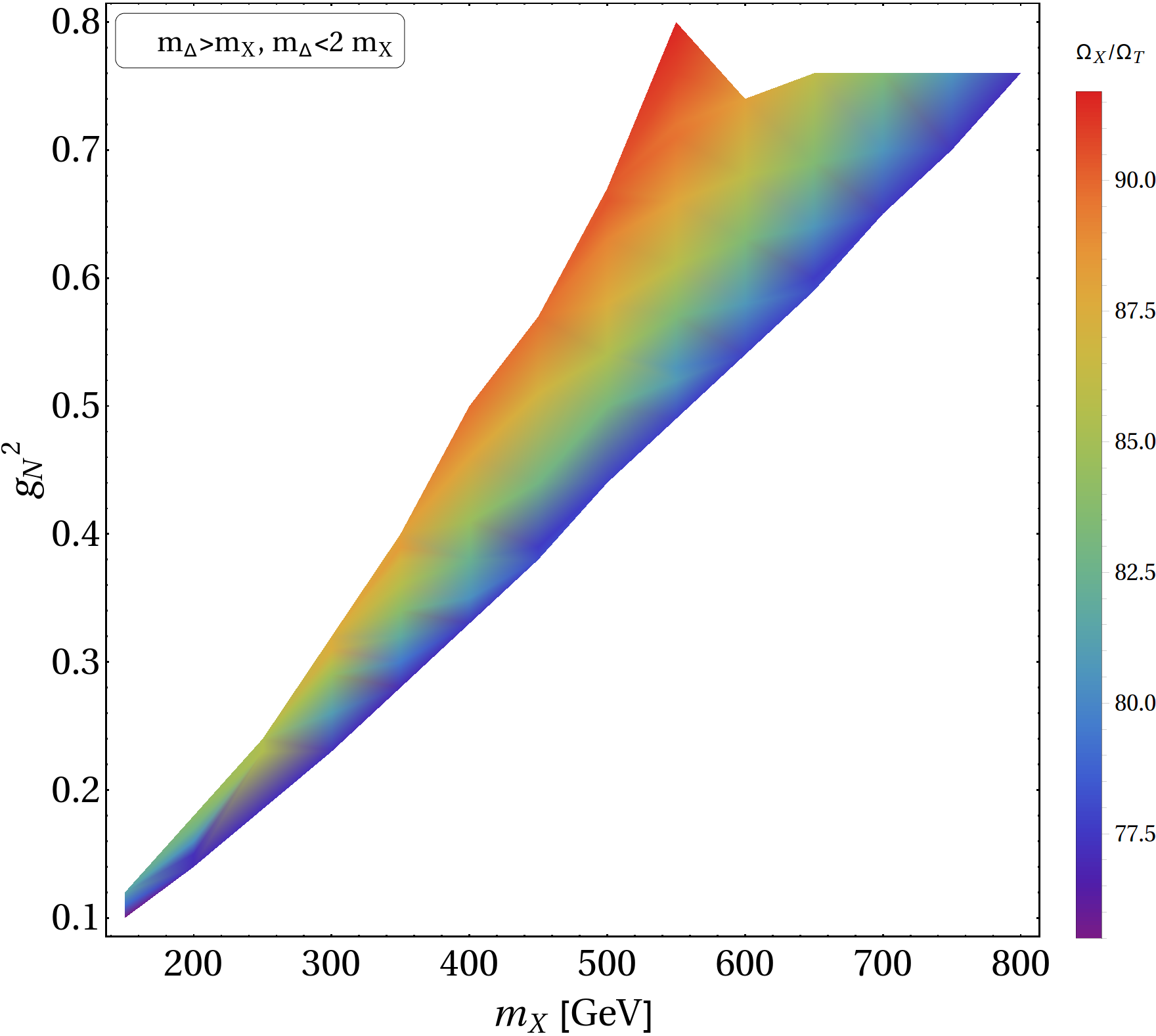}
 \includegraphics[scale=0.34]{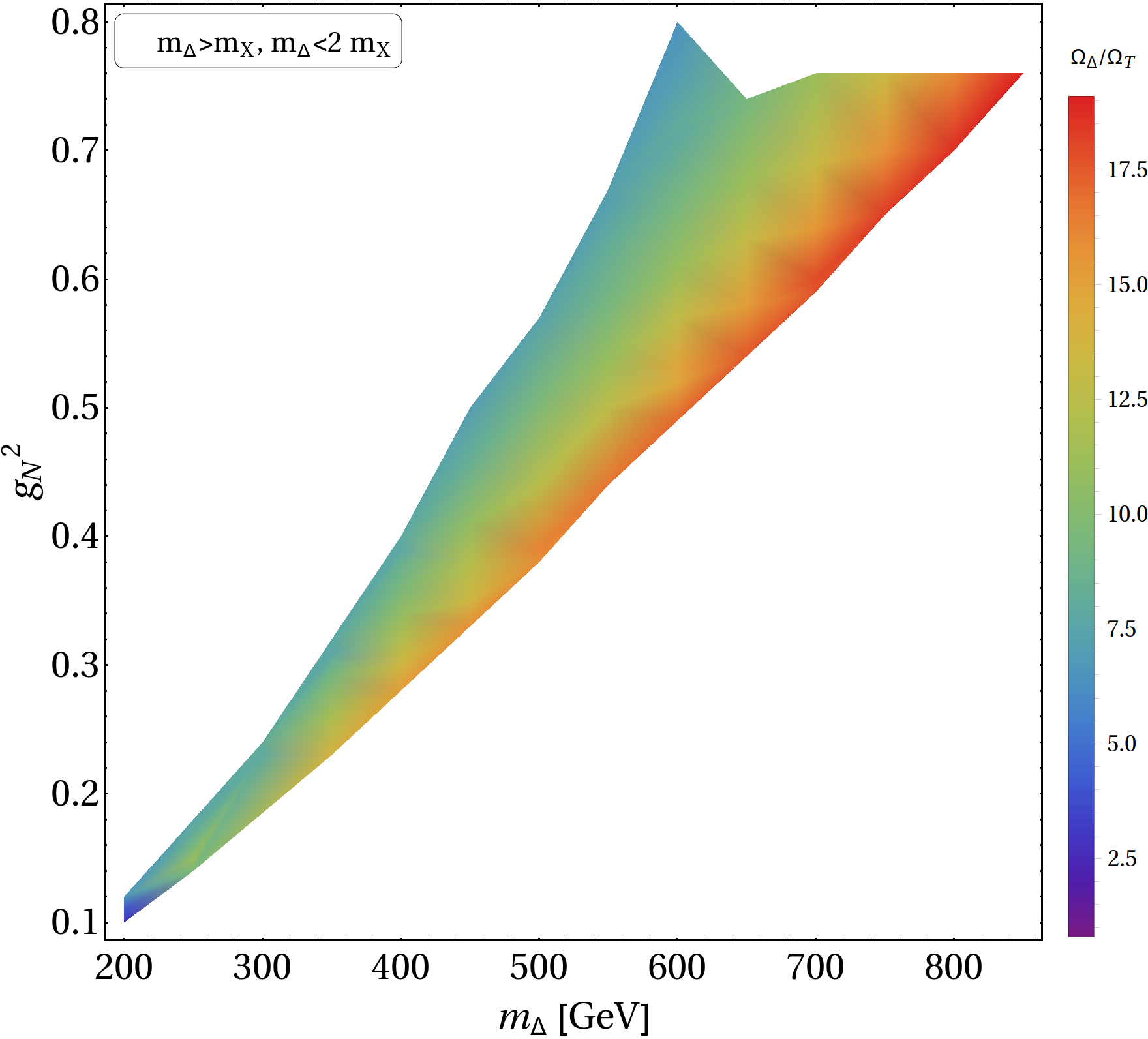}
 $$
 $$
 \includegraphics[scale=0.34]{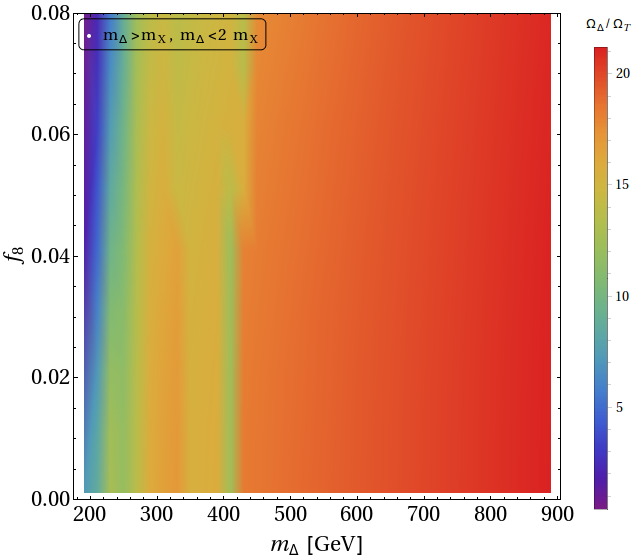}
 $$
 \caption{Top left: Relic density allowed parameter space for two component DM set-up in $m_X-g_N^2$ plane, where the colour shades indicate $\Omega_X/\Omega_T$. Top right: Same in $m_\Delta-g_N^2$ plane, where the colour shades indicate $\Omega_\Delta/\Omega_T$. Bottom panel: Same in $m_\Delta-f_8$ plane where colour shades indicate $\Omega_\Delta/\Omega_T$. Here $m_{\Delta}=m_{\zeta_1}=m_X+50$ and $m_{\zeta_2}=m_X-50$ has been chosen for illustration.}
 \label{fig:relicdelta}
\end{figure}


As we have already elucidated in Eq.~\ref{eq:beq4} of Sec.~\ref{sec:X DM}, the BEQ can be expressed in terms of the dimensionless quantity $x=m/T$, where $m$ is the mass of the DM. However, in the two-component case, we have a coupled Boltzmann equation, due to DM-DM interactions. Here, using a common $x$ is problematic since now there are two DM candidates with different masses: $\{m_{\Delta},m_X\}$. This issue didn't arise in the previous case of $\Delta_1, \Delta_2$ as they were degenerate and didn't have an  effective DM-DM interactions. The way-out for this non-degenerate scenario is to introduce a reduced mass: $\mu=\frac{m_{\Delta}m_X}{m_{\Delta}+m_X}$, in terms of which the BEQs read~\cite{Bhattacharya:2016ysw}:

\bea
\begin{split}
\nonumber \frac{dy_1}{dx} =& A\left[\langle\sigma v_{\Delta\Delta^{*}\to SM SM}\rangle\left(y_1^2-y_1^{EQ^2}\right)+\langle\sigma v_{\Delta\Delta^{*}\to X\bar{X}}\rangle\left(y_1^2-\frac{y_1^{EQ^2}}{y_2^{EQ^2}}y_2^2\right)\right], 
\end{split}
\label{eq:coupedbeq1}
\eea

\bea
\begin{split}
\frac{dy_2}{dx} =& A\left[\langle\sigma v_{X\bar{X}\to SM SM}\rangle\left(y_2^2-y_2^{EQ^2}\right)-\langle\sigma v_{\Delta\Delta^{*}\to X\bar{X}}\rangle\left(y_1^2-\frac{y_1^{EQ^2}}{Y_2^{EQ^2}}y_2^2\right)\right], 
\end{split}
\label{eq:coupedbeq1}
\eea

where $A=-0.264~M_{PL} \sqrt{g_{*}}\frac{\mu}{x^2}$ and the equilibrium distribution, recast in terms of $\mu$ has the form:

\bea
y_{i}^{EQ}\left(x\right)= 0.145 \frac{g}{g_{*}} x^{3/2} \left(\frac{m_{i}}{\mu}\right)^{3/2} e^{-x m_{i}/\mu},
\eea

with $i\in(X,\Delta)$. It has already been established in the case of multicomponent DM scenario, that the relic density of the heavier one is affected by the added annihilation to the other DM component, while the one for the lighter component remains the same~\cite{Bhattacharya:2016ysw}. Therefore, we can safely use an approximate analytical solution in determining the relic density for individual components as follows:

\bea
\nonumber \Omega_{X}h^2=\frac{854.45\times10^{-13}}{\sqrt{g_{*}}} y_{X}\left(x_{\infty}\right) &\simeq& \frac{0.1 ~\rm{pb}}{\langle\sigma v\rangle_{\text{eff}}},\\
\Omega_{\Delta}h^2=\frac{854.45\times10^{-13}}{\sqrt{g_{*}}} y_{\Delta}\left(x_{\infty}\right) &\simeq& \frac{0.1 ~\rm{pb}}{\langle\sigma v\rangle_{\Delta\Delta^{*}\to X\bar{X}}+\langle\sigma v\rangle_{\Delta\Delta^{*}\to SM~SM}},
\eea
where for annihilation of $X$, $\langle\sigma v\rangle_{\text{eff}}$ is given by Eq.~\ref{eq:coann}.

\begin{figure}[htb!]
$$
 \includegraphics[scale=0.52]{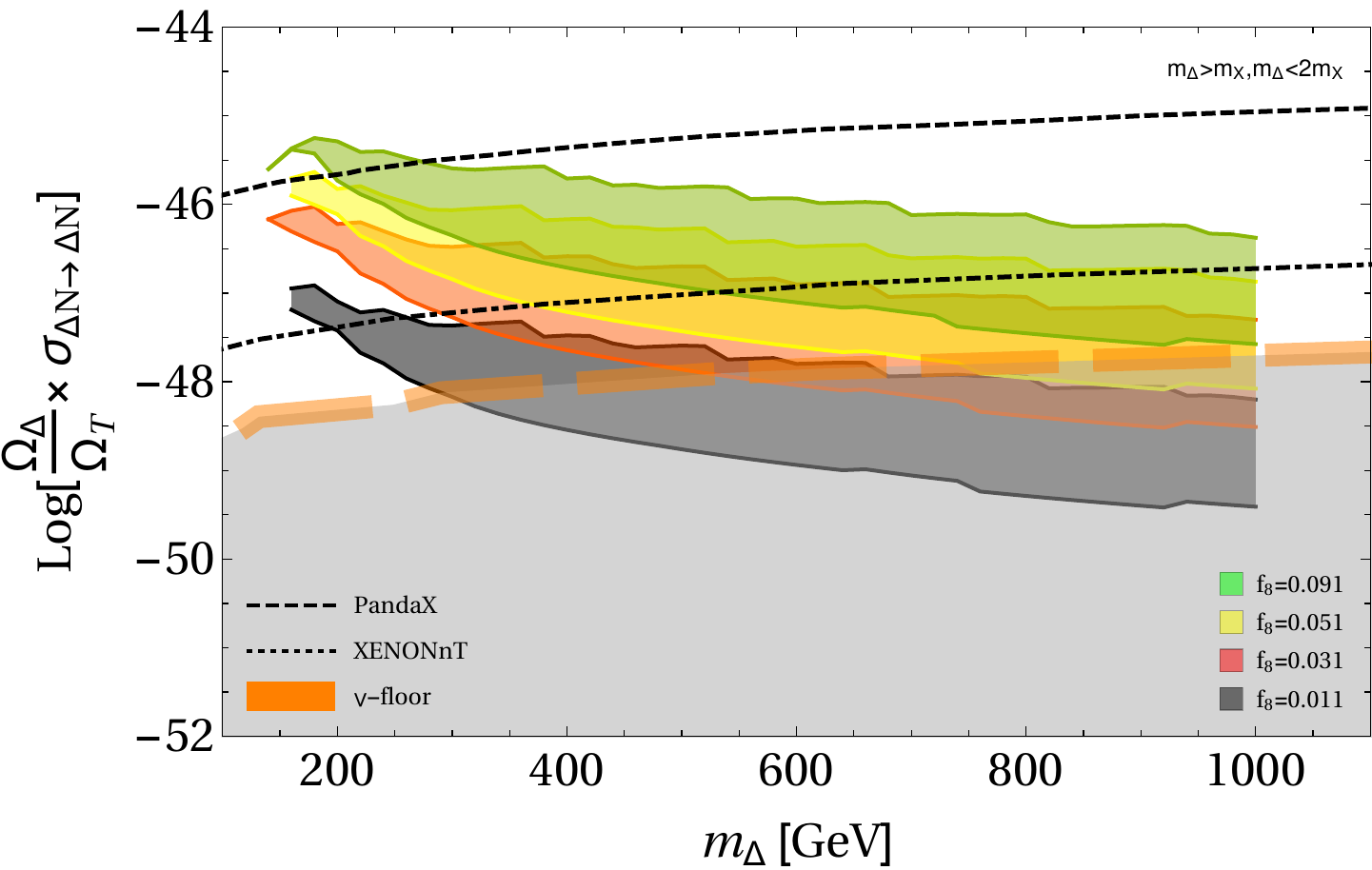}\hspace{0.3cm}
  \includegraphics[scale=0.52]{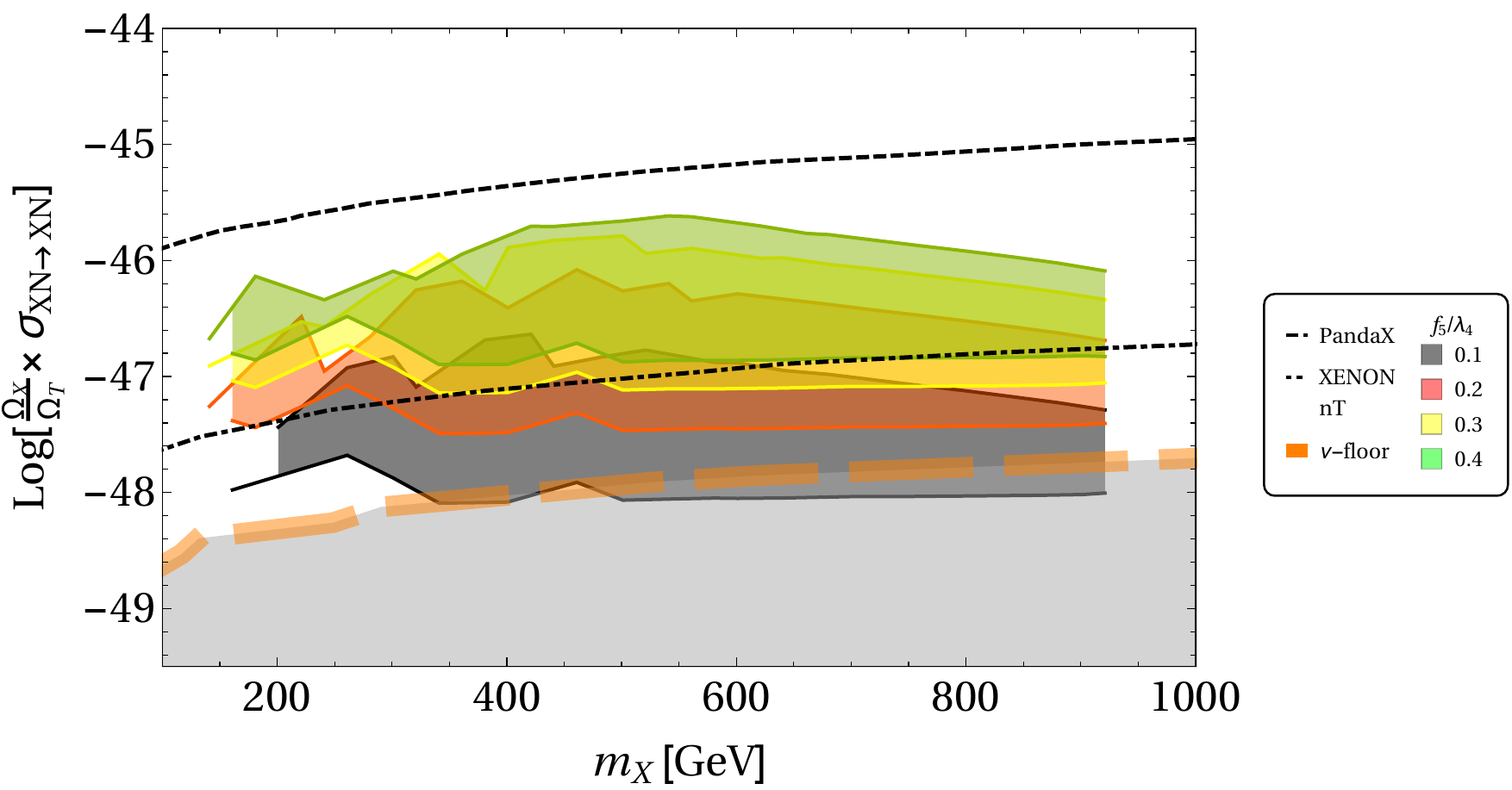}
 $$
 \caption{LHS: Spin independent effective direct search cross-section for $\Delta$, in terms of $m_{\Delta}$ vs. $Log{(\frac{\Omega_{\Delta}}{\Omega_T}\times\sigma_{\Delta N\to \Delta N})}$, when it is a part of two component dark matter scenario with $\rm{DM}:\{\Delta, X\}$. Allowed region of relic density parameter space have been divided into different $f_8$ values. RHS: Same for $X$ in terms of $m_X$ vs. $Log~(\frac{\Omega_X}{\Omega_T}\times\sigma_{X N\to X N})$, where different coloured regions correspond to different values of $f_5/\lambda_4$. In both the plots the bound from PANDA, future sensitivity of XENONnT and neutrino floor are depicted.}
 \label{fig:deltadir}
\end{figure}

The relic density allowed parameter space of this two component model is shown in Fig.~\ref{fig:relicdelta}. In the top left panel, we show the allowed region in $m_X-g_N^2$ plane while the colour shades indicate $\Omega_X/\Omega_T$. Top right panel shows a similar graph in $m_\Delta-g_N^2$ plane while the colour shades indicate the fraction of $\Delta$ DM in total abundance $\Omega_\Delta/\Omega_T$. The main take from these two graphs are that, the relic density is dominated here by the $X$ component. This can be explained simply as $X$ is the lighter component, it has less annihilations to the SM. On the contrary, larger annihilations of $\Delta$ compared to $X$, depletes the abundance of this component down to $20\%$ of the total.  In the bottom panel of Fig.~\ref{fig:relicdelta}, we show that $f_8$, when varied within this limit ($\{0.001-0.1\}$), does not constrain $m_\Delta$ at all in achieving the right relic density in this multipartite framework. For ease of the scan, we choose: $m_{\Delta}=m_{\zeta_1}=m_X+50$ and $m_{\zeta_2}=m_X-50$. However, they do not have a crucial role to play unless we change the hierarchy. 

Now the question is whether the relic density allowed parameter space of the two-component set-up ($\Delta,X$) is allowed by the direct search constraint. This is depicted in Fig.~\ref{fig:deltadir}. In the LHS we show the fate of $\Delta$ in direct search plane, where in the x-axis we have $m_{\Delta}\rm~(GeV)$ and along y-axis we have effective spin-independent direct search cross section $\left(\frac{\Omega_{\Delta}}{\Omega_T}\right)\times\sigma_{\Delta N\to\Delta N}$ in log-scale. Here different colour shades represent different values of $f_8:\{0.01,0.03,0.05,0.09\}$ chosen for the scan. We can see that except for the low mass region of $\Delta$ ($m_\Delta \le$ 250 GeV) for $f_8=0.09$, the whole relic density allowed parameter space is available through direct search constraints. It is easy to understand that the higher the values of $f_8$, the higher is the effective direct search cross-section is. Therefore, direct search crucially tames the coupling $f_8 \le 0.1$.

In the RHS of Fig.~\ref{fig:deltadir} we show the parameter space allowed by relic density and direct search for $X$ in the two-component DM scenario. All of the parameter space allowed by relic density lies below the PandaX limit and a part of it even goes below the neutrino floor~\cite{Liu:2017drf}. Different coloured regions in this plot correspond to different values of $f_5/\lambda_4$, which typically controls the direct detection cross-section of $X$ as discussed earlier. As expected, smaller values of $f_5/\lambda_4$ (shown for example by the black region) produces smaller cross-section, while a larger $f_5/\lambda_4$ is ruled out by XENONnT.


\begin{table}[htb!]
\small
\setlength\tabcolsep{1.5pt}
\setlength\thickmuskip{1.5mu}
\setlength\medmuskip{1.5mu}
\small
\begin{center}
\begin{tabular}{|c|c|c|c|c|c|c|c|c|c|c|}
\hline
Benchmark & $g_N$ & $\frac{f_5}{\lambda_4}$ & $m_X$  & $m_{\zeta_2}$ & $m_{\zeta_1}$ & $m_{\Delta}$ & $\Omega_X h^2$& $\Omega_{\Delta}h^2$  & $\sigma^{\Delta}_{DD}$ & $\sigma^{X}_{DD}$ \\ [0.5ex] 
Point &  &  & (GeV) &  (GeV) & (GeV) & (GeV) & &  & $(cm^2)$  & $(cm^2)$ \\ [0.5ex] 
\hline\hline

BP4 & 0.63 & 0.30 & 481 & 320 & 621 & 500 & 0.077 & 0.043 & $10^{-50}$ & $10^{-46.35}$ \\
\hline
BP5 & 0.70 & 0.10 & 541 & 380 & 701 & 560 & 0.079 & 0.037 & $10^{-50}$ & $10^{-47.20}$ \\
\hline
BP6 & 0.83 & 0.20 & 681 & 540 & 821 & 700 & 0.087 & 0.033 & $10^{-50}$ & $10^{-46.47}$ \\
\hline
\end{tabular}
\end{center}
\caption {Choices of the benchmark points for two-component \{$X$,$\Delta$\} DM scenario. Masses, couplings, relic density and direct search cross-sections for both the DM candidates are tabulated. } 
\label{tab:bp2}
\end{table}


 In Table.~\ref{tab:bp2} we have tabulated possible values of masses of the dark gauge boson and triplet scalar for different couplings satisfying relic density and direct search for two-component DM scenario \{$\Delta_1$,$X$\}. As the collider signature of the model is independent of the choice of $m_{\Delta}$ and dependent only on the masses of the charged scalars \{$\zeta_1^{\pm}$,$\zeta_2^{\pm}$\}, the model would give rise to the same final states as that of single component vector boson DM framework as the mass hierarchy between $X$ and $\zeta$ remains unaltered.

\subsection{Fate of the heavy neutrino as DM}
\label{sec:rhn}


\begin{figure}[htb!]
$$
\includegraphics[scale=0.5]{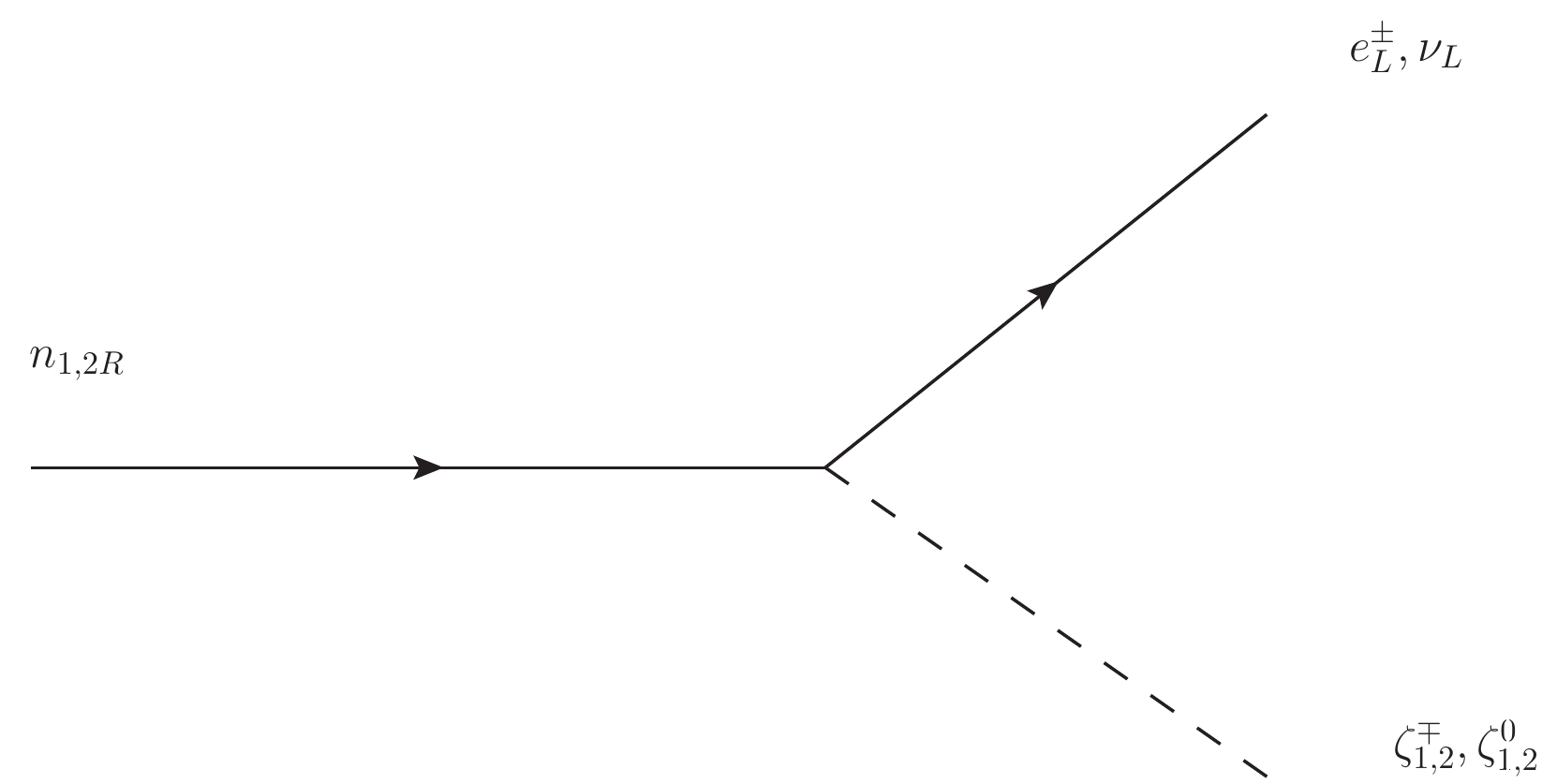}
$$
\caption{Decay of the right handed neutrinos.}
\label{fig:rhndecay1}
\end{figure}

The right handed neutrino (RHN) can decay into different final states through the Yukawa interaction mentioned in Eq.~\eqref{f-zeta} and shown in Fig.~\ref{fig:rhndecay1}. If we assume $m_{n_1}<m_{\zeta_1}$, then $n_{1R}$ is stable and contributes to the DM relic density. $n_{2R}$, on the other hand, can decay into leptons and  $\zeta_2$. As $\zeta_2$ mixes with SM Higgs, it can readily decay to SM and $n_{2R}$ can not qualify as DM. 


\begin{figure}[htb!]
\centering
\includegraphics[scale=0.6]{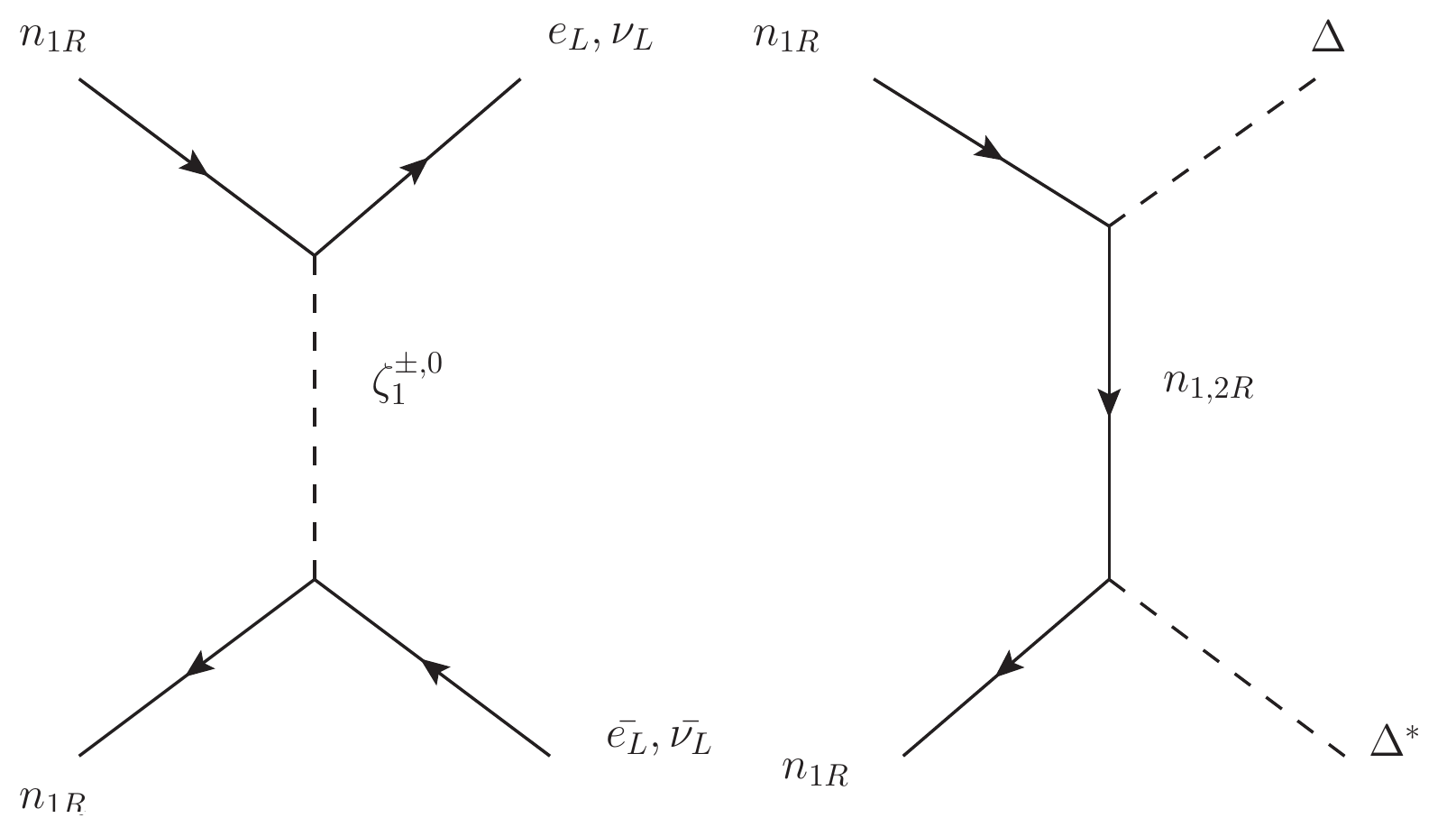}
\caption{Left: Annihilation of RHN into SM leptons via $t$-channel mediation of the heavy scalars $\zeta_1^{\pm,0}$. Right: Annihilation of RHN into exotic scalars $\Delta$ via the RHNs.}
\label{fig:rhnann}
\end{figure}

Therefore, $n_{1R}$, in this model, can contribute to the relic density if the following conditions are satisfied:
\begin{itemize}
 \item If $m_{n_1}<m_{\zeta_1}$, then $n_{1R}$ is stable.
\end{itemize}

\begin{figure}[htb!]
\centering
\includegraphics[scale=0.38]{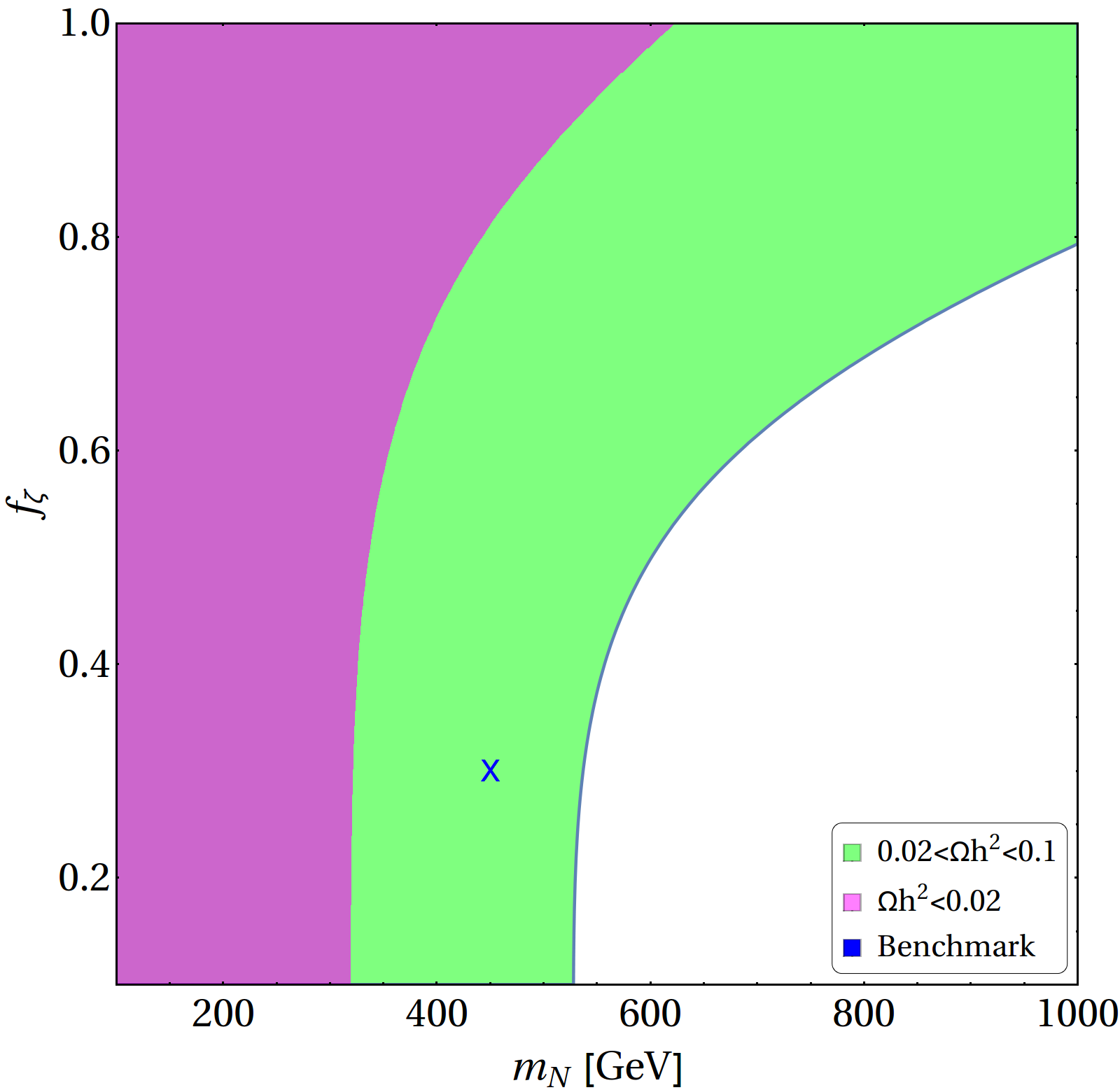}
\caption{The figure shows underabundant regions of $\Omega_n$ ($0.02<\Omega_n h^2<0.1$ in green and $\Omega_n h^2<0.02$ in pink) for different values of heavy neutrino mass $m_N$ (GeV) and Yukawa coupling $f_{\zeta}$. The typical choice of $m_N$ and coupling for the chosen BPs (Table.~\ref{tab:bp} and Table.~\ref{tab:bp2}) lies in the under abundant region shown by the blue cross.}
\label{fig:rhnrelic}
\end{figure}

As the interactions in Eq.~\ref{eq:yukawaint} suggest, the RHN can undergo annihilation via the channels shown in Fig.~\ref{fig:rhnann}. The thermally averaged cross section of these channels computed at $s=4~m_{n_{1R}}^2$ is given by:
\bea
\begin{split}
\langle\sigma v\rangle_{n_{1R}} =& \frac{f_{\zeta}^4}{32\pi}\frac{m_{n_{1R}}^2}{\left(m_{n_{1R}}^2+m_{\zeta_1}^2\right)^2}+\\ & \frac{f_{\Delta}^4}{64\pi}\left(1-\frac{m_{\Delta}^2}{m_{n_1}^2}\right)^{3/2}\left(\frac{m_{n_{1R}}^2}{\left(2 m_{n_{1R}}^2-m_{\Delta}^2\right)^2}+\frac{1}{2}\frac{m_{n_{1R}}^2}{\left(2 m_{n_2}^2-m_{\Delta}^2\right)^2}\right),
\end{split}
\eea

where we have assumed $m_{\Delta_1}=m_{\Delta_2}$ and $m_{n_{1R}}=m_{n_{2R}}$. Fig.~\ref{fig:rhnrelic} shows the under abundant region for $\Omega_n$ with $f_{\Delta}\sim\mathcal{O}(1)$ for different values of $f_{\zeta}$. $0.02<\Omega h^2<0.1$ region is shown in green, while $\Omega h^2<0.02$ is shown in pink. We can see, that our choice of the benchmark points (BP1-BP6) lies very much in the under abundant region. Therefore, we can safely ignore this contribution for our study. However, for smaller $f_\Delta~(\le 1)$, the annihilation can be smaller and the relic density will have sizable contribution. We would like to mention that adding the contribution of RHN as DM merely changes the relic density parameter space, and leaves the phenomenology of the other DM candidates unchanged. Also note that the left chiral components $(n_1,n_2)_L$ are always stable as they only have Yukawa interactions with the triplet scalar following Eq.~\eqref{f-delta}, so that they can also serve as DM. Their annihilation will be followed by the Feynman graph on the RHS of Fig.~\ref{fig:rhnann} with

\bea
\langle\sigma v\rangle_{n_{1,2L}}= \frac{f_{\Delta}^4}{64\pi}\left(1-\frac{m_{\Delta}^2}{m_{n_{1,2L}}^2}\right)^{3/2}\left(\frac{m_{n_{1,2L}}^2}{\left(2 m_{n_{1,2L}}^2-m_{\Delta}^2\right)^2}+\frac{1}{2}\frac{m_{n_{1,2L}}^2}{\left(2 m_{n_{2,1L}}^2-m_{\Delta}^2\right)^2}\right),
\eea

which is of the same order as that of $\langle\sigma v\rangle_{n_{1R}}$ in the limit of $f_{\Delta}\sim\mathcal{O}(1)$. So the relic density contribution is again small and can be neglected. The advantage of the heavy neutrinos as DM is that they lack a tree-level direct search interaction and do not alter the conclusions for the chosen BPs.


\section{Collider Phenomenology} 
\label{sec:collider pheno}

Out of all the BSM particles introduced in the model, only the scalar bi-doublet $\zeta$ transforms under SM $SU(2)_L$, and one can produce both the charged ($\zeta_{1,2}^{\pm}$) and neutral components ($\zeta_{1,2}^{0}$) at the collider. The Feynman graphs for the production of such particles in the Large Hadron Collider (LHC) is shown in Fig.~\ref{fig:scalarproduction}. These processes involve derivative couplings arising from the gauge kinetic term as detailed in Appendix-B. Here we have elucidated  two different processes which yield leptonic final states. One possibility is the charged current production of $\zeta_1^{\pm},\zeta_1^{0}$ shown in the left panel of Fig.~\ref{fig:scalarproduction}, and the other possibility is to have a neutral current production of  $\zeta_1^{\pm},\zeta_1^{\mp}$. Subsequent decays of these scalars to SM fermions and to RHN $n_{1R}$ via the Yukawa interactions enlisted in Eq.~\ref{eq:yukawaint}, are also shown in the figure. Here we assume the same mass hierarchy as chosen for DM phenomenology: $m_{\zeta_2}<m_X<m_{\zeta_1}$ and $m_{n_{1R}}<m_{\zeta_1}$. In such a hierarchy, the decay of the scalar bi-doublet components occur with 100 $\%$ branching ratios to the final states: $\zeta_1^{\pm} \to \ell^{\pm}n_{1R}$ and $\zeta_1^0 \to \nu n_{1R}$, as shown in the figure.

\subsection{Signals at LHC}
\label{sec:modelsignals}

\begin{figure}[htb!]
$$
\includegraphics[scale=0.7]{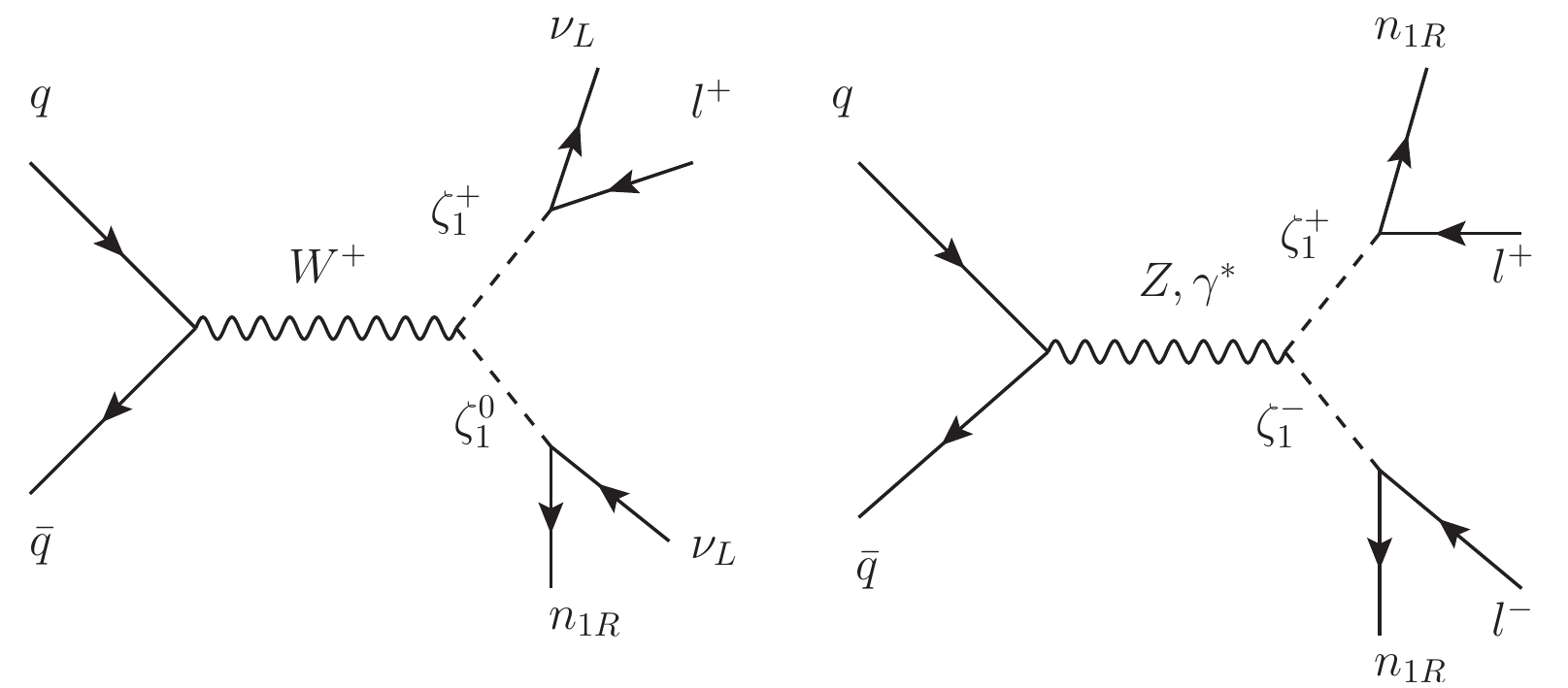}
$$
\caption{Figure showing production of heavy charged scalars and their subsequent decays into a hadronically quiet single lepton $\ell^{\pm}+ \slashed{E}_T~~$ channel (on left) and hadronically quiet opposite sign dilepton channel $\ell^+\ell^- + \slashed{E}_T ~~$ on right.}
\label{fig:scalarproduction}
\end{figure}

Following the mass hierarchy, the production of the scalar bi-doublets at LHC will end up with two different leptonic final states:

\begin{itemize}
 \item Single lepton plus missing energy $\left(1 \ell^{\pm}+ \slashed{E}_T\right)$ due to charged current interaction, as shown in the left panel of Fig.~\ref{fig:scalarproduction}.

 \item Opposite sign di-lepton plus missing energy (OSD+$\slashed{E}_T$) due to neutral current interaction, as shown in the right panel of Fig.~\ref{fig:scalarproduction}.
\end{itemize}

These channels are essentially hadronically quiet, as they contain no jets at the parton level, except for those which may arise due to initial state radiation (ISR). We will therefore focus only on the leptonic final states with zero jet veto, as we know, they are cleaner and suffer less from SM background contamination. We will analyze these two hadronically quiet lepton channels in details for the chosen benchmark points as in Table \ref{tab:bp}. We also note here, that the right handed neutrinos are stable for the chosen hierarchy, and hence contribute to missing energy. As has already been stated, given the interactions proposed in this model, the other two DMs, namely $ \Delta, X$ are harder to produce, if not impossible. Before proceeding further we would also like to note, since both $\zeta_1$ and $\zeta_2$ belong to the same bi-doublet, $\zeta_2^\pm, \zeta_2^0$ can also be produced in the collider via similar diagrams. But since $\zeta_2^{0}$ mixes with SM Higgs, it decays to $b\bar{b}$, while the charged companion will decay $\zeta_2^{\pm} \to \zeta_2^0 \ell^{\pm}\nu$ through off-shell $W$. The missing energy distribution for such a case is identical to those of SM and provides no way to distinguish the signal from background. Therefore, we will refrain from discussing $\zeta_2$ production in details here, although the outcomes are mentioned in subsection~\ref{sec:zetaatlhc}. 

\begin{figure}[htb!]
 $$
 \includegraphics[scale=0.7]{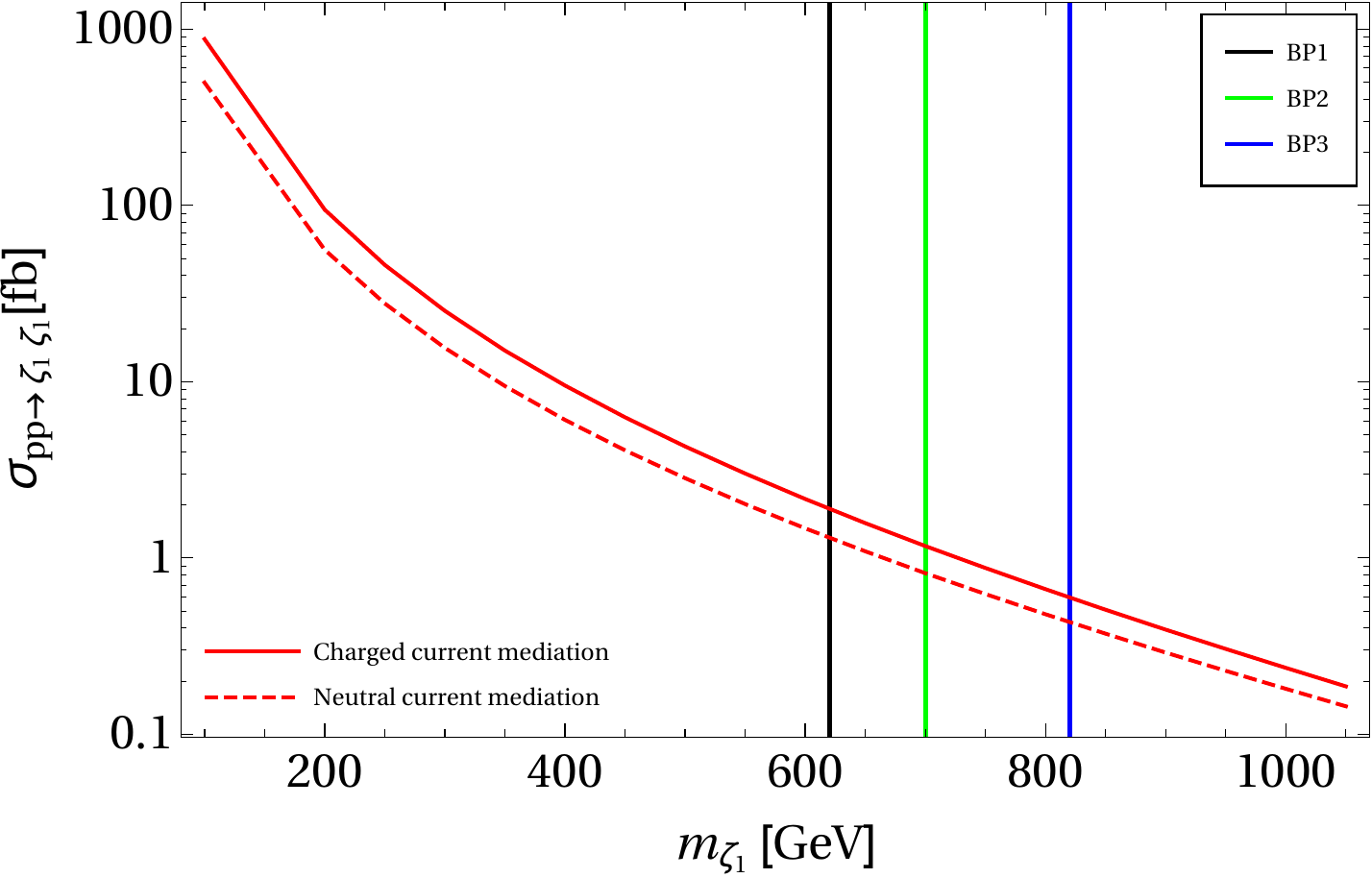}
 $$
 \caption{The variation of production cross section of $pp\to \zeta_1^{\pm}\zeta_1^{0},\zeta_1^{+}\zeta_1^{-}$ via charged current (red bold curve) and neutral current (red dashed curve) interaction at the LHC with $E_{cm}=$14 TeV. {\tt CTEQ6l} has been chosen as parton distribution function for generating the curves. The vertical lines in black, green and blue show the masses of $\zeta_1$ chosen for BP1, BP2 and BP3 respectively.}
 \label{fig:ccnc}
\end{figure}

The final state signal event rates are primarily dictated by the production of the scalar bi-doublet components at the LHC. In Fig.~\ref{fig:ccnc} we have shown the variation of the production cross section of $\zeta_1$ at the LHC with respect to its mass $m_{\zeta_1}$ with $E_{cm}=$14 TeV. We have not addressed the mass difference of the charged and neutral components which might appear from loop corrections and assumed them in the same ballpark. It is evident, with larger $m_{\zeta_1}$, the cross section falls due to phase space suppression, which is clearly visible from the plot. Noteworthy feature here is that the charged current interaction dominates over the neutral current one since the coupling strength is larger in the former case, contrary to SM fermions. Here, the ratio of the vertices in $Z$-mediation to that of $W$-mediation goes as $\sim\frac{cos 2\theta_{w}}{\sqrt{2}cos\theta_{w}}<1$. The difference in the production of charged current process versus neutral current process is also evident from Fig.~\ref{fig:ccnc}. Hence it is expected that this will give rise to larger $1 \ell^{\pm}+ \slashed{E}_T$ events over OSD+$\slashed{E}_T$ events. The chosen benchmark points, as in Table \ref{tab:bp}, are also indicated in Fig.~\ref{fig:ccnc}, from which we can clearly see that the production cross-section is already reduced to $\sim 1$ fb or less at $E_{cm}=$14 TeV LHC. We use {\tt CTEQ6l}~\cite{Placakyte:2011az} as a representative parton distribution function for generating this graph. The event simulation methodology is further detailed in the next subsection. 

\subsection{Simulation technique and event selection criteria}
\label{sec:simultech}

We implemented this model in {\tt CalcHEP}~\cite{Belyaev:2012qa} to generate the parton level events and then the events are fed into {\tt Pythia-6.4}~\cite{Sjostrand:2006za} for showering and hadronization. We have simulated all the events at $\sqrt{s}=14$ TeV using {\tt CTEQ6l} as the parton distribution function. To mimic the experimental environment of the LHC, we have reconstructed all the leptons and jets using the following criteria:

\begin{itemize}
 \item {\it Lepton ($l=e,\mu$):} Leptons are required to have a minimum transverse momentum $p_T>20$ GeV and pseudorapidity $|\eta|<2.5$. Two leptons are isolated objects if their mutual distance in the $\eta-\phi$ plane is $\Delta R=\sqrt{\left(\Delta\eta\right)^2+\left(\Delta\phi\right)^2}\ge 0.2$, while the separation between a lepton and a jet has to satisfy $\Delta R\ge 0.4$.
 
 \item {\it Jets ($j$):} All the partons within $\Delta R=0.4$ from the jet initiator cell are included to form the jets using the cone jet algorithm {\tt PYCELL} built in {\tt PYTHIA}. We require $p_T>20$ GeV for a clustered object to be considered as jet. Jets are isolated from unclustered objects if $\Delta R>0.4$.   
 
 \item {\it Unclustered Objects:}  All the final state objects which are neither clustered to form jets, nor identified as leptons, belong to this category. All particles with $0.5<p_T<20$ GeV and $|\eta|<5$, are considered as unclustered.
 
 \item {\it Missing Energy ($\slashed{E}_T$):} The transverse momentum of all the missing particles (those are not registered in the detector) can be estimated from the momentum imbalance in the transverse direction associated to the visible particles. Thus missing energy (MET) is defined as:
 
 \bea
 \slashed{E}_T = -\sqrt{(\sum_{\ell,j} p_x)^2+(\sum_{\ell,j} p_y)^2},
 \eea
 where the sum runs over all visible objects that include the leptons, jets and the unclustered components. 
 
 \item $H_T$: $H_T$ is defined as the scalar sum of all isolated jet and lepton $p_T$'s:
 
 \bea
 H_T = \sum_{\ell,j} p_T 	
 \eea

\end{itemize}

\begin{figure}[htb!]
$$
\includegraphics[scale=0.55]{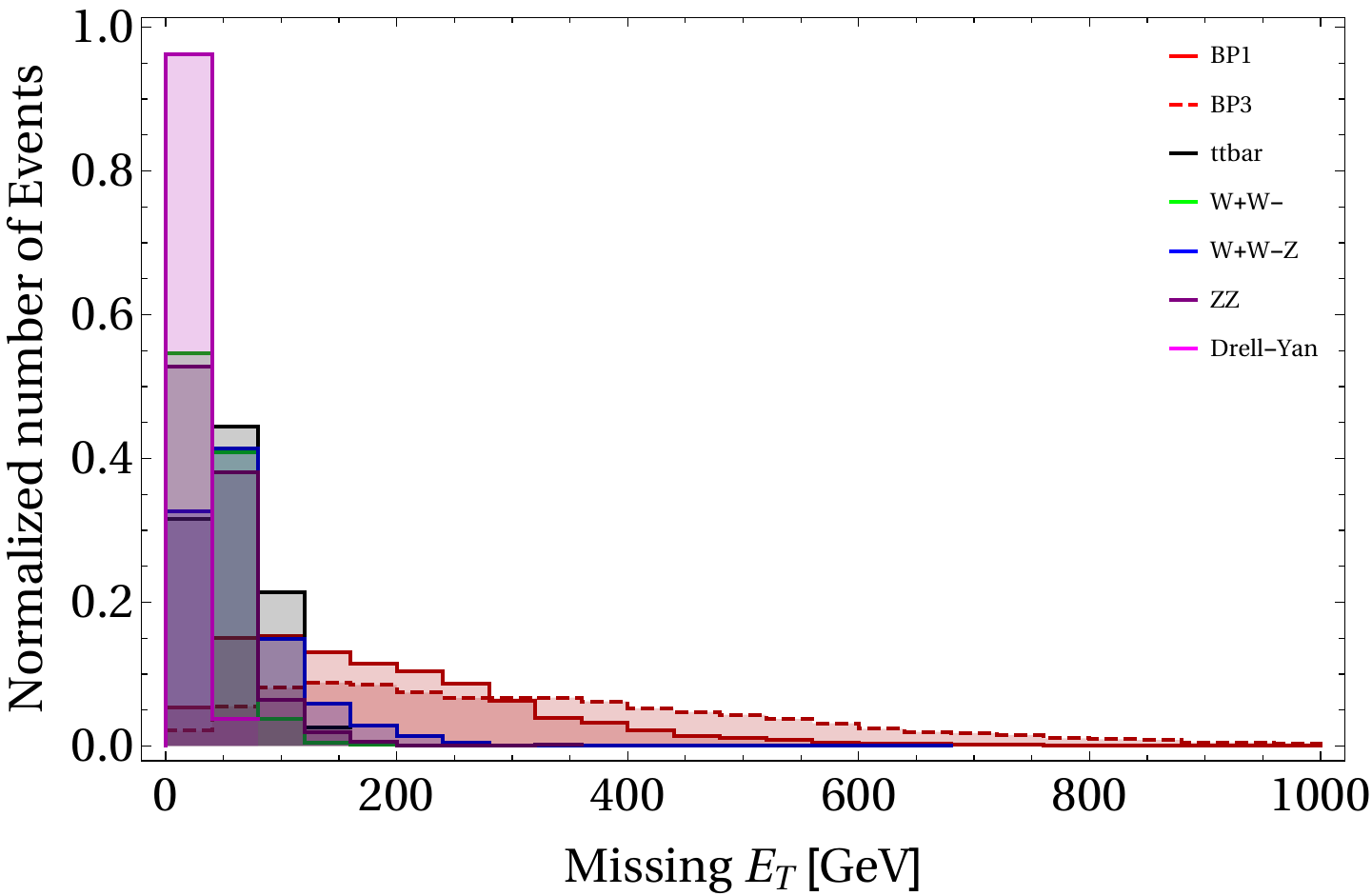}
\includegraphics[scale=0.55]{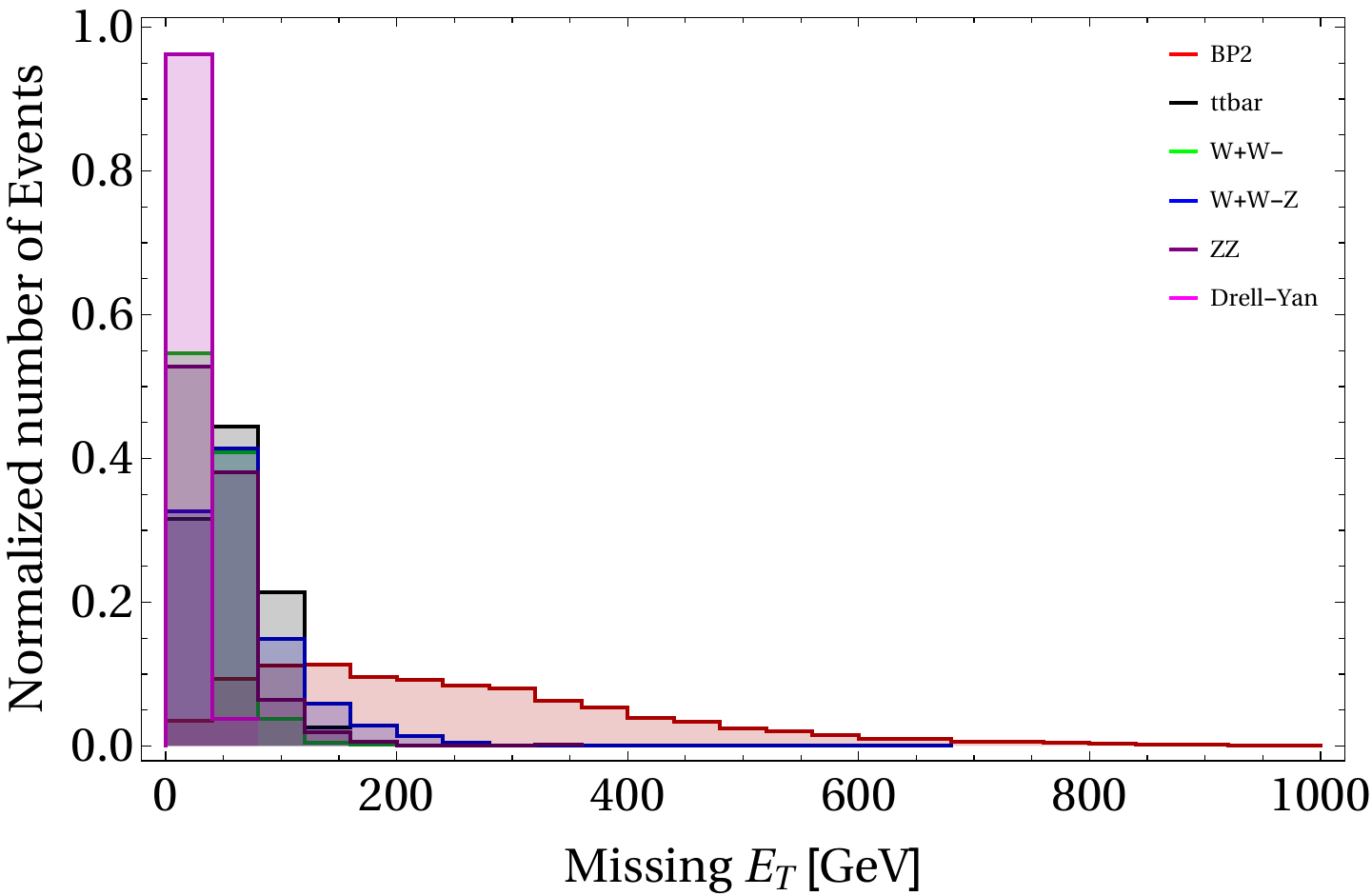}
$$
$$
\includegraphics[scale=0.55]{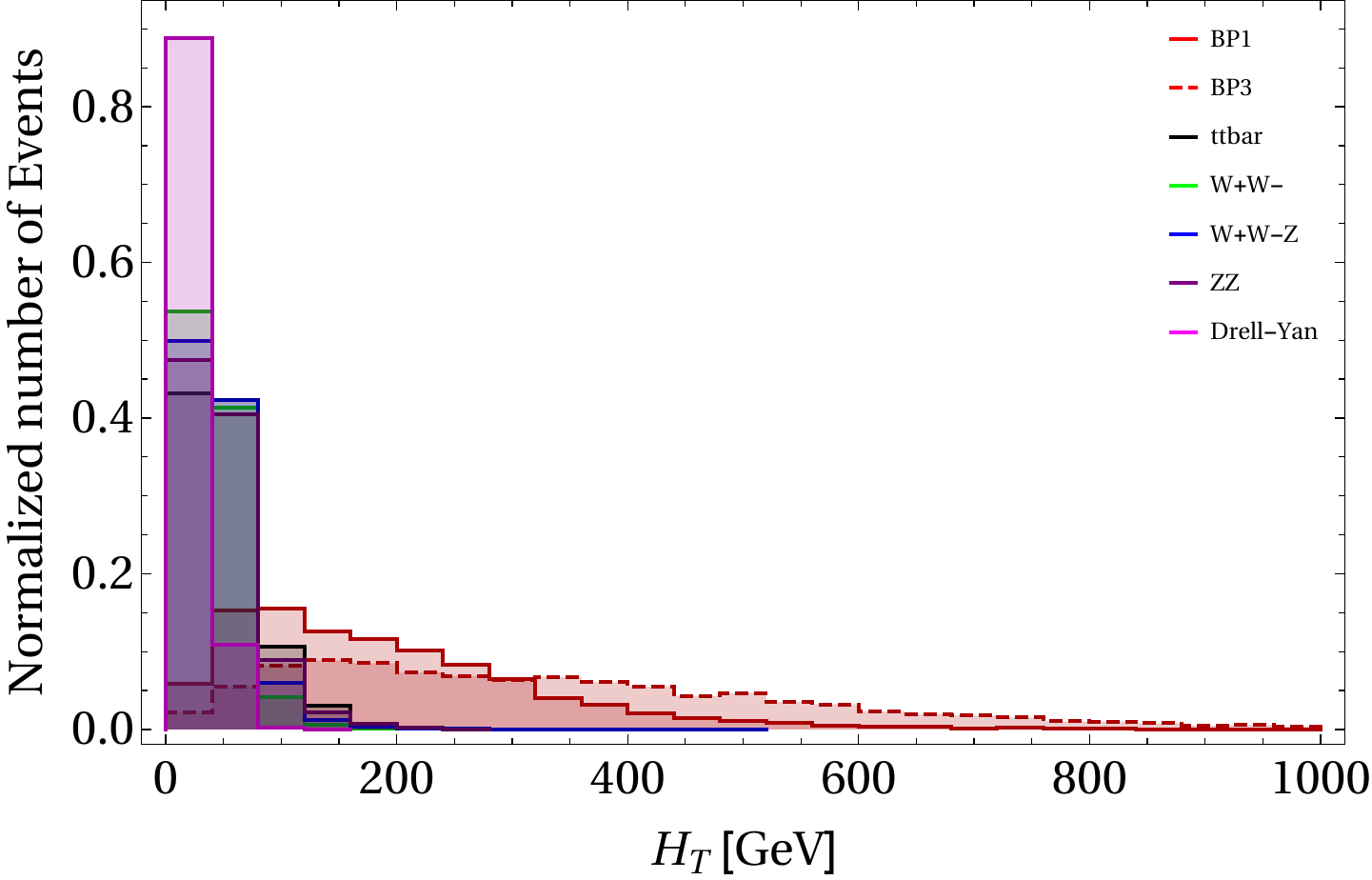}
\includegraphics[scale=0.55]{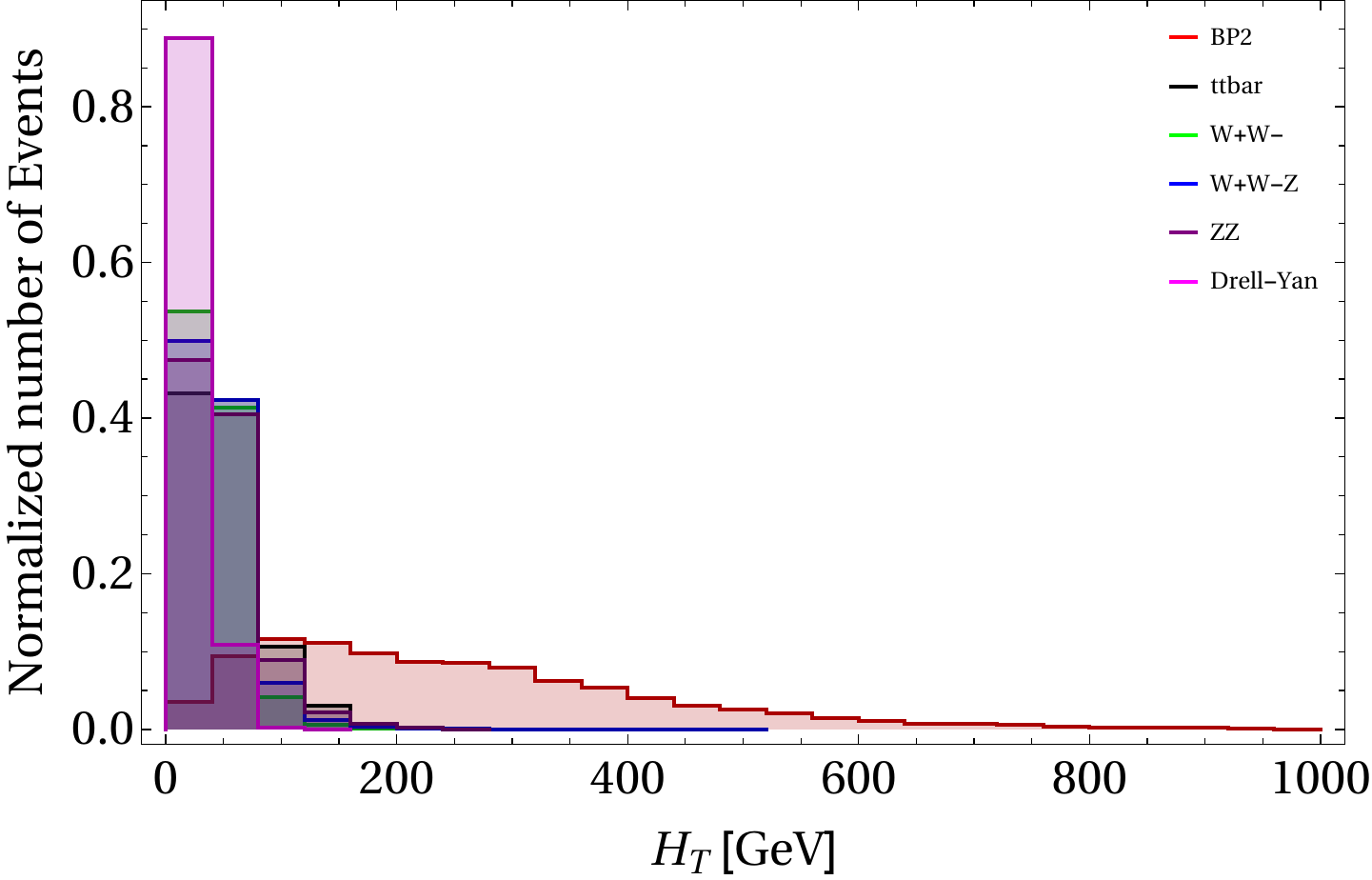}$$
\caption{Top: Missing energy distribution for $1 \ell^{\pm}+\slashed{E}_T$ final state for the benchmark points (BP1, BP2, BP3) are shown in red. Those of the dominant SM backgrounds are also shown. Bottom: $H_T$ distribution for the same. The simulation is done assuming LHC with $\sqrt s=14$ TeV.}
\label{fig:met1}
\end{figure}

\begin{figure}[htb!]
$$
\includegraphics[scale=0.55]{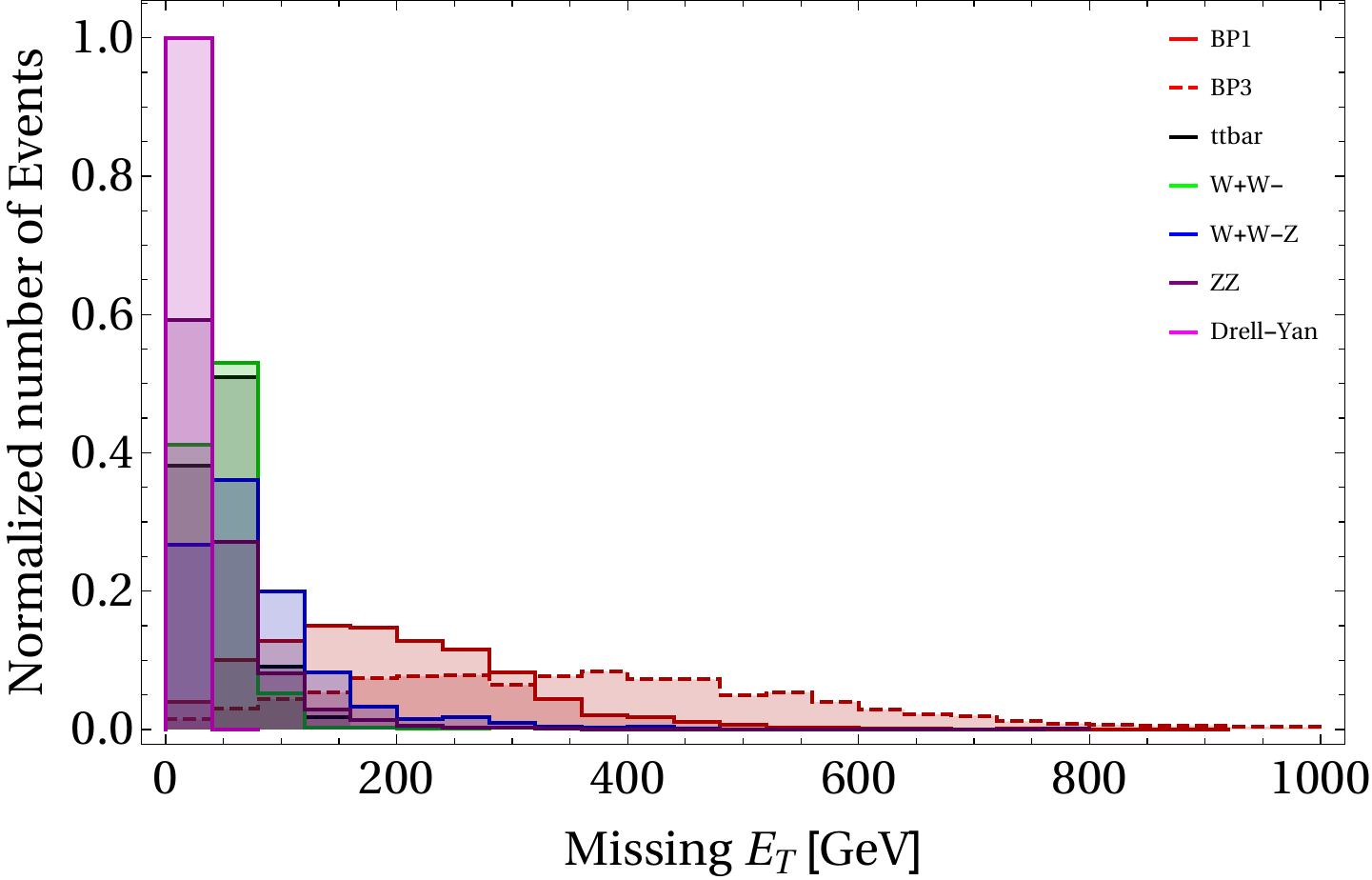}
\includegraphics[scale=0.55]{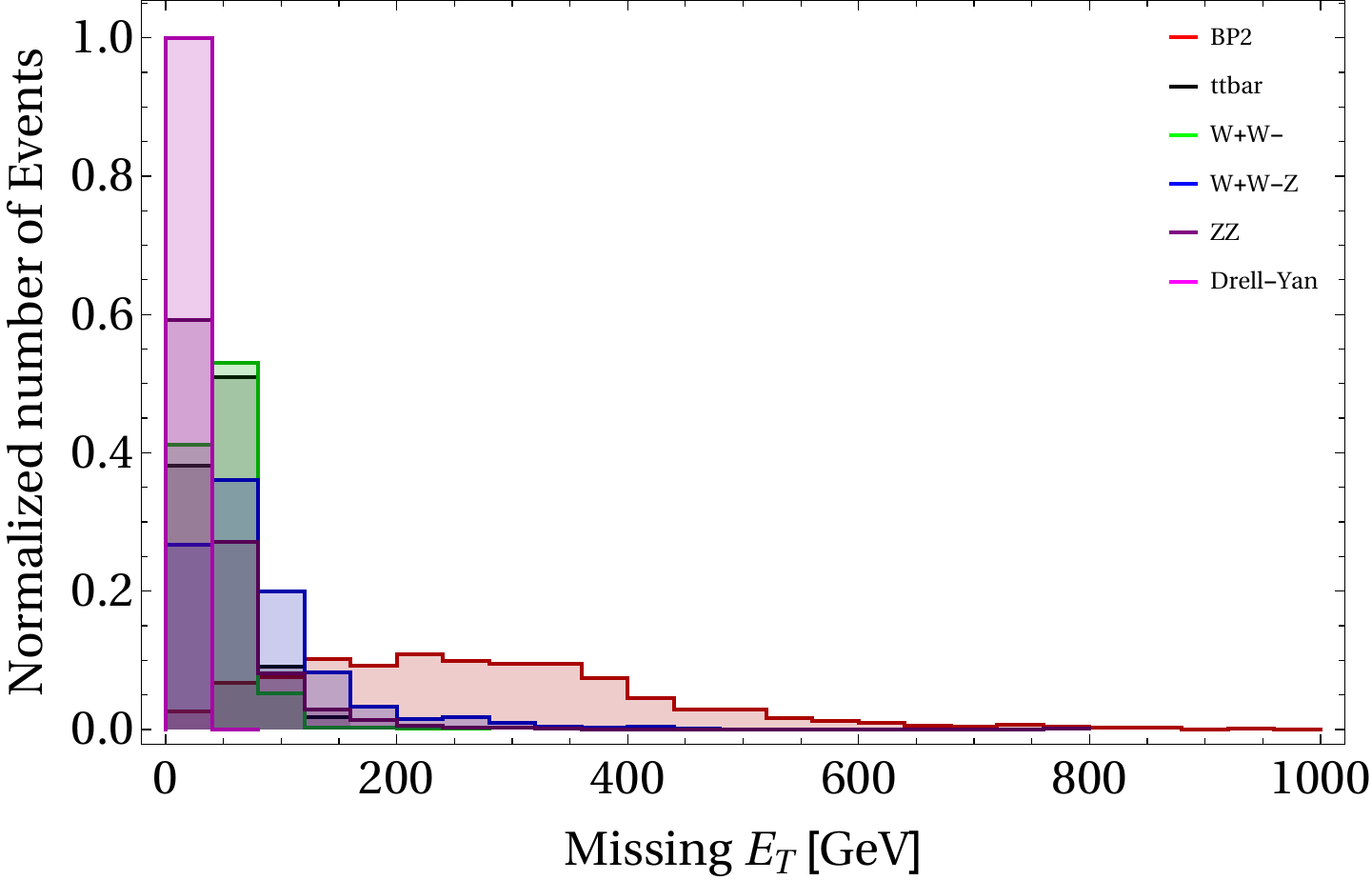}
$$
$$
\includegraphics[scale=0.55]{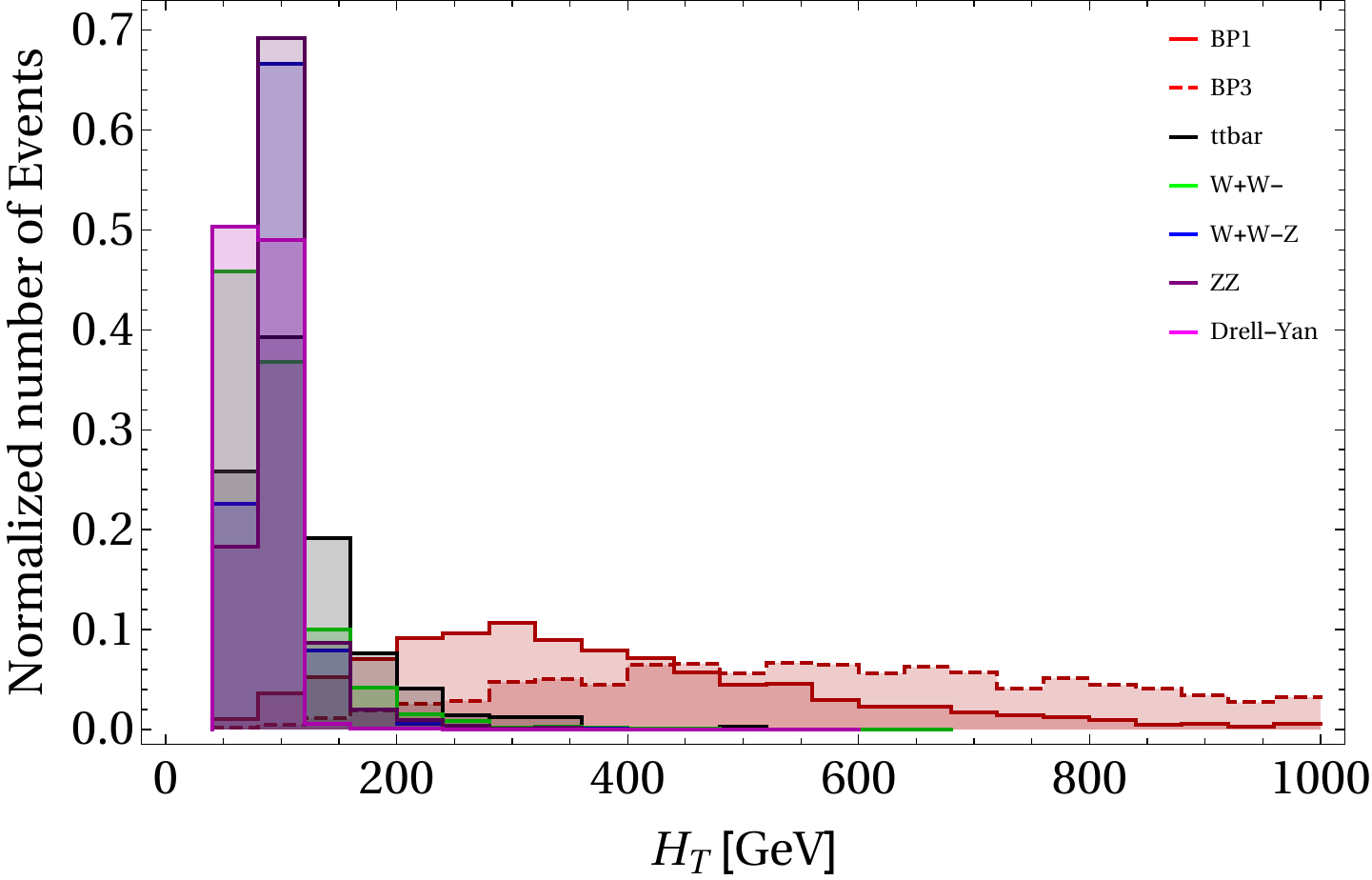}
\includegraphics[scale=0.55]{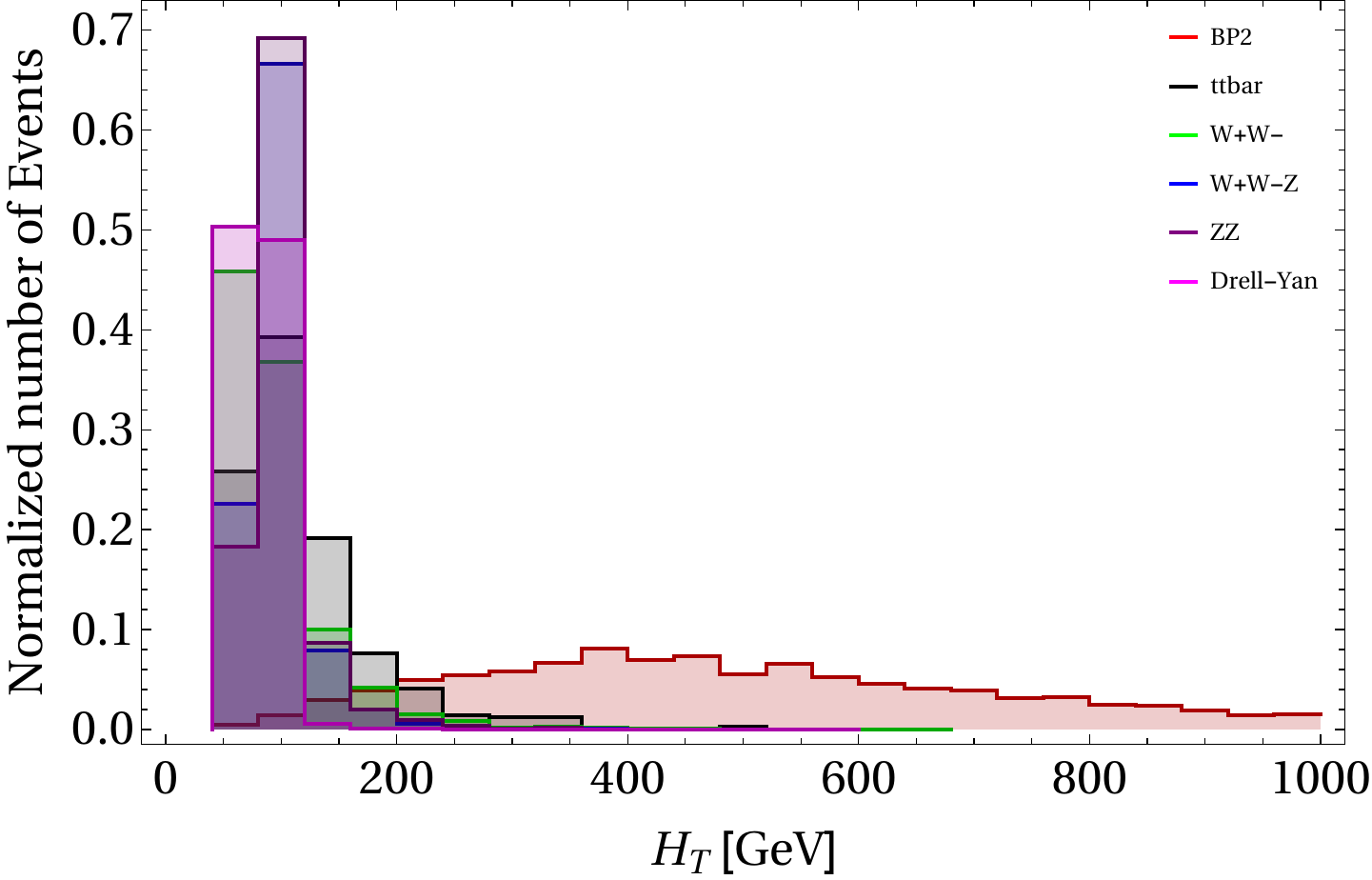}$$
\caption{Top: Missing energy distribution for $ \ell^{\pm}\ell^{\mp}+\slashed{E}_T$ final state for the benchmark points (BP1, BP2, BP3) are shown in red. Those of the dominant SM backgrounds are also shown. Bottom: $H_T$ distribution for the same. The simulation is done assuming LHC with $\sqrt s=14$ TeV.}
\label{fig:met2}
\end{figure}

The dominant SM backgrounds have been generated in {\tt MADGRAPH}~\cite{Alwall:2014hca} and then showered through {\tt PYTHIA}. Appropriate $K$-factors were used to incorporate the Next-to-Leading order (NLO) cross section for the backgrounds. Dominant SM backgrounds for the chosen signal are: $t\bar{t}$, $W^{+}W^{-}$, $W^{\pm}Z$, $ZZ$ and $Drell-Yan$. Since the backgrounds dominate over the signal, MET and $H_T$ cut has to be chosen in a sensible way such that it eliminates most of the backgrounds, while retaining the signal. For the backgrounds the contribution to MET comes from the neutrinos, while for the signal, it comes dominantly from the stable RHN (as shown in Fig.~\ref{fig:scalarproduction}). The MET distribution for the chosen BPs are plotted in the upper panel of Fig.~\ref{fig:met1} for single lepton and in Fig.~\ref{fig:met2} for OSD final states. Corresponding $H_T$ distributions are also shown in the lower panel of the same figures. In both cases, the dominant SM backgrounds are also shown. As it is clear from both Fig.~\ref{fig:met1} and Fig.~\ref{fig:met2}, a high MET cut can reduce SM background, while retaining some of the signal strength in both single and two lepton channels. This is also true for $H_T$-cut as well. Therefore, the final state event selection required to have the following selection criteria on top of the trigger level cuts:


\begin{itemize}
 \item Missing energy cut of $\slashed{E}_T>$ 100, 200 and 300 GeV have been employed in both single and two-lepton cases to reduce SM backgrounds.
 \item $H_T$ cut of 200 and 300 GeV are also applied on top of MET cut to reduce the backgrounds further.
 \item For OSD events, an invariant mass cut over the $Z$-window $|m_z-15|<m_{ll}<|m_Z+15|$ has been applied to get rid-off the $ZZ$ background to a significant extent.
\end{itemize}

\subsection{Event rates for the signal and the SM background}

Cross sections of $1 \ell^{\pm}+\slashed{E}_T$ and $ \ell^{\pm}\ell^{\mp}+\slashed{E}_T$ channels and corresponding number of events at a luminosity $\mathcal{L}=100~fb^{-1}$ for $E_{CM}=14$ TeV at the LHC are listed in Table.~\ref{tab:signal} for the Benchmark points (BP1, BP2, BP3). Here, the production cross-sections are also mentioned, so that we see the sensitivity of the missing energy cut as has been used with $\slashed {E}_T>100, ~200,~300$ GeVs.

\begin{table}[htb!]
\begin{center}
\scalebox{0.85}{
\begin{tabular}{|c|c|c|c|c|c|c|c|c|}
\hline

Benchmark Point & $\sigma_{\zeta_1^\pm\zeta_1^0}$ (fb) & $\sigma_{\zeta_1^{0}\bar{\zeta_1^0}}/\sigma_{\zeta_1^{+}\zeta_1^{-}}$ (fb) & $\slashed{E}_T$ (GeV) & $H_T$ (GeV) & $\sigma^{\ell\pm}$ (fb) & $N^{\ell\pm}_{\text{eff}}$ & $\sigma^{\text{OSD}}$ & $N^{\text{OSD}}_{\text{eff}}$ \\
\hline\hline 

& &  & $>$100 & $>$ 100 & 0.25  & 25 & 0.04 & 4 \\
&&&& $>$ 200 & 0.14 & 14 & 0.04 & 4 \\
&&&& $>$ 300 & 0.06 & 6 & 0.03 & 3 \\
\cline{4-9}

BP1 & 1.89 & 1.29 & $>$200 & $>$ 100 & 0.15 & 15 & 0.02 & 2 \\
&&&& $>$ 200 & 0.14 & 14 & 0.02 & 2 \\
&&&& $>$ 300 & 0.06 & 6 & 0.01 & 1 \\
\cline{4-9}

&  & & $>$300 & $>$ 100 & 0.06 & 6 & 0 & 0\\
&&&& $>$ 200 & 0.06 & 6 & 0 & 0\\
&&&& $>$ 300 & 0.06 & 6 &0 & 0\\
\cline{4-9}

\hline 

& &  & $>$100 & $>$ 100 & 0.17 & 17 & 0.03 & 3 \\
&&&& $>$ 200 & 0.12 & 12 & 0.03 & 3 \\
&&&& $>$ 300 & 0.07 & 7 & 0.02 & 2 \\
\cline{4-9}

BP2 & 1.16 & 0.81 & $>$200 & $>$ 100 & 0.12 & 12 & 0.02 & 2  \\
&&&& $>$ 200 & 0.11 & 11 & 0.02 & 2 \\
&&&& $>$ 300 & 0.07 & 7 & 0.02 & 2 \\
\cline{4-9}

&  & & $>$300 & $>$ 100 & 0.07 & 7  & 0.01 & 1 \\
&&&& $>$ 200 & 0.07 & 7 & 0.01 & 1 \\
&&&& $>$ 300 & 0.07 & 7 & 0.01 & 1 \\
\cline{4-9}

\hline 

& &  & $>$100 & $>$ 100 & 0.09 & 9 & 0.02 & 2 \\
&&&& $>$ 200 & 0.07 & 7 & 0.02 & 2 \\
&&&& $>$ 300 & 0.05 & 5 & 0.02 & 2 \\
\cline{4-9}

BP3 & 0.59 & 0.43 & $>$200  & $>$ 100 & 0.07 & 7 &  0.01 & 1\\
&&&& $>$ 200 & 0.07 & 7 & 0.01 & 1 \\
&&&& $>$ 300 & 0.05 & 5 & 0.01 & 1 \\
\cline{4-9}

&  & & $>$300 & $>$ 100 & 0.05  & 5 & 0.01 & 1 \\
&&&& $>$ 200 & 0.05 & 5 & 0.01 & 1 \\
&&&& $>$ 300 & 0.05 & 5 &  0.01 & 1 \\
\cline{4-9}
 
\hline
\end{tabular}
}
\end{center}
\caption {Signal events with $\sqrt{s}$ = 14 TeV at the LHC for luminosity $\mathcal{L} = 100~fb^{-1}$ for benchmark points (BP1, BP2, BP3). The variation of number of final state signal events with cut-flow are also tabulated.}
\label{tab:signal}
\end{table}

\begin{table}[htb!]
\begin{center}
\begin{tabular}{|c|c|c|c|c|c|c|c|}
\hline
Process & $\sigma_{\text{production}}$ (pb) & $\slashed{E}_T$ (GeV) & $H_T$ (GeV) & $\sigma^{\ell\pm}$ (fb) & $N^{\ell\pm}_{\text{eff}}$ & $\sigma^{\text{OSD}}$ & $N^{\text{OSD}}_{\text{eff}}$ \\
\hline\hline 

%
%

& & $>$100 & $>$100 & 22.80 & 2280 & 17.10 & 1710  \\

&&& $>$200 & 1.62 & 162 & 2.44 & 244\\

&&& $>$300 & $<$ 0.81 & $<$ 1 & $<0.81$ & $<$1 \\

\cline{3-8}

$t\bar {t}$ & 814.64 & $>$200  & $>$ 100 &  1.62 & 162   & $<0.81$  & $<$ 1 \\

&&& $>$200 & 0.81 & 81 & $<0.81$ & $<$ 1 \\

&&& $>$300 & $<0.81$ & $<$ 1 & $<0.81$ & $<1$ \\

\cline{3-8}

&  & $>$300  & $>$ 100 & $<$ 0.81 & $<$ 1  & $<$ 0.81  & $<$ 1 \\

&&& $>$200 & $<0.81$ & $<$ 1 & $<0.81$ & $<$ 1 \\

&&& $>$300 & $<0.81$ & $<$ 1 & $<0.81$ & $<$ 1 \\

\cline{3-8}
\hline

& & $>$100 & $>$ 100 & 54.48 & 5448 & 20.49 & 2049 \\

&&& $>$200 & 3.99 & 399 & 9.99 & 999 \\

&&& $>$300 &  0.49 & 49 & 1.99 & 199 \\
\cline{3-8}

$W^+ W^-$ & 99.98 & $>$200  & $>$ 100 & 1.99 & 199  & 1.99 & 199  \\

&&& $>$200 & 0.49 & 49 & 1.99 & 199 \\

&&& $>$300 & 0.49 & 49 & 0.49 & 49 \\
\cline{3-8}

&  & $>$300 & $>$ 100 & 0.49 & 49  & $<$ 0.49 & $<$ 1 \\ 

&&& $>$200 & 0.49 & 49 & $<0.49$ & $<$ 1 \\

&&& $>$300 & 0.49 & 49 & $<0.49$ & $<$ 1 \\
\hline

& & $>$100 &$>$ 100 & 0.14  & 14 & 0 & 0 \\

&&& $>$200 & 0.01 & 1 & 0 & 0 \\

&&& $>$300 & 0 & 0 & 0 & 0\\
\cline{3-8}

$W^\pm Z$ & 0.15 & $>$200 & $>$100 & 0.012  & 1 & 0 & 0 \\

&&& $>$200 & 0 & 0 & 0 & 0 \\

&&& $>$300 & 0 & 0 & 0 & 0 \\
\cline{3-8}

&  & $>$300  &$>$ 100 & 0  & 0 & 0 & 0 \\

&&& $>$200 & 0 & 0 & 0 & 0 \\

&&& $>$300 & 0 & 0 & 0 & 0 \\
\hline

& & $>$100 &$>$100 & 7.07 & 707 & 0.21 & 21\\

&&& $>$200 & 0.35 & 35 & 0.14 & 14 \\

&&& $>$300 & $<$ 0.07 & $<$ 1 & 0.07 & 7 \\
\cline{3-8}

$ZZ$ & 14.01 & $>$200  &$>$ 100 & 0.35 & 35 & $<$ 0.07 & $<$ 1 \\
&&& $>$200 & 0.28 & 28 & $<$ 0.07 & $<$ 1 \\

&&& $>$300 & $<$ 0.07 & $<$ 1 & $<$ 0.07 & $<$ 1 \\
\cline{3-8}

&  & $>$300  &$>$100 & $<$ 0.07 & $<$ 1  & $<$ 0.07  & $<$ 1 \\ 
&&& $>$200 & $<$ 0.07 & $<$ 1 & $<$ 0.07 & $<$ 1 \\

&&& $>$300 & $<$ 0.07 & $<$ 1 & $<$ 0.07 & $<$ 1 \\ 
\hline
\end{tabular}
\end{center}
\caption {SM background events at $\sqrt{s}$ = 14 TeV for luminosity $\mathcal{L} = 100~fb^{-1}$ at the LHC. The cross sections have been multiplied by the appropriate $K$-factors to match with their NLO order cross-section (see text for details). The variation of number of final state background events with cut-flow are also tabulated.} 
\label{tab:background}
\end{table}

The first obvious thing to notice is, with heavier masses for the charged bi-doublet scalars for the benchmark points, the cross-section in both the final states diminishes accordingly due to larger phase space suppression. Secondly, the number of events in OSD is smaller than that of single lepton ones, owing to (i) the hierarchy of the charge current and neutral current production cross-section itself and (ii) those cases, where the neutral current production may also yield an effective single lepton event, if one of the leptons is soft and fails to register with the desired $p_T$ cut. The production cross-section is small for all the benchmark points, and so is the numbers of expected events at a luminosity as high as $\sim 100 ~\rm{fb}^{-1}$.  Here, the effective number of final state events is given as:

\bea
N_{\text{eff}} = \frac{\sigma_{\text{p}}\times n}{N}\times\mathcal{L},
\eea

where $n$ is the simulated number of events obtained by simulating $N$ events corresponding to production cross-section of $\sigma_p$ and $\mathcal{L}$ is the integrated luminosity. 

Using the same selection criteria, the number of final state events for dominant SM backgrounds are tabulated in Table.~\ref{tab:background}. Here we have multiplied the cross section in leading order (LO) with the appropriate $K$-factors to obtain the cross section in the NLO approximation. The $K$-factors for the different SM processes are chosen as~\cite{Alwall:2014hca}:  for $t\bar{t}:~K=1.47$, $WW:~K=1.38$, $WZ:~K=1.61$, $ZZj:~K=1.33$, $Drell$-$Yan: ~K=1.2$. Again, in Table.~\ref{tab:background}, we show that we can reduce SM backgrounds to a significant extent, by employing the MET cut.  However, for $\slashed{E}_T> 100, 200$ GeV, a significant number of events are still left from $W^+W^-$ final state. This can only be reduced with $\slashed{E}_T> 300$ GeV for two lepton but single lepton case will still be submerged under a large $W^+W^-$ background events.   

\begin{figure}[htb!]
$$
\includegraphics[scale=0.62]{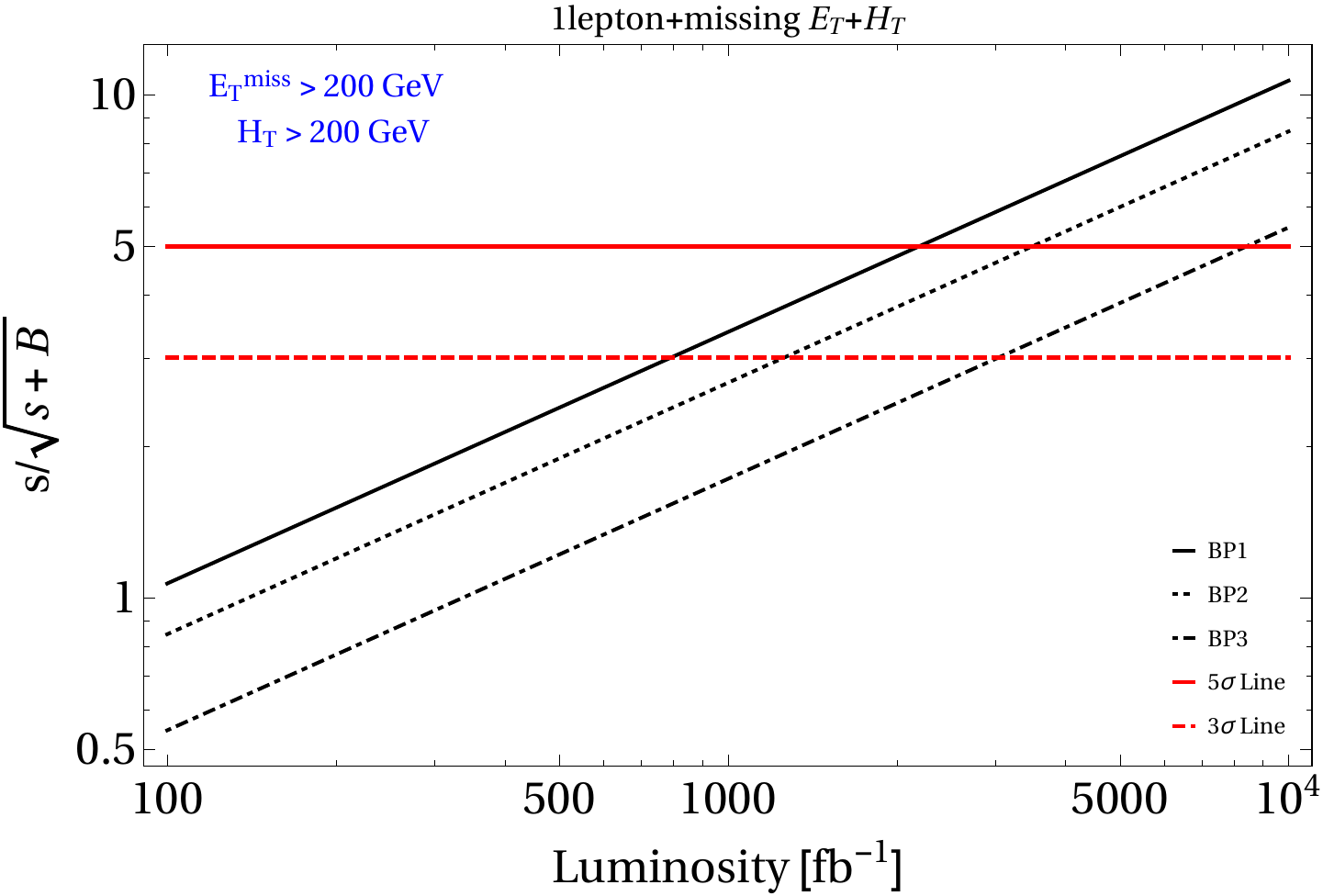}
$$
\caption{Significance plot for the signal $1 \ell^{\pm}+ \slashed{E}_T$ events for the chosen benchmark points in terms of luminosity. The red solid (dashed) line shows the $5\sigma~(3\sigma)$ discovery limit.}
\label{fig:signalsig}
\end{figure}

The main take of this analysis is that, although it is possible to eliminate or at least reduce the SM background with judicious choice of different cuts, but as the production cross section for the signals itself is very low, it is only possible to see such a signal at the LHC at a very high integrated luminosity for both single lepton and opposite-sign-dilepton cases. And as it is evident from Table.~\ref{tab:signal}, the OSD case is even harder to probe at LHC. In Fig.~\ref{fig:signalsig}, we show the significance for the single lepton channel only. As it is seen, with $\slashed{E}_T>$ 200 GeV and $\slashed{H}_T>$ 200 GeV (which removes most of the backgrounds), the significance reaches the discovery limit (5$\sigma$) at a luminosity of  $\gtrsim$ 1000 $fb^{-1}$. As the OSD case fails to reach at least $3\sigma$ confidence even at $\gtrsim 1000~\rm fb^{-1}$ luminosity, we are not showing its discovery potential.



\subsection{Fate of $\zeta_2$ at the LHC}
\label{sec:zetaatlhc}

\begin{figure}[htb!]
 $$
 \includegraphics[scale=0.65]{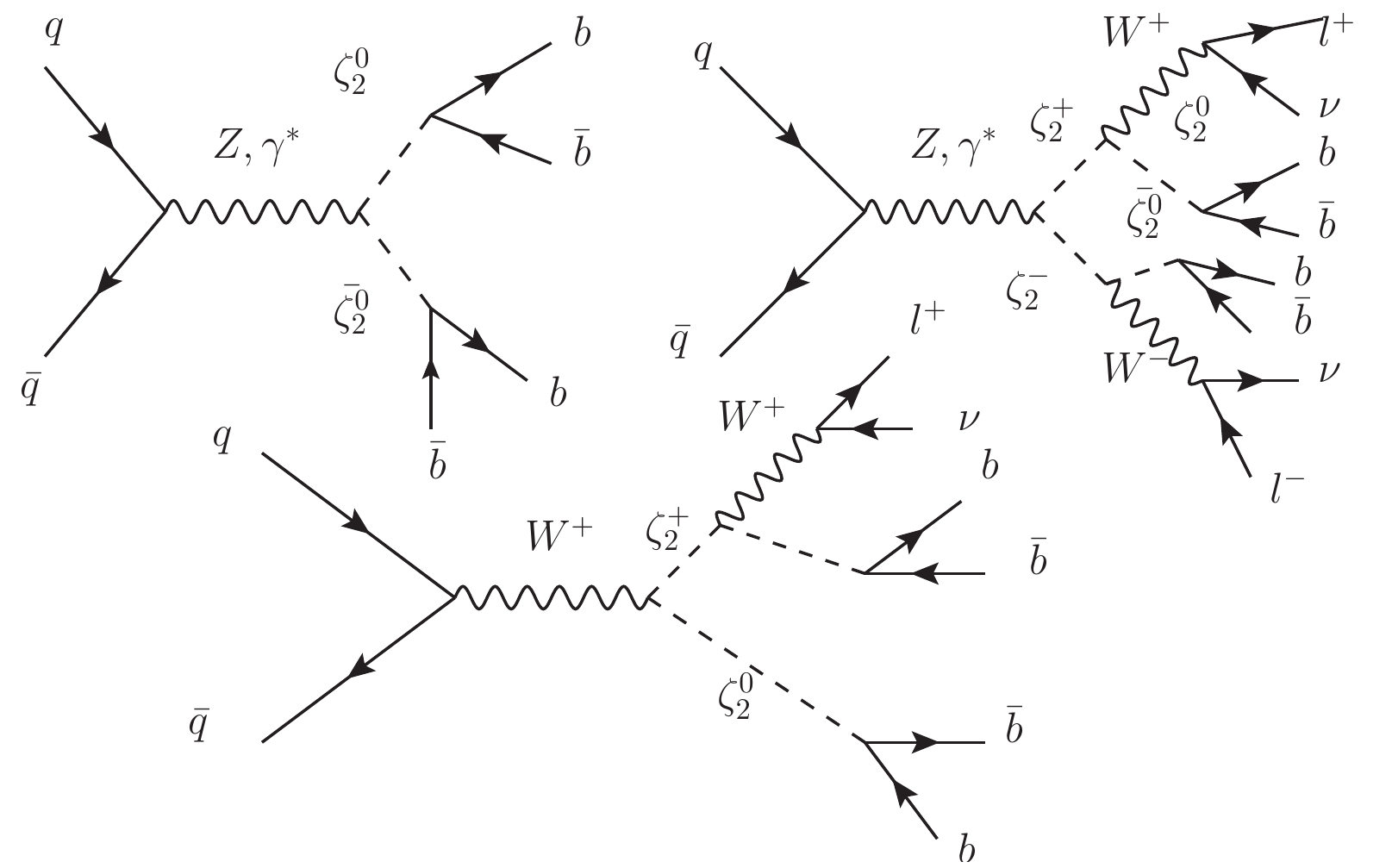}
 $$
 
 \caption{Top Left: Production of $\zeta_2^{0}\bar{\zeta_2^{0}}$ at the LHC via neutral current interaction and their subsequent decays to four $b$-jet final state; Top Right: $\zeta_2^{+}\zeta_2^{-}$ production at the LHC via neutral current interaction leading to opposite sign dilepton plus four $b$-jet final state; Bottom: $\zeta_2^{+}(\zeta_2^{-})\zeta_2^{0}$ production via charged current interaction through $W^{+}(W^{-})$ leading to four $b$-jet final state with single lepton. }
 \label{fig:zetaproduction}
\end{figure}

As $\zeta_2$ is also part of the scalar bi-doublet introduced in the model, it has the same interaction vertices with SM as that of $\zeta_1$. Therefore, it is possible to produce $\zeta_2$ at the LHC through the same channels as $\zeta_1$. However, the decay channels of $\zeta_2$ are different from that of $\zeta_1$. We already have mentioned in Sec.~\ref{sec:model}, that $\zeta_2^{0}$ mixes with the SM Higgs through electroweak (EW) symmetry breaking. As a result, the neutral component, $\zeta_2^0$ will readily decay to SM particles, for example, $b$-jets at the LHC. The charged components $\zeta_2^{\pm}$ will only decay to $\zeta_2^0$ through charged current interactions $\zeta_2^{\pm} \to \zeta_2^0 W^{*\pm} \to \zeta_2^0 \ell^{\pm} \nu$. This is due to the specific hierarchy chosen for the DM analysis of the model : $m_{\zeta_1}>m_{nR}>m_{\zeta_2}$, where right handed neutrinos are assumed to be heavier than $\zeta_2$. Also, note here, due to the small mass difference between $\zeta_2^{\pm}$ with $\zeta_2^0$ (which can happen through loop corrections) the decay of the charged components will always occur through off-shell $W$ bosons. Therefore, different combinations of charged and neutral current productions of $\zeta_2$ in pairs at the LHC will yield the following final states:

\begin{itemize}
\item 4$b$-jets plus no missing energy,
\item 1$\ell^{\pm}$+4$b$-jets+missing energy,
\item $\ell^{\pm}\ell^{\mp}$+4$b$-jets+missing energy.
\end{itemize}
Different productions that lead to such final states are shown in the Feynman graphs in Fig.~\ref{fig:zetaproduction}. One important point to note here, is that the missing energy in the above channels, essentially appear from SM neutrinos and not from the DMs assumed in the framework. So, it is very difficult to segregate these channels from the SM background with a missing energy cut. This is even more true, because of the off-shell decays of $\zeta_2^{\pm}$, which leaves no way to separate the signal from SM missing energy distribution. Therefore, we do not elaborate on  the event level analysis of these channels for the chosen benchmark points at the LHC.  

However, the small mass difference between the charged and the neutral scalar component may aid to a large decay lifetime of $\zeta_2^{\pm}$. This can lead to the observation of one or two displaced vertex signatures or stable charged tracks within the detector.


\section{Conclusion}
\label{sec:conclusion}

In the absence of direct and collider search evidences for WIMP-like DM, an important emerging issue is to address the existence of such DM candidates. While LHC puts a milder limit on heavy WIMP-like DMs due to huge SM background, spin-independent direct search puts a stronger limit on such DM-nucleon interaction. The challenge is to produce right relic density even in the absence of a direct detection signal. Segregation of annihilation processes from that of direct search interaction plays a key role in this context. In particular, annihilation of DM to non-SM particles, DM-DM interaction and co-annihilation serve as crucial features to save WIMP-like DM candidates. The paper exemplifies one such case with a detailed parameter space scan.

The model of our interest is an $SU(2)_N$ gauge extension of SM as proposed in~\cite{Fraser:2014yga}. The lightest vector boson $X$ is stabilized by an unbroken $S$ charge arising from $SU(2)_N \times S^{'} \to S$ through spontaneous symmetry breaking. We assume symmetry $S$ remains intact up to the Planck scale to avoid constraints coming from CMB, Gamma ray, neutrino flux etc due to the DM decay. The model offers a multipartite DM framework, involving the scalar triplet $\Delta$ and heavy neutrinos, depending on the kinematics.  We highlighted the case of DM-DM interactions to govern thermal freeze-out of the heavier component. For example, when $\{\Delta,X\}$ can both be DM, we show that $X$ always has the larger part of relic density while obeying DD bound (thanks to its annihilation to non-SM particles). $\Delta$ being heavier in such a case, can annihilate to $X$ yielding a larger annihilation cross-section and smaller relic density. However, as the freeze-out is mainly governed by DM-DM interaction, it is saved from direct search.


We also have explored the possibility of collider search of such models and we see that this model might manifest itself through hadronically quiet leptonic final states along with missing energy in the LHC. This can come from the production and subsequent decays of the scalar bi-doublets assumed in the theory. Although, the missing energy (MET) and $H_T$ allows us to separate the signal from the SM background, the small strength of EW production cross-section does not allow the model to be accessible in the next run of the LHC, postponing it to a large luminosity regime. We also note here, that as no SM particle are charged under the additional $SU(2)_N$, the phenomenology is in sharp contrast to what we obtained in~\cite{Barman:2017yzr}. This is even more true for collider signatures, as the model studied here, could not even produce the vector DM at LHC.

The model also addresses the generation of neutrino masses through inverse see-saw mechanism by assuming the presence of heavy chiral neutrinos $(n_1,n_2)$. The proportionality of neutrino mass to the vev $\langle \Delta_3\rangle$, requires it to have a small value thus making the vector boson ($X$) degenerate with $X_3$. This predicts additional contribution to the thermal freeze-out of $X$ DM through $X-X_3$ co-annihilation. Hence the model offers a very interesting connection between the neutrino sector and DM phenomenology. Secondly, the presence of inverse seesaw mechanism to generate light neutrino masses, also allows one to assume the heavy neutrinos to be within $\sim \mathcal{O}$(TeV). This yields the only possibility of exploring the model in the LHC through multi lepton channels. As the scalar bi-doublet can only decay to the right handed neutrinos along with SM leptons (through Yukawa interactions)
, a very heavy neutrino can stop such a decay chain and allows one to see displaced vertex signature or stable charge tracks only.           

\section{Acknowledgements}
\label{sec:ack}
We would like to acknowledge discussions with Joydeep Chakrabortty at different stages of this work. SB would like to acknowledge the DST-INSPIRE research grant IFA-13 PH-57. BB would acknowledge the hospitality at IIT Kanpur, where a major part of the work was carried out. BB would also like to thank Sunando Patra for technical help with Mathematica and Amit Duttabanik for useful discussions.  

\section{Appendix A: Scalar Potential}

The scalar potential of the model is given by:
\begin{equation}
\begin{split}
V &= \mu_\zeta^2 Tr(\zeta^\dagger \zeta) + \mu_\Phi^2 \Phi^\dagger \Phi + \mu_\chi^2 
\chi^\dagger \chi + \mu_\Delta^2 Tr(\Delta^\dagger \Delta) + (\mu_1 
\tilde{\Phi}^\dagger \zeta \chi + \mu_2 \tilde{\chi}^\dagger \Delta \chi + H.c.)\\ 
&+ {1 \over 2} \lambda_1 [Tr(\zeta^\dagger \zeta)]^2 + {1 \over 2} \lambda_2 (\Phi^\dagger \Phi)^2 + {1 \over 2} \lambda_3 Tr(\zeta^\dagger \zeta \zeta^\dagger \zeta) + {1 \over 2} \lambda_4 (\chi^\dagger \chi)^2 + 
{1 \over 2} \lambda_5 [Tr(\Delta^\dagger \Delta)]^2 \\ 
&+ {1 \over 4} \lambda_6 Tr(\Delta^\dagger \Delta - \Delta \Delta^\dagger)^2  + f_1 \chi^\dagger \tilde{\zeta}^\dagger \tilde{\zeta} \chi + f_2 \chi^\dagger \zeta^\dagger \zeta \chi + f_3 \Phi^\dagger \zeta \zeta^\dagger \Phi + f_4 \Phi^\dagger \tilde{\zeta} \tilde{\zeta}^\dagger \Phi   \\ 
&+ f_5 (\Phi^\dagger \Phi)(\chi^\dagger \chi) + f_6 (\chi^\dagger \chi) Tr(\Delta^\dagger \Delta) + f_7 \chi^\dagger 
(\Delta \Delta^\dagger - \Delta^\dagger \Delta) \chi + f_8 (\Phi^\dagger \Phi) Tr(\Delta^\dagger \Delta)\\ 
&+ f_9 Tr(\zeta^\dagger \zeta) Tr(\Delta^\dagger \Delta) + f_{10} Tr[\zeta(\Delta^\dagger \Delta - \Delta \Delta^\dagger) \zeta^\dagger],
\end{split}
\end{equation}
where
\begin{equation}
\tilde{\Phi}^\dagger = (\phi^0, -\phi^+), ~~~ \tilde{\chi}^\dagger = 
(\chi_2, -\chi_1), ~~~ \tilde{\zeta} = \begin{pmatrix}
\zeta_2^+ & - \zeta_1^+ \cr 
-\bar{\zeta}_2^0 & \bar{\zeta}_1^0\end{pmatrix}.
\end{equation}

\section{Appendix B: Gauge Interactions and masses of the scalar triplets}

Covariant derivative for the scalar bi-doublet: 

\bea
\mathcal{D}_{\mu}\zeta=\partial_{\mu}\zeta-ig_{L}\vec{W_{L_\mu}}\frac{\vec{\tau}}{2}\zeta+ig_{N}\zeta\frac{\vec{\tau}}{2}\vec{X_\mu}.
\eea

Covariant derivative for the $SU(2)_N$ triplet: 

\bea
\mathcal{D}_{\mu}\Delta= \partial_{\mu}\Delta-\frac{i g_N}{2}\left[\vec{\tau}.X_{\mu},\Delta\right].
\eea

Relevant diagrams come from the gauge kinetic terms: 

\bea
\mathcal{L}_{gauge}\supset Tr\left[\left(\mathcal{D}_{\mu}\zeta\right)^{\dagger}\left(\mathcal{D}_{\mu}\zeta\right)\right]+Tr\left[\left(\mathcal{D}_{\mu}\Delta\right)^{\dagger}\left(\mathcal{D}_{\mu}\Delta\right)\right].
\eea

The masses of the scalar triplet $\Delta$ are given by:

\bea
\begin{split}
m^2(\Delta_3)\simeq \mu^2_{\Delta}+(f_6-f_7)u_2^2+f_8 v_1^2\\
m^2(\Delta_2)\simeq \mu^2_{\Delta}+f_6 u_2^2+f_8 v_1^2\\
m^2(\Delta_1)\simeq \mu^2_{\Delta}+(f_6+f_7)u_2^2+f_8 v_1^2.
\end{split}
\eea

\section{Appendix C: Decay of $\Delta_3$ to $b\bar{b}$}

In the rest frame of $\Delta_3$:

\bea
\Gamma_{\Delta_3\to b\bar{b}} = \frac{m_{\Delta_3}f_8^2}{8\pi}\left(1-\frac{m_b^2}{m_{\Delta_3}^2}\right)^{\frac{3}{2}}.
\eea

Now, for $m_b=4.18$ GeV and age of the universe $\sim 4.1\times 10^{17}$ sec, we obtain: $f_8\simeq 2.008\times 10^{-22}$.


\end{document}